\documentclass[aps,10pt,prb,twocolumn,amsmath,amssymb,longbibliography,superscriptaddress,eqsecnum,floatfix]{revtex4-2}

\usepackage{epsfig}
\usepackage{dcolumn}
\usepackage{bm}
\usepackage{bbm}
\usepackage[normalem]{ulem}
\usepackage{color}
\usepackage[colorlinks=true,linkcolor=blue,citecolor=blue,urlcolor=blue,breaklinks=true]{hyperref}%
\usepackage{verbatim}
\usepackage{subfigure}

\usepackage[ruled,vlined]{algorithm2e}
\usepackage{tikz}
\usetikzlibrary{calc}
\usetikzlibrary{shapes}
\usetikzlibrary{arrows.meta}
\tikzset{
  fractal line/.style args={#1 and #2}{%
    to path={
      let
      \p1=(\tikztostart), %
      \p2=(\tikztotarget), %
      \n1={veclen(\x1-\x2,\y1-\y2)}, %
      \p3=($(\p1)!.5!(\p2)$), %
      \p4=(rand*#1*\n1,rand*#1*\n1), %
      \p5=(\x3+\x4,\y3+\y4) %
      in \pgfextra{
        \pgfmathtruncatemacro\mytest{(\n1<#2)?1:0}
        \ifnum\mytest=1 %
        \tikzset{fractal line/.style args={#1 and #2}{line to}}
        \fi
      } to[fractal line=#1 and #2] (\p5) to[fractal line=#1 and #2] (\p2)
    },
  },
}

\usepackage{braket}
\usepackage{listings}
\usepackage{color}
\definecolor{lightgray}{gray}{1}

\begin{document}
\graphicspath{{Figs/}}

\title{Superconducting States and Intertwined Orders in Metallic Altermagnets}

\author{Xuan Zou}
\affiliation{Institute for Advanced Study, Tsinghua University, Beijing 100084, China} 
\affiliation{Department of Physics and Anthony J. Leggett Institute for Condensed Matter Theory, The Grainger College of Engineering, University of Illinois Urbana-Champaign, Urbana, Illinois 61801, USA}

\author{Rafael M. Fernandes}
\affiliation{Department of Physics and Anthony J. Leggett Institute for Condensed Matter Theory, The Grainger College of Engineering, University of Illinois Urbana-Champaign, Urbana, Illinois 61801, USA}

\author{Eduardo Fradkin}
\affiliation{Department of Physics and Anthony J. Leggett Institute for Condensed Matter Theory, The Grainger College of Engineering, University of Illinois Urbana-Champaign, Urbana, Illinois 61801, USA}

\begin{abstract}
Altermagnets are a newly identified class of magnets with nodal spin-split band structures, providing a fertile platform for studying unconventional superconductivity and intertwined orders. Here we investigate multicomponent superconductivity and fluctuation-induced intertwined orders in an interacting $d$-wave metallic altermagnet that is invariant under a combination of a fourfold rotation $C_4$ and time-reversal symmetry $\mathcal{T}$. Within mean-field theory, the superconducting ground-state manifold is described in terms of two equal-spin two-component $p$-wave gap functions  $(\Delta_A^x,\Delta_B^y)$ and $\left( \Delta_A^y,\Delta_B^x\right)$, where $A$ and $B$ refer to the two spin-polarized Fermi surfaces related by $C_4\mathcal{T}$ symmetry. Because these two sets of gap functions condense at different temperatures, a rich phase diagram with multiple superconducting phase transitions emerges. Distinct fluctuations of sub-leading normal-state instabilities that compete with altermagnetism lift the degeneracy of the multicomponent pairing state in different ways. While nematic fluctuations enhance competition between distinct superconducting components and stabilize nematic superconducting phases, spin current-loop fluctuations promote coexistence and select a pair of chiral states. Our results uncover the pairing structure and elucidate how intertwined sub-leading fluctuations shape superconducting order in altermagnetic metals, suggesting a route toward realizing nematic and topological superconductivity.
\end{abstract}

\date{\today}
\maketitle

\section{Introduction}
\label{intro}

The concept of intertwined orders provides an important perspective for understanding the rich phase diagrams of strongly correlated electron systems. Unlike the conventional picture of competing orders, where independent order parameters suppresses each other, the intertwined-order perspective emphasizes a more complex relationship between distinct orders that can lead to the emergence of new phases of matter \cite{Berg_2009,fradkin2012,Fradkin2015,Fradkin2025,Fernandes2019,Vojta2009}.
Intertwined orders and their associated vestigial phases have been extensively found to play a key role in many strongly correlated quantum materials \cite{agterberg2008,berg2009a,Radzihovsky2009,Radzihovsky2011,Fernandes2012,Wang2015,Cai2017,Nie2017,Chatterjee2017, Fernandes2019,Agterberg2020,jian2021,Fernandes2021,liu2024nematic,Liu2023,Hecker2023,poduval2024,Gali2024,zou2025,verghis2025}. One prominent example are the cuprate high-$T_c$ superconductors, where the energy scales of superconductivity and various charge, spin, and nematic orders are comparable. In this context, the intertwined-order framework has been widely used to analyze the relations among these orders, including theoretical proposals for pair-density-wave (PDW) states and their associated vestigial orders \cite{Agterberg2020,agterberg2008,berg2009a,berg2009b,Radzihovsky2009,Radzihovsky2011}. Another example is the phase diagram of the iron-based superconductors \cite{Fernandes2022}, whose nematic phase has been identified as a vestigial phase of the stripe antiferromagnetic state \cite{Fang2008,Xu2008,Fernandes2012}. 

Recent theoretical and experimental studies have identified a new class of magnetic materials, known as altermagnets (AMs), that exhibit nodal spin-split electronic bands without  having a net magnetization and in the absence of spin-orbit coupling (SOC) \cite{Smejkal2020,Libor2022b,Libor2022c}.  Such spin-split Fermi surfaces are a consequence of the symmetries of the altermagnetic state, which is invariant under a combination of time-reversal symmetry and a point group symmetry operation that is not inversion \cite{jungwirth2024,jungwirth2024b}. Such spin-polarized Fermi surfaces can fundamentally reshape superconducting instabilities, enabling unconventional pairing channels and multicomponent superconducting states, and thus provide a natural platform for intertwined-order phenomena.
Experimental evidence for AM order has been reported in several materials \cite{Feng2022,Betancourt2023,Zhou2023,Fedchenko2023,Lee2023,Osumi2023,Kluczyk2023,Reimers2023,Ding2024,Yang2024,Lusignature2025,lin2024,krempasky2024,Amin2024,Babu2024,Jiang2024discovery,Zhang2024crystal,Wei2025,li2025exp}. 
On the theoretical side, minimal models on the Lieb lattice and other two-sublattice structures have been developed to investigate the consequences of AM as well as the microscopic interactions that stabilize it  \cite{Sudbo2023,antonenko2024,Roig2024,Cano2024,Valenti2024,Das2024,Durrnagel2025,Kaushal2025,Takahashi2025,li2025dwave,Chang2025,ghosh2026}.
Related spin-split electronic structures can also arise in Fermi liquids near an even-parity spin-triplet Pomeranchuk instability \cite{Wu2007}, albeit with a different microscopic origin \cite{jungwirth2024b}.

The study of superconducting instabilities in spin-split Fermi surfaces predates the formulation of altermagnetism \cite{Soto-Garrido2014,Lee2021}. Early works focused on nematic–spin-nematic (NSN) metallic states driven by Pomeranchuk instabilities in the spin-triplet channel \cite{Wu2007,Pomeranchuk1958}. In such NSN metals, spin-singlet pairing can lead to finite-momentum superconducting states \cite{Soto-Garrido2014}, including Fulde–Ferrell–Larkin–Ovchinnikov (FFLO) phases \cite{FF1964,LO1964} and PDW superconductivity \cite{Agterberg2020} \footnote{\textcolor{blue}{Strictly speaking the superconducting states described in Ref.\cite{Soto-Garrido2014} cannot be spin singlets since the altermagnetism breaks the spin symmetry leading to spin-split bands. In general, there are four PDW states of the type described in Ref.\cite{Soto-Garrido2014}. Unidirectional order arises when suitable biquadrating terms in the Landau theory are repulsive.} }. 
Closely related finite-momentum and multicomponent superconducting states have also been investigated more recently in AM systems \cite{zhang2024,Sumita2023,Chakraborty2024,Knolle2025,Hong2025,Sumita2025,froldi2025,parthenios2025,hu2025,liu2026}.
Alternatively, odd-parity spin-split Fermi surfaces that arise from SOC in non-magnetic and non-centrosymmetric systems, such as transition-metal dichalcogenides, can host Ising superconducting states \cite{lu2015,xi2016}.
In contrast, the spin splitting in the electronic bands of AMs arises without SOC and reflects the simultaneous breaking of time-reversal and rotational symmetries.
This distinctive Fermi-surface structure naturally favors equal-spin triplet pairing \cite{Mazin-2023}, which is stabilized even in the weak-coupling regime by the fact that the electronic states with opposite momenta have equal spin components, in sharp contrast to the spin-singlet pairing typically realized in time-reversal-symmetric systems. As a result, the superconducting pairing instabilities in AMs are fundamentally distinct than in standard metals \cite{Wu_Wang2025,parshukov2025,Khodas2025,rasmussen2025}.

In addition, spin-triplet pairing channels in AMs have also attracted growing interest, as they may provide a platform for realizing $p$-wave topological superconductivity \cite{Ghorashi2024,Zhu2023,Li_Liu2023,Wu_Wang2025,Heung2025,hodge2025,mcardle2026}. These developments establish altermagnetic systems as a promising platform for exploring unconventional superconductivity and related emergent phenomena \cite{Ouassou2023,sun2023,Brekke2023,Chakraborty2024,Ghorashi2024,Banerjee2024,Bose2024,Sumita2025,Fukaya_2025,parshukov2025,fukaya2025,froldi2025,li2025Jose,heinsdorf2025,ma2025,rasmussen2025,Leraand2025,monkman2025,Wu_Wang2025,parthenios2025,hu2025,hodge2025,lu2025}. 
Moreover, the coupling of superconductivity to fluctuations associated with normal-state instabilities that are sub-leading to the altermagnetic instability, such as nematic and loop order quantum fluctuations \cite{Sun2009}, provides fertile ground for realizing fluctuation-mediated intertwined phenomena, which remain largely unexplored in AM.

In this work, we investigate multi-component superconducting states in AM metals and examine how they intertwine with nematic and spin current-loop fluctuations.
Starting from a $d$-wave altermagnetic model \cite{antonenko2024,Roig2024}, we analyze the symmetry and structure of the superconducting ground state. 
Within a mean-field theory of a specific microscopic model \cite{antonenko2024} for an altermagnetic metallic phase, we show that the spin-split Fermi surfaces of a d-wave AM metal favor $p_x \pm i p_y$ superconductivity in both spin-up and spin-down sectors.
While the amplitudes of the $p_x$ and $p_y$ components on the same spin-sector are unequal and onset at different temperatures, the amplitudes of the $p_x$ and $p_y$ components on opposite spin sectors are enforced to be the same by the $C_4 \mathcal{T}$ symmetry of the altermagnetic phase. While related $p_x \pm i p_y$ states on the opposite-spin Fermi surfaces have been previously proposed to emerge in altermagnets \cite{monkman2025,heinsdorf2025,rasmussen2025}, our work goes beyond the leading instability analysis of previous works by analyzing the free-energy minima below $T_c$. This results in a rich phase diagram with  multiple successive superconducting transitions. Although we consider a specific model, the essential results we obtained are expected to hold for systems with an altermagnetic metallic state.

We further examine the coupling between superconductivity and fluctuations in the nematic and spin current-loop channels, which in the Lieb lattice are sub-leading in the parameter regime where the altermagnetic instability takes place \cite{Sun2009,Tsai_2015}. Nematic fluctuations enhance competition between superconducting components and promote nematic superconducting phases that break the combined $C_4 \mathcal{T}$  symmetry, whereas spin current-loop fluctuations favor their coexistence and lift the ground-state degeneracy by selecting a pair of chiral superconducting states.  These results demonstrate how superconductivity in AM metals can intertwine with other electronic fluctuations, establishing a promising platform for realizing nematic and topological superconducting phases.

This paper is organized as follows. In Sec.~\ref{model}, we introduce the $d$-wave altermagnetic model and analyze its superconducting instabilities by explicitly calculating the corresponding superconducting susceptibilities. In Sec.~\ref{pairing states}, we investigate the anisotropic spin-triplet pairing states and elucidate their underlying symmetry structure within a mean-field framework. Section~\ref{Landau} is devoted to the construction of the Landau free energy for the multi-component superconducting order parameters. Within this framework, we analyze how fluctuations associated with nematic and spin current-loop channels generate additional couplings and qualitatively reshape the superconducting phase diagram. Finally, Sec.~\ref{conclusions} discusses the possible emergence of vestigial phases and summarizes our main results. Technical details of the microscopic derivations are presented in the Appendices.

\begin{figure}
    \centering
    \hfill
    \parbox{0.53\linewidth}{
        \centering
        \includegraphics[width=\linewidth]{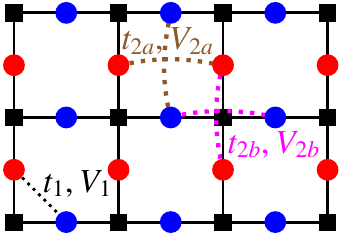}\\
        (a)
    }
    \parbox{0.37\linewidth}{
        \centering
        \includegraphics[width=\linewidth]{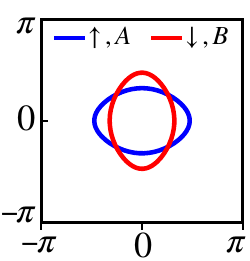}\\
        (b)
    }
    \parbox{1.0\linewidth}{
        \centering
        \includegraphics[width=\linewidth]{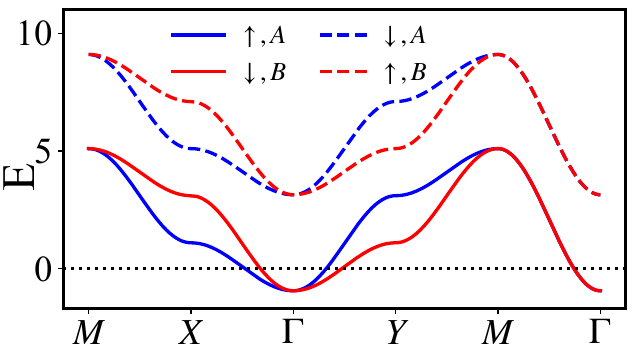}\\
        (c)
    }
    \caption{(a) Lieb lattice model of the $d$-wave altermagnet \cite{antonenko2024}. 
    Blue and red dots denote sublattices 1 and 2, respectively,
    while black squares indicate the crystalline environment. $t_1$, $t_{2a}$, and $t_{2b}$ represent the nearest- and next-nearest-neighbor hoppings, with corresponding interactions $V_1$, $V_{2a}$, and $V_{2b}$. (b) Fermi surface and (c) band structure for $t_1=0.1$, $t_{2a}=1.0$, $t_{2b}=0.5$, and $\mu=-2.1$. The labels $A$ and $B$ denote the bands (eigenstates), and up (down) arrows indicate spin-up (spin-down) states. }
    \label{fig1}
\end{figure}

\section{Altermagnet model and its superconducting instabilities}
\label{model}

We consider a Lieb lattice consisting of magnetic atoms with opposite spins on Wyckoff position $2f$ of the square lattice, and thus related by a $90^\circ$ rotation, as shown in Fig. \ref{fig1}. There are also non-magnetic atoms on the sites of the square lattice (Wyckoff position $1a$), which enforce the symmetries of the Lieb lattice. Hereafter, we assume their energy states to be high enough so that the effective low-energy model includes only the states from sublattices 1 (blue dots) and 2 (red dots). The $d_{x^2-y^2}$-wave altermagnetic (AM) metal is described by
$H_0=\sum_{\boldsymbol{k}} c_{\boldsymbol{k}}^\dagger \mathcal{H}_0(\boldsymbol{k}) c_{\boldsymbol{k}}, $
where the spin–sublattice basis is $c_{\boldsymbol{k}}=(c_{\boldsymbol{k}\uparrow 1}, c_{\boldsymbol{k}\uparrow 2}, c_{\boldsymbol{k}\downarrow 1}, c_{\boldsymbol{k}\downarrow 2})^T$. The single-particle Hamiltonian of the altermagnetic state (mean-field) on a Lieb lattice is given by  \cite{antonenko2024}
\begin{equation}
\label{eq:H0}
\begin{aligned}
\mathcal{H}_0(\boldsymbol{k})={}&-4t_1\cos\frac{k_x}{2}\cos\frac{k_y}{2}\tau_x
-2t_2\left(\cos k_x+\cos k_y\right)\tau_0 \\
&-2t_d\left(\cos k_x-\cos k_y\right)\tau_z
-\tilde{\mu}\tau_0
-N_{\rm am}\sigma_z\tau_z,
\end{aligned}
\end{equation}
As indicated in Fig.~\ref{fig1}(a), $t_1$ denotes the nearest-neighbor (NN) hopping, while the next-nearest-neighbor (NNN) amplitudes are parameterized by $t_2=(t_{2a}+t_{2b})/2$ and $t_d=(t_{2b}-t_{2a})/2$ in terms of the anisotropic NNN hoppings $t_{2a}$ and $t_{2b}$. Here $\tilde{\mu}$ is the chemical potential and $N_{\rm am}$ is the \textit{d-wave altermagnetic order parameter for the Lieb lattice}, with the sublattice spin polarization chosen along $\sigma_z$ without loss of generality, since we do not include SOC in our analysis. 
The Pauli matrices $\sigma_i$ and $\tau_i$ act on spin and sublattice spaces, respectively. The interactions shown in Fig. \ref{fig1}(a) include a nearest-neighbor density-density repulsive interaction $V_1$ and next-nearest-neighbor attractive interactions $V_{2a}$ and $V_{2b}$. The attractive interactions $V_{2a}$ and $V_{2b}$ drive the equal-spin superconducting instabilities discussed in Sec. \ref{pairing states}, while the repulsive interaction $V_1$ generates nematic and spin current-loop fluctuations, which play an important role in shaping the superconducting phase diagram analyzed in Secs. \ref{nematic} and \ref{loop}.

The model respects the $C_4\mathcal{T}$ symmetry characteristic of $d$-wave altermagnets. Eq.~\eqref{eq:H0} can be derived from symmetry considerations \cite{antonenko2024,Roig2024} or obtained as the mean-field Hamiltonian of a Lieb lattice with on-site repulsive Hubbard interactions \cite{Das2024}.
Deep in the AM phase, where spin-up (down) electrons predominantly occupy sublattice $1$ ($2$), only two energy bands cross the Fermi level. The resulting Fermi surface is shown in Fig.~\ref{fig1}(b), while the corresponding band dispersion is displayed in Fig.~\ref{fig1}(c). In this case, the low-energy Hamiltonian reduces to a two-band model
\begin{equation}
\mathcal{H}_0(\boldsymbol{k})=-2t_2 (\cos k_x + \cos k_y) s_0 
- 2\phi (\cos k_x - \cos k_y) s_z 
- \mu,
\end{equation}
in the basis $c_{\boldsymbol{k}} = (c_{\boldsymbol{k}\uparrow A}, 
c_{\boldsymbol{k}\downarrow B})^{T}$, where $c_{\boldsymbol{k}\uparrow A}$ is predominantly composed of spin-up states on sublattice 1 and $c_{\boldsymbol{k}\downarrow B}$ is predominantly composed of spin-down states on sublattice 2 (see Appendix \ref{app:proj} for details).  
The shifted chemical potential is $\mu=\tilde{\mu}+N_{\rm am}$.
For $|N_{\rm am}|\ll 4|t_1|$, corresponding to weak AM order, a low-energy expansion near the $\Gamma$ point yields an effective $d$-wave spin splitting $\phi \simeq N_{\rm am} t_d/(4t_1)$~\cite{jungwirth2024b}. In contrast, deep in the AM phase where the order is strong and the low-energy states are nearly spin-sublattice polarized, the effective spin-splitting strength becomes $\phi \simeq t_d \mathrm{sgn}(N_{\rm am})$.
In this representation the exchange $A\leftrightarrow B$ is equivalent to a global spin flip, $\uparrow \leftrightarrow \downarrow$. This is the characteristic feature of AM state, which in this case has  $C_4\mathcal{T}$ symmetry. In what follows, we will assume that the sign of the AM order parameter $N_{\rm am}$ is positive, such that band $A$ has $\uparrow$ spin polarization and  band $B$ has $\downarrow$ spin polarization. Changing the sign of $N_{\rm am}$ exchanges the bands and the spin polarizations. The resulting band structure closely resembles the NSN $\alpha$ phase of Refs.\cite{Wu2007} and \cite{Kivelson-2003}, where the Fermi-surface dispersion in the continuum takes the form $E(\boldsymbol{k}_F) \propto \xi - \delta \cos(l_\alpha\theta)s_z$ with angular momentum $l_\alpha=2$ and isotropic dispersion $\xi = k^2/2m - \mu$ (see Appendix \ref{app:A0} for details). 

We now turn to superconductivity in AM metals. Due to the spin-split Fermi surface, conventional spin-singlet pairing lacks perfect nesting and instead favors finite-momentum pairing, which produces a finite peak (rather than a divergence) in the superconducting susceptibility~\cite{Soto-Garrido2014}. 
In the weak-coupling limit, and deep enough in the altermagnetic state, equal-spin $p$-wave pairing is expected to be the most favorable, since the states at $\mathbf{k}$  and $-\mathbf{k}$  have the same spin/band quantum numbers \textcolor{blue}{\cite{Mazin-2023}}. On the other hand, for weak distortions of the native Fermi surface, spin singlet superconducting states can compete with the $p$-wave states \cite{Soto-Garrido2014}. In what follows we will assume that we are deep enough in the altermagnet phase so that these more complex situations can be ignored.

More specifically, the pairing interaction has the form
\begin{equation}
H_p = -g \sum_{\mathbf{k},\mathbf{k}',\mathbf{q}} \sum_{s,s'}
\gamma(\mathbf{k}) \gamma(\mathbf{k}')
c_{\mathbf{k} + \frac{\mathbf{q}}{2}, s}^\dagger
c_{-\mathbf{k} + \frac{\mathbf{q}}{2}, s'}^\dagger
c_{-\mathbf{k}' + \frac{\mathbf{q}}{2}, s'}
c_{\mathbf{k}' + \frac{\mathbf{q}}{2}, s},
\label{eq:Hp}
\end{equation}
where $g$ is the effective coupling constant, which depends on the interactions included in the model, and $\gamma(\mathbf{k})$ encodes the form factor of the pairing channel. In the reduced two-band description, spin and band degrees of freedom are locked, such that the indices are 
$\{\uparrow A, \downarrow B\}$. For brevity we omit explicit spin labels and write $\{s,s'\}\in\{A,B\}$. Equal-spin triplet pairing corresponds to $s=s'$, whereas spin-singlet pairing requires $s\neq s'$. The center-of-mass momentum $\mathbf{q}$ is kept general to allow for finite-momentum pairing favored by spin splitting. The normalized form factors are given by $\gamma_s(\mathbf{k}) = 1$ for $s$-wave pairing, $\gamma_d(\mathbf{k}) = \sqrt{2}\cos(2\theta_{\mathbf{k}})$ for $d$-wave pairing, $\gamma_{p_x}(\mathbf{k}) = \sqrt{2}\cos\theta_{\mathbf{k}}$ for $p_x$ pairing,
and $\gamma_{p_y}(\mathbf{k}) = \sqrt{2}\sin\theta_{\mathbf{k}}$ for $p_y$ pairing. 

\begin{figure}
    \centering
    \includegraphics[width=1.0\linewidth]{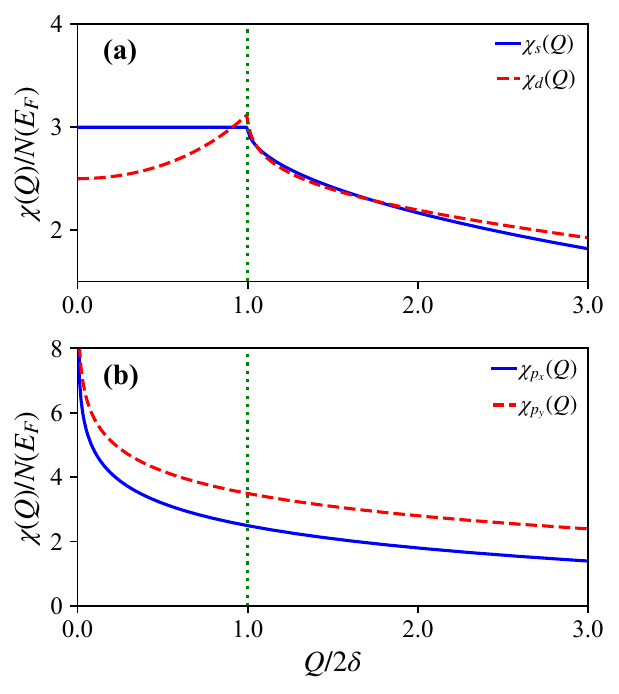}
    \caption{Superconducting susceptibility in the (a) singlet and (b) triplet channels as a function of pair momentum $Q$. In the singlet channel, $\chi_{\text{SC}}(Q)$ exhibits only a finite-momentum peak, reflecting the lack of perfect nesting. In contrast, the triplet $p$-wave susceptibility diverges at $Q=0$, identifying equal-spin $p$-wave pairing as the leading instability. See text for details.}

    \label{fig2}
\end{figure}

The superconducting susceptibility at zero external frequency is given by
\begin{equation}
\chi_{\text{SC}}(\mathbf{Q}) =
\sum_{\mathbf{k}} |\gamma(\mathbf{k})|^2
\frac{1 - n_F\big(\epsilon_{\mathbf{k}+\frac{\mathbf{Q}}{2}}\big)
      - n_F\big(\epsilon_{-\mathbf{k}+\frac{\mathbf{Q}}{2}}\big)}
     {\epsilon_{\mathbf{k}+\frac{\mathbf{Q}}{2}}+\epsilon_{-\mathbf{k}+\frac{\mathbf{Q}}{2}}}.
\end{equation}
Here $\epsilon_{\mathbf{k}}$ denotes the band dispersion and $n_F$ is the Fermi–Dirac distribution.
The behavior of $\chi_{\text{SC}}$ can be understood analytically in the continuum limit (see Appendix \ref{app:A} for details). We evaluate $\chi_{\text{SC}}(\mathbf{Q})$ along the $(Q,0)$ direction for different pairing channels. As shown in Fig.~\ref{fig2} (a), for spin-singlet pairing, $\chi_{\text{SC}}$ exhibits only a finite peak at $Q=2\delta$, where $2\delta$ corresponds to the momentum splitting between the two Fermi surfaces along the $x$ direction in the altermagnet d-wave state, consistent with the results of Ref.\cite{Soto-Garrido2014}. This finite peak, whose strength scales with the inverse of the altermagnet order parameter $N_{\rm am}$, reflects the absence of perfect nesting on the spin-split Fermi surface in the altermagnet. By contrast, in the $p$-wave channel, the susceptibility diverges at $Q=0$ as shown in Fig.~\ref{fig2} (b), signaling that intra-band $p$-wave superconductivity can become the leading instability in AM metals provided that there is an effective attractive interaction in this channel. We therefore focus on the $p$-wave superconducting state and analyze the properties of the ground state.

\section{Mean-Field Theory of spin-triplet pairing states in Metallic Altermagnets}
\label{pairing states}

In this section we develop the mean-field theory of triplet superconductivity in d-wave metallic altermagnets. Details are given in Appendix \ref{app:B}.

The $d$-wave altermagnet preserves the $C_4\mathcal{T}$ symmetry: a $C_4$ rotation followed by time reversal maps one spin sector onto the other. Within a single spin sector, only $C_2$ rotational symmetry is required rather than full $C_4$. We do not specify the microscopic origin of pairing and instead we assume a symmetry consistent intra-sublattice density–density attraction of the form
\begin{equation}
H_{\text{int}} = -\sum_{i,s} V_{x,s}  n_{i,s}  n_{i+x,s}
                 - \sum_{i,s} V_{y,s}  n_{i,s}  n_{i+y,s},
\end{equation}
where $n_{i,s}$ is the density operator on site $i$ with sublattice index $s=1,2$. 
We take $V_{x,1}=V_{2b},\ V_{y,1}=V_{2a}$ and $V_{x,2}=V_{2a},\ V_{y,2}=V_{2b}$, and define the isotropic and anisotropic components $V_2=(V_{2a}+V_{2b})/2$ and $V_d=(V_{2b}-V_{2a})/2$. Although an anisotropic interaction component $V_d$ is not symmetry forbidden, we emphasize that the qualitative features discussed below, including anisotropic pairing and the emergence of two distinct superconducting transition temperatures, are already present for $V_d=0$ provided the altermagnetic band anisotropy $\phi$ is nonzero. We therefore take $V_d=0$ in the main text, and analyze the finite-$V_d$ case in Appendix \ref{app:D}.
Pairing occurs within the spin-polarized Fermi surfaces and in all cases has the form $\uparrow \uparrow$  or $\downarrow \downarrow$, respectively. Since bands $A$ and $B$ are approximately sublattice-polarized, we replace the sublattice indices $1$ and $2$ by $A$ and $B$ in what follows.
In a given sector $s=A, B$, which also corresponds to a definite spin polarization, the gap function is
\begin{equation}
\Delta_s(\boldsymbol{k}) = -\frac{1}{N_L} \sum_{\boldsymbol{k}'} V_s(\boldsymbol{k} - \boldsymbol{k}') \langle c_{\boldsymbol{k}',s} c_{-\boldsymbol{k}',s} \rangle,
\label{sc_amplitudes}
\end{equation}
with 
\begin{equation}
V_s(\boldsymbol{k} - \boldsymbol{k}') = -2V_{x,s}  g_x(\boldsymbol{k}) g_x^*(\boldsymbol{k}') - 2V_{y,s}  g_y(\boldsymbol{k}) g_y^*(\boldsymbol{k}')
\label{pairing-interactions}
\end{equation}
are the pairing interactions and 
where $g_{x}(\boldsymbol{k}) = \sin k_{x}$  and $g_{y}(\boldsymbol{k}) = \sin k_{y}$ are the p-wave form factors, respectively. Here, $N_L$ denotes the number of lattice sites.
Introducing the pairing amplitudes (i.e. the superconducting order parameters) $\Delta_s^\eta$,  
\begin{equation}
\Delta_s^{\eta} = -\frac{2}{N_L} \sum_{\boldsymbol{k}} V_{\eta,s} g_\eta(\boldsymbol{k}) \langle c_{\boldsymbol{k}} c_{-\boldsymbol{k}} \rangle, \quad \eta = x, y,
\end{equation}
where $s=A, B$ labels the bands and $\eta=x, y$ labels the $p$-wave component,
the gap becomes 
\begin{equation}
\Delta_s(\boldsymbol{k}) = - g_x(\boldsymbol{k}) \Delta_s^x -   g_y(\boldsymbol{k}) \Delta_s^y.
\end{equation}
For convenience, we refer to $\Delta_s^x$ and $\Delta_s^y$ as the $p_x$ and $p_y$ pairing components, respectively, corresponding to the basis functions $g_x(\boldsymbol{k})$ and $g_y(\boldsymbol{k})$.
Note that these two $p$-wave components are not degenerate, since the site symmetry of a given sublattice is that of an orthorhombic point group.

To characterize the pairing symmetry, we parametrize the two-component superconducting order parameter in each spin sector as
\begin{equation}
\label{parametrize}
(\Delta_s^x, \Delta_s^y) = \Delta_s e^{i\theta_s} (\cos\alpha_s, e^{i\beta_s}\sin\alpha_s),
\end{equation}
with $\theta_s\in[0,2\pi)$, $\beta_s\in[-\pi,\pi)$, and $\alpha_s\in[0,\pi/2]$. 
Here, $\alpha_s$ measures the relative weight of the $p_x$ and $p_y$ components, while $\beta_s$ represents their relative phase. 
Minimizing the condensation energy with respect to $(\Delta_s,\alpha_s,\beta_s)$ determines the superconducting ground state.

We emphasize that, because the two components $\Delta^x_s$ and  $\Delta^y_s$ are not degenerate, they generally condense at different temperatures and assume different magnitudes, $\alpha_s \neq \pi/4$. This can be understood from the fact that, for a given spin sector, the corresponding Fermi surface is elliptical and thus only have two-fold symmetry. As a result, the $p_x$ and $p_y$ form factors are not degenerate, since the $C_4$ symmetry connecting them is broken for a given spin sector. There is, however, a degeneracy enforced by the  $C_4\mathcal{T}$ symmetry of the altermagnetic phase, which transforms $\Delta_A^x$  onto $\Delta_B^y$ and  $\Delta_A^y$  onto $\Delta_B^x$. Therefore, there are two sets of degenerate $p$-wave two-component gaps, $\left( \Delta_A^x, \Delta_B^y\right)$ and $\left( \Delta_A^y, \Delta_B^x\right)$, which must condense at different temperatures. Which of the two pairs condenses first depends on the sign of the altermagnetic order parameter. This is a consequence of the fact that switching the sign of the altermagnetic order parameter maps $\left( \Delta_A^x, \Delta_B^y\right)$ onto $\left( \Delta_A^y, \Delta_B^x\right)$, and implies that the four gaps are degenerate in the non-altermagnetic state \cite{Wu_Wang2025}. This motivates us to consider below the interplay between the two sets of independent order parameters, despite the fact that one of them is always a sub-leading channel. We note that other works considered the four components on an equal footing even in the altermagnetic phase \cite{monkman2025,heinsdorf2025}. Our approach with two separate sets of two-component gap functions agrees with the formal spin-group classification of pairing states performed in Ref. \cite{Khodas2025}. Using that paper's notation, the two sets of gaps introduced above transform as the spin-group irreducible representations $\boldsymbol{\Gamma}_{B_{1u}}^{+1}$ and $\boldsymbol{\Gamma}_{B_{1u}}^{-1}$. 

In the remainder of this paper, we employ this mean-field theory to determine the superconducting ground states of this altermagnetic system. The actual ground state depends on the choice of hopping parameters, interactions, and chemical potential. In the main text, we focus on the simplified case of vanishing pairing anisotropy, $V_d=0$, while a more general analysis allowing for $V_d\neq 0$ is presented in Appendix~\ref{app:spinless}. Moreover, unless otherwise stated, we use $t_2=0.75$, $V_2=2.25$, and  $\mu=-2.1$. 

As explained above, the altermagnetic order parameter $\phi$ governs the relationship between the $p_x$ and $p_y$ gap components   $\Delta_s^x$ and $ \Delta_s^y$ on the same spin sector. For $\phi=0$, the two components are equal, $|\Delta_s^x|=|\Delta_s^y|$, and $\alpha_s=\pi/4$, corresponding to the conventional $p_x\pm i p_y$ form.
However, for $\phi>0$, the $p_y$ component dominates on the $A$ band, leading to $\alpha_A>\pi/4$, while the $p_x$ component dominates on the $B$ band. The opposite trend occurs for $\phi<0$. Despite the finite anisotropy induced by nonzero $\phi$, the time-reversed pair of states $p_x + i p_y$ and $p_x - i p_y$ remain degenerate.
Throughout the entire parameter regime, as discussed above, the gaps on the two spin sectors are related by the combined $C_4\mathcal{T}$ symmetry, enforcing $\alpha_A+\alpha_B=\pi/2$ and $|\Delta_A^{x}|=|\Delta_B^{y}|$, $|\Delta_A^{y}|=|\Delta_B^{x}|$.

In the absence of inter-sublattice interactions and in the limite of a large $N_{\mathrm{am}}/t_1$, the superconducting components of sublattices $1$ and $2$, and thus of bands $A$ and $B$, can be treated as decoupled. 
To generate coupling without explicitly breaking additional symmetries, in the next section we will consider the effects of fluctuations of two normal-state order parameters that are subleading to the altermagnetic instability in the Lieb lattice (in their unbroken states): nematic fluctuations and spin current-loop fluctuations.

\section{Landau Free Energy and the Symmetries of the P-wave Superconducting States}
\label{Landau}

In the previous section, we showed that the spin triplet $p$-wave superconducting states of the two-dimensional altermagnet are characterized by \textit{four complex order parameter fields} $\Delta_A^x$ and $\Delta_A^y$ (both with spin up) and $\Delta_B^x$ and $\Delta_B^y$ (both with spin down), each having an amplitude and a phase. Superficially, this structure implies four independent global $U(1)$ symmetries. 
However, interaction terms generically constrain the relative phases between different components, so that only a reduced set of continuous symmetries remains.

\subsection{General Form of the Landau Free Energy}
\label{sec:general-Landau}

The Landau free energy consistent with the symmetries of the AM systems and the tetragonal lattice is constructed from gauge-invariant quadratic, quartic, and biquadratic terms, and can be written as

\begin{align}
F = & r_1 (|\Delta_A^x|^2 + |\Delta_B^y|^2)+ r_2 (|\Delta_A^y|^2 + |\Delta_B^x|^2) \label{eq:Landau} \\
+& u_1 (|\Delta_A^x|^4 + |\Delta_B^y|^4)+ u_2 (|\Delta_A^y|^4 + |\Delta_B^x|^4) \nonumber \\
+& v_{xy} (|\Delta_A^x|^2|\Delta_A^y|^2 + |\Delta_B^x|^2|\Delta_B^y|^2) \nonumber \\
+& v_{AB} (|\Delta_A^x|^2|\Delta_B^x|^2 + |\Delta_A^y|^2|\Delta_B^y|^2)\nonumber \\
+& v_{AB}^{xy} |\Delta_A^x|^2|\Delta_B^y|^2 +  v_{AB}^{yx} |\Delta_A^y|^2|\Delta_B^x|^2 \nonumber \\
+& w_{xy}\left[(\Delta_A^x\Delta_A^{y*})^2 + (\Delta_B^x\Delta_B^{y*})^2 + \text{c.c.}\right] \nonumber \\
+& w_{AB} (\Delta_A^{x*}\Delta_B^x\Delta_A^y\Delta_B^{y*} + \text{c.c.})
\nonumber \\
+& w_{AB}^{xy} \left[(\Delta_A^x\Delta_B^{y*})^2 + \text{c.c.}\right] + w_{AB}^{yx} \left[(\Delta_A^y\Delta_B^{x*})^2 + \text{c.c.}\right] \nonumber
\end{align}

\begin{figure}[t]
    \centering
    \includegraphics[width=1.0\linewidth]{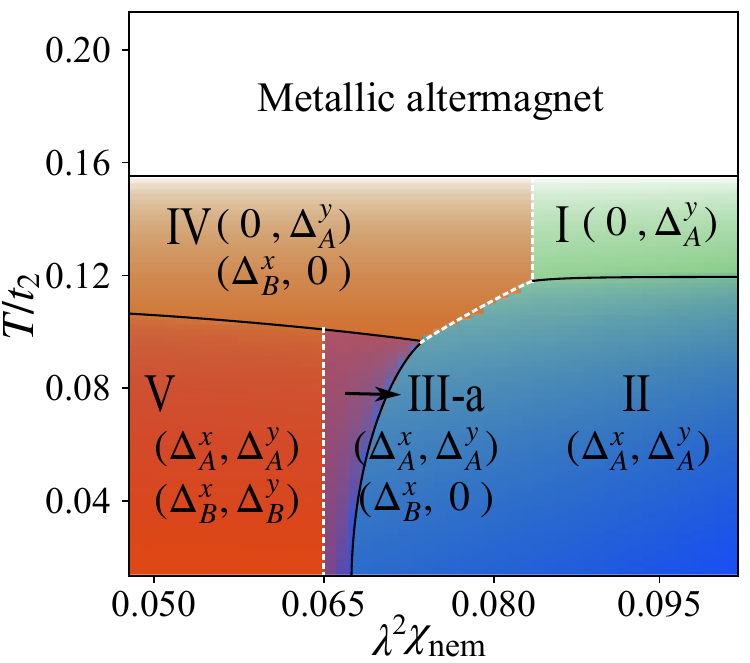}
\caption{Typical superconducting phase diagram as a function of the dimensionless temperature $T/t_2$ and the nematic susceptibility $\lambda^2 \chi_{\text{nem}}$ for $\phi = 0.05$, where $\lambda$ denotes the average of $\lambda_1$ and $\lambda_2$.
Black solid lines indicate continuous transitions, while the white dashed line marks a first-order transition. For clarity we focus on specific a range of values of the nematic susceptibility.}
    \label{nem}
\end{figure}

The effective coupling constants of these allowed terms given in Eq.\eqref{eq:Landau} were obtained by a standard Gor'kov-type expansion of the mean-field Hamiltonian near the highest critical temperature (details in Appendix \ref{ref:app-SC}). 
Before going into the fluctuation analysis, we first comment on the phase diagram and free energy obtained in the limit where nematic and spin current-loop fluctuations are turned off. The sequence of phase transitions is the same as that shown in the phase diagram of Fig. \ref{fig3} for the smallest value of $\chi_{\mathrm{nem}}$. In this limit, the quadratic coefficients take the standard form:
\begin{equation}
r_i(T)=\frac{1}{2}[1/V_{2}-\chi_{i}(T)],\quad i=1,2
\end{equation}
where $\chi_{1,2}(T)$ are the superconducting susceptibilities in the corresponding channels (explicit expressions are given in Appendix \ref{ref:app-SC}). Recall that in the main text we consider $V_d=0$. The two mean-field transition temperatures are determined by $r_i(T_i^c)=0$. $T_1^c$ corresponds to the superconducting instability in the $(\Delta_A^x,\Delta_B^y)$ channel, while $T_2^c$ corresponds to the 
$(\Delta_A^y,\Delta_B^x)$ channel. For the parameters used in Fig.~\ref{nem} and Fig.~\ref{lp}, with $\phi>0$, we obtain $T_1^c<T_2^c$. If $\phi<0$, the hierarchy is reversed, with $T_1^c>T_2^c$.

The splitting $|T_1^c-T_2^c|$ is governed by the dispersion anisotropy encoded in the $d$-wave spin splitting $\phi$, which makes the two channel susceptibilities unequal. For weak anisotropy one finds $\chi_{1}(T)-\chi_{2}(T)\propto \phi$, implying that $T_1^c-T_2^c$ is linear in $\phi$ to leading order (see Appendix \ref{Tc_splitting}). Since $ \phi$ is controlled by the AM order parameter $N_{\rm am}$ (see Sec. \ref{model}), the $T_c$ splitting inherits a corresponding parametric dependence on $N_{\rm am}$: for small $N_{\rm am}$ it grows approximately linearly with $N_{\rm am}$, while deep in the AM phase it saturates in magnitude and changes sign under $N_{\rm am}\to -N_{\rm am}$.

Below the lower critical temperature, all four order parameters have a non-zero expectation value. As shown in Appendix \ref{ref:app-SC}), we find a positive quartic coupling $w_{xy}>0$. Then, energetics requires that the \textit{relative phases} of $\Delta_A^x$ and $\Delta_A^y$ can be locked to each other to be $\pm \pi/2$ and the same applies  for  $\Delta_B^x$ and $\Delta_B^y$, corresponding to $p_x \pm i p_y$ configurations.  Note, however, that the relative amplitudes of the two $p$-components on the same spin sector are different, such that these configurations are more precisely described as $p \pm i \epsilon p$, in agreement with the results of Ref. \cite{rasmussen2025}. We use this notation in the remainder of the paper. 
This phase locking reduces the apparent four $U(1)$ symmetries to two $U(1)$ symmetries, together with two discrete ``spinless'' time-reversal symmetries (i.e., that formally do not act on the spin degrees of freedom) of the form $\tilde{\mathcal T}^A: (\Delta_A^x,\Delta_A^y)\to(\Delta_A^x,\Delta_A^y)^*$ and similarly for the $B$ band. Within our mean-field model, the Landau coefficients in Eq. (\ref{eq:Landau}) that couple the $A$ and $B$ sectors turn out to be zero, namely, $v_{AB} = v_{AB}^{xy}=v_{AB}^{yx} = 0$ and $w_{AB} = w_{AB}^{xy}=w_{AB}^{yx} = 0$. We attribute this artifact to the simplified form of our microscopic Hamiltonian and to the absence of inter-band interactions in our model. As we show below, however, most of these terms become non-zero once fluctuations with intertwined subleading instabilities are included.   

Upon increasing the temperature above the lower critical temperature, the system enters a phase in which one superconducting component in each band vanishes.  As a result, the fully gapped $p \pm i \epsilon p$ state becomes a nodal $p$-wave state, and the two spinless time-reversal symmetries $\tilde{\mathcal T}^{A,B}$ are restored.  A further increase in temperature drives the system into the normal metallic phase.  This two-step sequence of transitions can be clearly seen in the leftmost part of the phase diagrams in Fig.~\ref{nem} and Fig.~\ref{lp}, where fluctuations remain small,  corresponding to phases IV and V.

In the next subsections we will also include the contributions of nematic and spin current-loop fluctuations to the Landau coefficients arising from nearest-neighbor density-density repulsion ($V_1=1.0$; see Appendix \ref{app:C}), and we show that these fluctuations naturally generate symmetry-allowed $A$–$B$ coupling terms.

\begin{figure}
    \centering
    \includegraphics[width=0.98\linewidth]{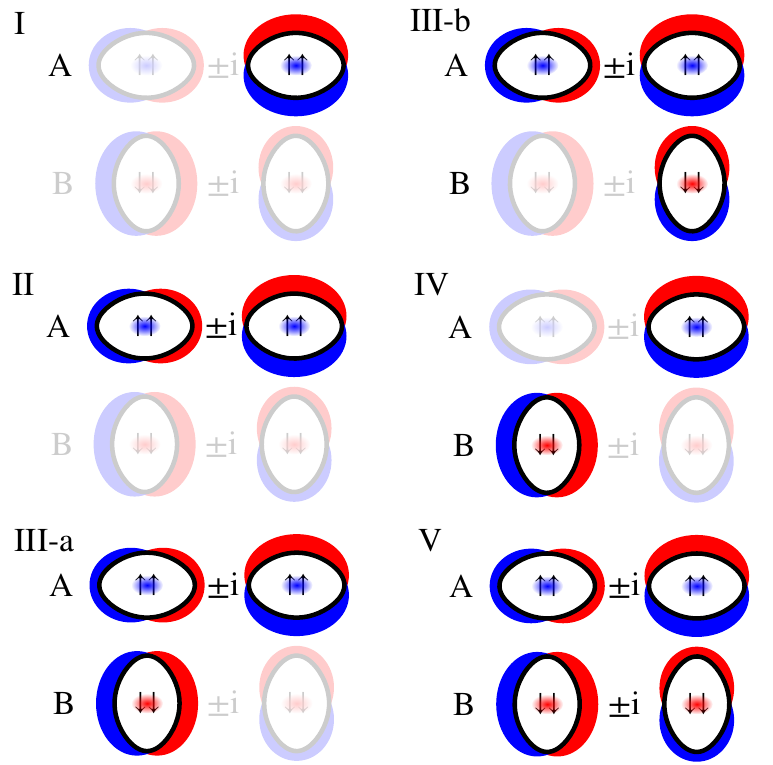}
\caption{
Schematic pairing structures corresponding to superconducting phases~I–V in Fig.~\ref{nem}, shown separately for the $A$ and $B$ bands.  
Each band hosts a $(p_x, p_y)$ order parameter, with the black contour indicating the underlying Fermi surface. 
Light, faded regions indicate that the corresponding component is absent in that phase.
Phases~IV and V satisfy $|\Delta_A^x|=|\Delta_B^y|$ and $|\Delta_A^y|=|\Delta_B^x|$.  
In contrast, phases~I-III are characterized by $|\Delta_A^x|\neq|\Delta_B^y|$ and $|\Delta_A^y|\neq|\Delta_B^x|$, reflecting the imbalance between competing superconducting components induced by strong nematic fluctuations.
}
    \label{pairing}
\end{figure}

\subsection{Role of nematic fluctuations}
\label{nematic}
The interacting Lieb lattice model has additional normal-state instabilities besides the altermagnetic one \cite{Sun2009,Tsai2015}. Even after the condensation of altermagnetic order, the fluctuations of these other particle-hole instabilities should remain present and, thus, intertwine with the particle-particle (i.e., superconducting instabilities). In this subsection, we focus on the impact of nematic fluctuations. 

Symmetry dictates that the leading coupling between the nematic field $\varphi$, which breaks $C_4$ symmetry, and the $p$-wave superconducting order parameters $\Delta_s^\eta$ is expressed as
\begin{equation}
F_{\text{n-SC}} = \varphi (\lambda_1 S_1 + \lambda_2 S_2),
\end{equation}
where $S_1 = |\Delta_A^x|^2 - |\Delta_B^y|^2$ and $S_2 = |\Delta_A^y|^2 - |\Delta_B^x|^2$ are local nematic fluctuations of the superconducting order parameters in the metallic altermagnet.
In the Lieb lattice model, nematic fluctuations originate from the nearest-neighbor density-density repulsive interaction between the $1$ and $2$ sublattices, described by $V_1 n_{i,1} n_{i,2}$, from which the coupling coefficients $\lambda_{1,2}$ can be derived microscopically, yielding $\lambda_{1,2}\propto V_1$ (see Appendix~\ref{ref:app-nematic}). Although we assume that no static nematic order is present, we  show now that nematic fluctuations play an important role in modifying the superconducting state.

The free energy density of the nematic fluctuations is given by
\begin{equation}
F_{\text{nem}} = \frac{1}{2\chi_{\text{nem}}} \varphi^2,
\end{equation}
where $\varphi$ is a local field representing the local nematic fluctuations and $\chi_{\text{nem}}$ is the nematic susceptibility, which diverges at the nematic transition and remains finite and positive in the altermagnetic phase. Here we assume that the system is outside the nematic phase, such that $\langle \varphi \rangle = 0$ in the parameter regime of interest.

Minimizing the free energy with respect to $\varphi$ yields
\begin{equation}
\varphi=-\chi_{\text{nem}}(\lambda_1S_1+\lambda_2S_2).
\end{equation}
Substituting this result back into the free energy gives
\begin{equation}
\label{eq:n_SC}
F(\varphi)=F_{\text{nem}} + F_{\text{n-SC}}
   = -\frac{\chi_{\text{nem}}}{2} (\lambda_1 S_1 + \lambda_2 S_2)^2.
\end{equation}
Equation~\eqref{eq:n_SC} generates additional quartic terms in the superconducting free energy.  
Expanding the square produces corrections to the coefficients of Eq.~\eqref{eq:Landau} and, in particular, induces inter-band couplings: 
\begin{equation}
\label{eq:n_SC2}
\begin{aligned}
F(\varphi)
= &-\frac{\chi_{\text{nem}}}{2}\big[\lambda_1^2\big(|\Delta_A^x|^4+|\Delta_B^y|^4\big)
  +\lambda_2^2\big(|\Delta_A^y|^4+|\Delta_B^x|^4\big)\big]\\
&-\chi_{\text{nem}}\lambda_1\lambda_2\big(|\Delta_A^x|^2|\Delta_A^y|^2+|\Delta_B^x|^2|\Delta_B^y|^2\big)\\
&+\chi_{\text{nem}}\lambda_1\lambda_2\big(|\Delta_A^x|^2|\Delta_B^x|^2+|\Delta_A^y|^2|\Delta_B^y|^2\big)\\
&+\chi_{\text{nem}}\lambda_1^2|\Delta_A^x|^2|\Delta_B^y|^2
+\chi_{\text{nem}}\lambda_2^2|\Delta_A^y|^2|\Delta_B^x|^2.
\end{aligned}
\end{equation}
While the first two terms simply renormalize Landau coefficients that were already nonzero in Eq.~\eqref{eq:Landau}, the last three introduce terms that were initially absent in the Landau expansion. These fluctuation-induced terms $v_{AB}=\chi_{\text{nem}}\lambda_1\lambda_2$, $v_{AB}^{xy}=\chi_{\text{nem}}\lambda_1^2$, and $v_{AB}^{yx}=\chi_{\text{nem}}\lambda_2^2$ , are all positive for the parameter sets considered here; the explicit expressions for $\lambda_{1,2}$ are given in Appendix \ref{ref:app-nematic}.
These terms disfavor the simultaneous development of distinct superconducting components on the two bands, thereby enhancing the competition between order-parameter sectors and promoting $C_{4}\mathcal{T}$-breaking superconducting states.
Even though the system has no pure $C_4$ symmetry, we still refer to these superconducting phases as nematic, as they spontaneously break the combined $C_4\mathcal T$ symmetry of the AM normal state and thus induce a lattice distortion.

The resulting superconducting phase diagrams as a function of temperature and nematic susceptibility $\chi_{\text{nem}}$ are shown in Fig.~\ref{nem}. 
In this figure, the transition lines indicated by white dashed lines are first order, while the black solid lines correspond to continuous transitions. The non-vanishing superconducting components in each phase are labeled by numbers in the diagrams, and the normal state is a metallic altermagnet. The pairing structures corresponding to the different numbered phases (I to V) are schematically illustrated in Fig.~\ref{pairing}. 

For weak nematic fluctuations (small $\chi_{\text{nem}}$), the fluctuation-induced couplings are insufficient to differentiate the four superconducting components on the two bands. Consequently, the superconducting amplitudes still satisfy
$|\Delta_A^x|=|\Delta_B^y|$ and $|\Delta_A^y|=|\Delta_B^x|$.
In this regime, upon heating, the system undergoes two successive thermal transitions, as explained in the previous subsection: a fully developed $p \pm i \epsilon p$ superconducting state on both bands (phase~V) first evolves into a nodal $p_x$ or $p_y$ phase (phase~IV), and then into the normal state.
As $\chi_{\text{nem}}$ increases, nematic fluctuations enhance the biquadratic couplings
$v_{AB}$, $v_{AB}^{xy}$, and $v_{AB}^{yx}$, leading to strong competition among the four superconducting components. This competition drives a spontaneous breaking of $C_4\mathcal{T}$ symmetry, producing nematic superconducting phases in which
$|\Delta_A^x|\neq|\Delta_B^y|$ and $|\Delta_A^y|\neq|\Delta_B^x|$ (phases I, II, and III-a). 

The pairing structures shown in Fig.~\ref{pairing} provide an intuitive visualization of the superconducting states discussed above. 
In phase~V, which is the ground-state phase at weak nematic fluctuations, the order parameter in each spin–band sector exhibits a $p \pm i \epsilon p$ structure. The black curve denotes the Fermi surface, while the red and blue lobes indicate the sign of the real form factors $p_x\propto\sin k_x$ and $p_y\propto\sin k_y$, which change sign across $k_x=0$ and $k_y=0$, respectively. The lobes and their colors represent the nodal sign structure of the $p_x$ and $p_y$ components, and the overall $\pm i$ relative phase between them is indicated explicitly. Vanishing components are shown using faded colors.

Phases~IV–V correspond to the regime in which the $A$ and $B$ bands carry equivalent $p \pm i \epsilon p$ combinations up to a $C_4$ rotation, consistent with the relations $|\Delta_A^x|=|\Delta_B^y|$ and $|\Delta_A^y|=|\Delta_B^x|$. 
In contrast, phases~I–III exhibit unequal superconducting amplitudes on the two bands, with $|\Delta_A^x|\neq|\Delta_B^y|$ and $|\Delta_A^y|\neq|\Delta_B^x|$, reflecting the imbalance generated by strong nematic fluctuations. 

All phases shown in Fig.~\ref{nem} are obtained for $\phi>0$ and $V_d=0$. Changing the sign of $\phi$ simply interchanges the roles of the $x$ and $y$ directions, corresponding to the transformation $\Delta_s^x\leftrightarrow\Delta_s^y$ on each band. As a result, the phase diagram remains qualitatively the same, with, for example, phase~III-a transforming into phase~III-b.
When a finite pairing anisotropy $V_d$ is included, phases~III-a and III-b can coexist within the same phase diagram, as discussed in Appendix~\ref{app:D}. Apart from the emergence of phase~III-b, the qualitative structure of the phase diagram remains unchanged compared to the $V_d=0$ case.

\begin{figure}[t]
    \centering
    \includegraphics[width=1.0\linewidth]{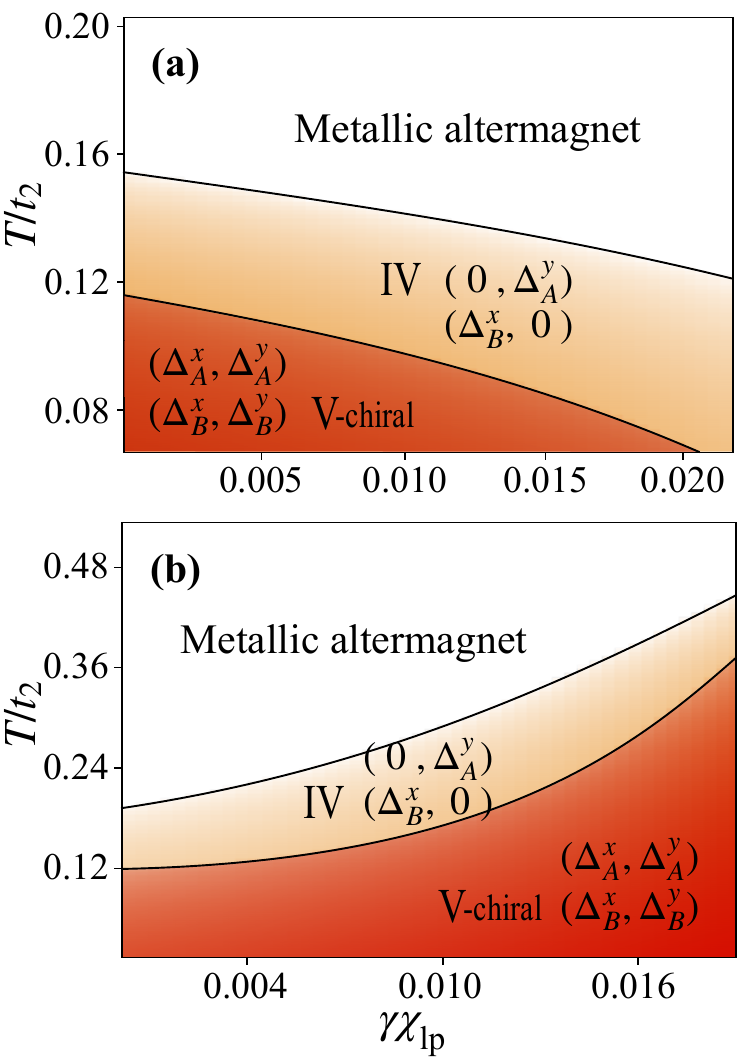}
    \caption{
    Superconducting phase diagrams including spin current-loop fluctuations for $\phi=0.05$, where $\gamma$ denotes the average of $\gamma_1$ and $\gamma_2$.
(a) Same parameter set as in Fig. \ref{nem}: $\mu=-2.1$, $V_2=2.25$.
(b) Alternative parameter set: $\mu=-2.8$, $V_2=3.75$.
In (a) [(b)], spin current-loop fluctuations suppress (enhance) the superconducting transition temperature $T_c$.
At low temperatures, all four superconducting components coexist; upon increasing $T$, the $(\Delta_A^x,\Delta_B^y)$ components first vanish.
    }
    \label{lp}
\end{figure}

\subsection{Role of spin current-loop fluctuations}
\label{loop}

Besides a subleading nematic instability, the interacting Lieb lattice model also has a subleading spin current-loop instability \cite{Sun2009,Tsai2015}. We now turn to the effect of spin current-loop fluctuations, and thus introduce a complex order parameter $\phi_l$ representing the spin current-loop order.
The coupling between $\phi_l$ and the superconducting order parameters $\Delta_s^\eta$ is given by
\begin{align}
F_{\phi_l\text{–SC}} &= 
\gamma_1 |\phi_l|^2 (|\Delta_A^x|^2 + |\Delta_B^y|^2)
+ \gamma_2 |\phi_l|^2 (|\Delta_A^y|^2 + |\Delta_B^x|^2)  \nonumber \\
&\quad + \big[\gamma_3 \phi_l^2(\Delta_A^{x*}\Delta_B^x + \Delta_A^{y*}\Delta_B^y) + \text{c.c.}\big] .
\end{align}
These couplings originate from the nearest-neighbor interaction $V_1 n_{i,A}n_{i,B}$ decoupled in the spin current-loop channel, and the corresponding coefficients $\gamma_i$ can be microscopically derived (see Appendix \ref{ref:app-loop}) .

Assuming no pre-existing spin current-loop order, the free energy of the spin current-loop field can be approximated by a quadratic form 
\begin{equation}
F_{\text{lp}} = \tfrac{1}{2\chi_{\text{lp}}}|\phi_l|^2
\label{eq:loop}
\end{equation}
so that the total contribution to the free energy density involving $\phi_l$ is $F_{\text{lp}} + F_{\phi_l\text{–SC}}$. Here, $\chi_{\mathrm{lp}}$ is the finite spin current-loop susceptibility and we assume that $\langle \phi_l\rangle=0$.
Integrating out $\phi_l$ yields an effective free energy containing the inter-band coupling terms $v_{AB}$, $v_{AB}^{xy}$, $v_{AB}^{yx}$, and $w_{AB}$ in Eq.~\ref{eq:Landau}. 
Unlike nematic fluctuations, spin current-loop fluctuations generate these terms with negative coefficients, 
favoring coexistence among the superconducting components. 

The term proportional to $w_{AB}$ originates only from spin current-loop fluctuations, as it is absent from the free-energy renormalized by nematic fluctuations. It acts as a phase-locking interaction between bands, contributing a term of the form $w_{AB}\cos(\beta_A - \beta_B)$, with $w_{AB}<0$. 
Here $\beta_s = \arg(\Delta_s^y) - \arg(\Delta_s^x) = \pm \pi/2$  denotes the relative phase between the $p_x$ and $p_y$ components of the superconducting order parameter on band $s$, which are set to $\pm \pi/2$ by the bare $w_{xy}>0$ coefficient of the Landau expansion. Because $w_{AB}<0$, the additional term induced by spin current-loop fluctuations energetically favors $\beta_A = \beta_B$. This corresponds to a chiral configuration (phase V-chiral in Fig.~\ref{lp}) and reduces the ground-state degeneracy  $(\epsilon p\pm ip)_A\otimes(p\pm i\epsilon p)_B$ to $(\epsilon p+ip)_A\otimes(p+i\epsilon p)_B$ and $(\epsilon p-ip)_A\otimes(p-i\epsilon p)_B$. Note that time-reversal that is broken here is the ``spinless'' time-reversal symmetry that, due to the absence of SOC, acts only on the orbital magnetic moments generated by the superconducting degrees of freedom. Related chiral states have been proposed in previous works \cite{monkman2025,heinsdorf2025,rasmussen2025}.
Note that, even with both nematic and spin current-loop fluctuations, the Landau coefficients $w_{AB}^{xy}=w_{AB}^{yx}=0$. As a result, the effective free energy contains no global phase–locking term between the overall phases $\theta_A$ and $\theta_B$, since the $\phi_l$ field here represents fluctuations rather than a symmetry breaking order. If, however, $\phi_l$ acquires a finite expectation value and mirror symmetry is broken, a direct coupling term of the form $\Delta_A^{x*}\Delta_B^{y}+\text{c.c.}$ can emerge in the effective theory, locking the relative phase between two spin sectors. 

The resulting phase diagram under the spin current-loop fluctuations is shown in Fig.~\ref{lp}. 
The superconducting transition temperatures decrease with increasing $\chi_{\mathrm{lp}}$ for $\gamma_{1,2}>0$, as shown in Fig.~\ref{lp}(a), but are enhanced for $\gamma_{1,2}<0$ using a different set of parameters, as shown in Fig.~\ref{lp}(b).
The low-temperature phase V hosts four superconducting components, whereas the intermediate-temperature phase IV involves only two. Throughout the diagram, the relations $|\Delta_A^x|=|\Delta_B^y|$ and $|\Delta_A^y|=|\Delta_B^x|$ hold, suggesting that the system preserves the combined $C_4\mathcal{T}$ symmetry -- recall that the chiral state breaks what we dubbed the ``spinless'' time-reversal symmetry. In Fig.~\ref{lp}, the two superconducting transition lines do not cross within the parameter range shown. 
This absence of crossing is not generic and depends on the choice of interactions. 
In particular, when a finite inter-orbital coupling $V_d$ is present, the two transition lines can cross, as shown in Appendix~\ref{app:D}.

\section{Discussion and Conclusions}
\label{conclusions}

To conclude, in this paper we have investigated superconducting instabilities of metallic altermagnets, starting from a minimal $d$-wave altermagnetic model with spin-split Fermi surfaces that preserve the $C_4\mathcal{T}$ symmetry. The spin splitting of the Fermi surfaces suppresses zero-momentum spin-singlet pairing, since electronic states with opposite spins are no longer at opposite momenta, and this momentum splitting increases as the spin splitting increases. In this sense, $p$-wave pairing is naturally expected to dominate under appropriate interaction conditions. Within mean-field theory, the ground state corresponds to a $p_x \pm i p_y$ superconducting state on each spin-split Fermi surface. However, the amplitudes of the $p_x$ and $p_y$ components are generally unequal, reflecting an intrinsic anisotropy of the pairing related to the existence of a combined $C_4 \mathcal{T}$ symmetry instead of separate $C_4$ and $\mathcal{T}$ symmetries. This naturally leads to multiple successive superconducting transitions upon increasing temperature, as first one of the gap components vanishes and then at a higher temperature the remaining component vanishes.

We further examined how sub-leading fluctuations couple the different superconducting components, focusing on two types of collective fluctuations that are naturally present in the interacting Lieb lattice model: nematic and spin current-loop fluctuations. While weak nematic fluctuations leave the phase diagram qualitatively unchanged, stronger nematic fluctuations promote competition among the superconducting components and generate a rich set of nematic superconducting phases.
In contrast, spin current-loop fluctuations favor coexistence among different superconducting components, lift the ground-state degeneracy, and select a pair of chiral states.

Our results establish the pairing structure of superconductivity in metallic altermagnets and demonstrate how intertwining with the fluctuation landscape naturally gives rise to a hierarchy of unconventional superconducting phases.
These findings highlight altermagnetic metals as promising platforms for realizing both nematic and topological superconductivity.

Although the phase diagrams were obtained within a Landau free-energy description, the multi-component nature of the superconducting order is generic for metallic altermagnets. The multi-component nature of these superconducting states implies that these phases also harbor ``daughter orders'' characterized by composite order parameters. This phenomenon has been discussed extensively in the context of pair-density-wave superconductors \cite{agterberg2008,berg2009a,Agterberg2020} and multi-component uniform superconductors \cite{Fernandes2021,Hecker2023}. In the present case, the natural composite orders are nematic order, given by 
\begin{equation}
N_{\rm nem}=(|\Delta^x_{A}|^2+|\Delta^x_{B}|^2)-(|\Delta^y_{A}|^2+|\Delta^y_{B}|^2)
\label{eq:nematic-op}
\end{equation}
and time-reversal symmetry breaking, given by
\begin{equation}
\kappa_s = i(\Delta^{x*}_{s}\Delta^y_{s}-\Delta^{x}_{s}\Delta^{y*}_{s}), \qquad s=A,B
\label{eq:trs}
\end{equation}
However, this bilinear composite order does not fix the relative phase between the two time-reversal breaking orders. This feature can be captured by the quartic composite operator
\begin{equation}
\tau \equiv \kappa_A\kappa_B=\Delta_A^{x*}\Delta_B^x\Delta_A^y\Delta_B^{y*}-\Delta_A^{x*}\Delta_B^{x*}\Delta_A^y\Delta_B^{y} + \text{c.c.}
\label{eq:Tau}
\end{equation}
which determines the relative sign.

The existence of composite daughter orders in some of the superconducting phases also suggest the existence of nematic and/or time-reversal symmetry breaking \textit{vestigial} orders which can arise at higher temperatures where superconducting coherence is lost \cite{berg2009b,Nie2017,Fernandes2019,Fradkin2015} but nematic and/or time-reversal symmetry breaking may persist. As in other superfluid/superconducting states with multiple order parameters we  expect that the condensed phases will also imply the  existence of higher charge condensates and of fractionalized vortices \cite{berg2009b,Zhou-2001,Mukerjee-2006,babaev2002,agterberg2008,Fernandes2021,jian2021,Gali2024,zou2025}. Such phases are characterized by particle-particle composite order parameters of the form $\Delta_{s}\Delta_{s'}$, and can arise when the two superconducting sectors are coupled in an effectively attractive manner while single-pair coherence is suppressed. Charge-$4e$ superconducting phases have been extensively discussed in the context of pair-density-wave and nematic superconductors. We will return to these problems in a separate publication.

On the other hand, phase-sensitive inter-band bilinears of the form $\Delta^{*}_{s}\Delta_{s'}$ with $s\neq s'$, which depend explicitly on the relative global phase between the two spin-band sectors, cannot acquire a finite expectation value here, since the relative phase between the two sectors remains unlocked. In real materials, however, inversion symmetry breaking or other interactions not included in our model, such as pair-hopping terms, can generate an effective Josephson coupling that locks the relative phase between the two sectors, in which case such phase-sensitive composite orders may become possible.

Taken together, these considerations suggest that the thermal melting of superconductivity in metallic altermagnets may proceed in a multi-step manner, yielding nematic and/or time-reversal symmetry breaking metallic vestigial phases, as well as possible higher-charge superconducting states. A complete characterization of these vestigial phases requires incorporating phase fluctuations beyond the present Landau free-energy analysis, which we leave for future work.

\begin{acknowledgments}
{\it Acknowledgments.}—We thank Jin-Chao Zhao, Shao-Kai Jian, Zhongbo Yan, Zhigang Wu, Wen Huang, Zhou-Quan Wan, Yu-Xuan Wang, Yi-Ming Wu, Erez Berg, Daniel Agterberg, Hong Yao and Yi-Ting Hsu for helpful discussions. X.Z. acknowledges support from the Tsinghua Visiting Doctoral Students Foundation during the stay at the Anthony J. Leggett Institute for Condensed Matter Theory (AJL-ICMT) at the University of Illinois Urbana–Champaign and thanks the AJL-ICMT for its hospitality. 
This work was supported in part by the US National Science Foundation grant DMR 2225920 at the University of Illinois (EF). R.M.F. acknowledges support from the Research Corporation for Science Advancement through the Cottrell SEED Award CS-SEED-2025-012. 
\end{acknowledgments}

\appendix
\numberwithin{equation}{section} 
\renewcommand\theequation{\thesection\arabic{equation}}

\section{Derivation of the Effective Two-Band Model}
\label{app:proj}
In this section, we explicitly demonstrate the relation between the four-band model and the effective lower two-band model, and clarify how a large $N_{\rm am}$ influences the basis transformation. 

In the main text, we consider a $d$-wave altermagnet (AM) on a tetragonal lattice described by
\begin{equation}
H_0 = \sum_{\mathbf k} c_{\mathbf k}^\dagger \mathcal H_0(\mathbf k) c_{\mathbf k},
\quad
c_{\mathbf k} = (c_{\mathbf k\uparrow 1},c_{\mathbf k\uparrow 2},c_{\mathbf k\downarrow 1},c_{\mathbf k\downarrow 2})^{T},
\end{equation}
where the Pauli matrices $\tau_{x,z}$ act in the $(1,2)$ sublattice space and $\sigma_z$ acts in spin space. The single-particle Hamiltonian reads
\begin{equation}
\begin{aligned}
\mathcal{H}_0(\boldsymbol{k}) ={}& -4t_1\cos\frac{k_x}{2}\cos\frac{k_y}{2}\tau_x
-2t_2(\cos k_x+\cos k_y)\tau_0 \\
&-2t_d(\cos k_x-\cos k_y)\tau_z
-\tilde{\mu}\tau_0
-N_{\rm am}\sigma_z\tau_z,
\end{aligned}
\end{equation}
where the physical meaning of the hopping parameters is discussed in the main text. In the absence of interactions, the spin sectors are decoupled in the single-particle Hamiltonian.

For a fixed spin projection $\sigma_z$ ($+1$ for $\uparrow$ and $-1$ for $\downarrow$), the spin-resolved $2\times2$ block takes the form
\begin{equation}
H_\sigma(\mathbf k)
= h_0(\mathbf k)\tau_0 + h_s(\mathbf k)\tau_x + \bigl[h_d(\mathbf k)-N_{\rm am}\sigma_z\bigr]\tau_z,
\end{equation}
where
\begin{align}
h_s(\mathbf k) &= -4t_1\cos\frac{k_x}{2}\cos\frac{k_y}{2}, \\ \nonumber
h_0(\mathbf k) &= -2t_2(\cos k_x+\cos k_y)-\tilde \mu, \\ \nonumber
h_d(\mathbf k) &= -2t_d(\cos k_x-\cos k_y).
\end{align}

To diagonalize this $2\times2$ block, we rewrite it explicitly as
\begin{equation}
H_\sigma(\mathbf k)=
\begin{pmatrix}
h_0(\mathbf k)+m_\sigma(\mathbf k) & h_s(\mathbf k) \\
h_s(\mathbf k) & h_0(\mathbf k)-m_\sigma(\mathbf k)
\end{pmatrix},
\end{equation}
with $m_\sigma(\mathbf k)=h_d(\mathbf k)-N_{\rm am}\sigma_z$.

For a given spin sector, the eigenvector corresponding to the upper band is
$(u_\sigma(\mathbf k), v_\sigma(\mathbf k))^T$,
whereas the eigenvector of the lower band is $(- v_\sigma(\mathbf k), u_\sigma(\mathbf k))^T$.
The factors are given by
\begin{align}
u_\sigma(\mathbf k) &= \frac{E_\sigma(\mathbf k)+m_\sigma(\mathbf k)}{\sqrt{\bigl(E_\sigma(\mathbf k)+m_\sigma(\mathbf k)\bigr)^2+h_s^2(\mathbf k)}}, \\
v_\sigma(\mathbf k) &= \frac{h_s(\mathbf k)}{\sqrt{\bigl(E_\sigma(\mathbf k)+m_\sigma(\mathbf k)\bigr)^2+h_s^2(\mathbf k)}},
\end{align}
where $E_\sigma(\mathbf k)=\sqrt{h_s^2(\mathbf k)+m_\sigma^2(\mathbf k)}.$

The corresponding unitary transformation is
\begin{equation}
U_\sigma(\mathbf k)=
\begin{pmatrix}
u_\sigma(\mathbf k) & -v_\sigma(\mathbf k) \\
v_\sigma(\mathbf k) & u_\sigma(\mathbf k)
\end{pmatrix}.
\end{equation}
After diagonalization, the band dispersions are
\begin{align}
\varepsilon_{\uparrow,A}(\mathbf k) &= h_0(\mathbf k)-E_\uparrow(\mathbf k), \\ \nonumber
\varepsilon_{\uparrow,B}(\mathbf k) &= h_0(\mathbf k)+E_\uparrow(\mathbf k), \\ \nonumber
\varepsilon_{\downarrow,A}(\mathbf k) &= h_0(\mathbf k)+E_\downarrow(\mathbf k), \\ \nonumber
\varepsilon_{\downarrow,B}(\mathbf k) &= h_0(\mathbf k)-E_\downarrow(\mathbf k).
\end{align}

The corresponding band operators are related to the original sublattice operators by
\begin{align}
c_{\mathbf k\uparrow A}
&= -v_\uparrow(\mathbf k)c_{\mathbf k\uparrow 1}
+ u_\uparrow(\mathbf k)c_{\mathbf k\uparrow 2}, \\ \nonumber
c_{\mathbf k\uparrow B}
&= u_\uparrow(\mathbf k)c_{\mathbf k\uparrow 1}
+ v_\uparrow(\mathbf k)c_{\mathbf k\uparrow 2}, \\ \nonumber
c_{\mathbf k\downarrow A}
&= u_\downarrow(\mathbf k)c_{\mathbf k\downarrow 1}
+ v_\downarrow(\mathbf k)c_{\mathbf k\downarrow 2}, \\ \nonumber
c_{\mathbf k\downarrow B}
&= -v_\downarrow(\mathbf k)c_{\mathbf k\downarrow 1}
+ u_\downarrow(\mathbf k)c_{\mathbf k\downarrow 2}.
\end{align}
Deep in the AM phase, $|N_{\rm am}|$ is large. In the regime $|N_{\rm am}|\gg |h_s(\mathbf k)|, |h_d(\mathbf k)|$, we obtain
\begin{align}
E_\sigma(\mathbf k)&=\sqrt{h_s^2(\mathbf k)+\bigl(h_d(\mathbf k)-\sigma_z N_{\rm am}\bigr)^2} \\ \nonumber
&\simeq |N_{\rm am}|-\sigma_z \mathrm{sgn}(N_{\rm am}) h_d(\mathbf k)+\frac{h_s^2(\mathbf k)}{2|N_{\rm am}|}+\mathcal O(N_{\rm am}^{-2}).
\end{align}
The corresponding band splitting is
\begin{align}
 E_\downarrow(\mathbf k)- E_\uparrow(\mathbf k)
&\simeq 2 \mathrm{sgn}(N_{\rm am}) h_d(\mathbf k)+\mathcal O(N_{\rm am}^{-1})\\ \nonumber
&= -4 t_d \mathrm{sgn}(N_{\rm am}) (\cos k_x-\cos k_y)+\cdots,
\end{align}
which implies that the effective spin-splitting strength in the effective two band model is $\phi \simeq t_d \mathrm{sgn}(N_{\rm am})$, and its sign reverses under $N_{\rm am}\to -N_{\rm am}$.

For large $|N_{\rm am}|$ and $\sigma_z N_{\rm am}>0$, the coherence factors admit the expansions
\begin{align}
u_\sigma(\mathbf k) &= \frac{h_s(\mathbf k)}{2|N_{\rm am}|} + \frac{h_s(\mathbf k)h_d(\mathbf k)}{2N_{\rm am}^2} + \mathcal O(N_{\rm am}^{-3}), \\
v_\sigma(\mathbf k) &= 1 - \frac{h_s^2(\mathbf k)}{8N_{\rm am}^2} + \mathcal O(N_{\rm am}^{-3}).
\end{align}
For $\sigma_z N_{\rm am}<0$, one instead obtains
\begin{align}
u_\sigma(\mathbf k) &= 1 - \frac{h_s^2(\mathbf k)}{8N_{\rm am}^2} + \mathcal O(N_{\rm am}^{-3}), \\
v_\sigma(\mathbf k) &= \frac{h_s(\mathbf k)}{2|N_{\rm am}|} - \frac{h_s(\mathbf k)h_d(\mathbf k)}{2N_{\rm am}^2} + \mathcal O(N_{\rm am}^{-3}).
\end{align}

For $N_{\rm am}>0$, the low-energy two-band basis is $c_{\boldsymbol{k}} = (c_{\boldsymbol{k}\uparrow A}, c_{\boldsymbol{k}\downarrow B})^{T}$, where the former is predominantly sublattice 1 for spin up and the latter is predominantly sublattice 2 for spin down.
In the new basis, the effective two band model reads
\begin{equation}
\mathcal{H}_0(\boldsymbol{k})=-2t_2 (\cos k_x + \cos k_y) s_0 
- 2\phi (\cos k_x - \cos k_y) s_z 
- \mu,
\end{equation}
where $\mu=\tilde{\mu}+N_{\rm am}$.

\section{Altermagnetic phases and nematic-spin-nematic phases}
\label{app:A0}

In the main text, we presented the results for the superconducting susceptibility in altermagnetic systems, which can be viewed as the analogue of the $\alpha$ phase in the nematic–spin–nematic (NSN) framework \cite{Wu2007}. 
In this Appendix, we first briefly review the structure of the NSN phases and clarify the correspondence between the $\alpha$ and $\beta$ phases and their altermagnetic counterparts. 

Fermi-surface instabilities have been studied in the spin-triplet channel with high orbital partial waves \cite{Wu2007}. Two primary phases were identified, referred to as the $\alpha$ and $\beta$ phases. The mean-field Hamiltonian is given by
\begin{equation}
\begin{aligned}
H = & \sum_{\boldsymbol{k}} 
c_{\boldsymbol{k},\alpha}^{\dagger}
\Big\{
\epsilon_{\boldsymbol{k}}
-\big[\boldsymbol{n}_1 \cos (l\theta)
+\boldsymbol{n}_2 \sin (l\theta)\big]
\cdot \boldsymbol{\sigma}_{\alpha\beta}
\Big\}
c_{\boldsymbol{k},\beta} \\
& + \frac{|\boldsymbol{n}_1|^2 + |\boldsymbol{n}_2|^2}{2|f_l^a|},
\end{aligned}
\label{eq:nem-spin-nem}
\end{equation}
where $\boldsymbol{n}_1$ and $\boldsymbol{n}_2$ are the order parameters describing the spin-triplet phase. The corresponding Landau free energy takes the form~\cite{Wu2007}
\begin{equation}
\begin{aligned}
F(\boldsymbol{n}_1,\boldsymbol{n}_2) = &\quad  r\left(|\boldsymbol{n}_1|^{2}+|\boldsymbol{n}_2|^{2}\right) + v_1\left(|\boldsymbol{n}_1|^{2}+|\boldsymbol{n}_2|^{2}\right)^{2}\\
&+ v_2|\boldsymbol{n}_1\times\boldsymbol{n}_2|^{2}+\cdots
\end{aligned}
\label{eq:landau-nem}
\end{equation}
The structure of the ordered state is determined by the quartic term $v_2|\boldsymbol{n}_1\times\boldsymbol{n}_2|^{2}$. Depending on the sign of $v_2$, two distinct phases emerge:
\begin{itemize}
    \item \textbf{$\beta$ phase ($v_2 < 0$):} 
    In this configuration, the two order parameters have equal magnitude,$|\boldsymbol{n}_1| = |\boldsymbol{n}_2| = \bar{n}$, and are mutually orthogonal, with $\boldsymbol{n}_1 = \bar{n} \hat{\boldsymbol{x}}$ and $\boldsymbol{n}_2 = \bar{n} \hat{\boldsymbol{y}}$. The spin-polarization axis winds by $\pm 2\pi l$ around the Fermi surface, leading to a momentum-dependent spin texture.

    \item \textbf{$\alpha$ phase ($v_2 > 0$):} 
    Here the order parameters satisfy $\boldsymbol{n}_1 = \bar{n} \hat{\boldsymbol{z}}$ and $\boldsymbol{n}_2 = 0$, resulting in spin-up and spin-down anisotropic Fermi surfaces. However, the combined operation of a spatial rotation by $\pi/l$ followed by a global spin flip remains a symmetry of the system. For even $l$, this symmetry coincides with that of altermagnets.
\end{itemize}

In the main text, we considered a $d$-wave altermagnetic metal. Deep in the AM phase, where spin-up (spin-down) electrons predominantly occupy sublattice $1$ ($2$), only two energy bands, $A$ and $B$, cross the Fermi level. The low-energy Hamiltonian reduces to a two-band form
\begin{equation}
\begin{split}
\mathcal{H}_0(\boldsymbol{k})
= {}& -2t_2 \big(\cos k_x + \cos k_y\big) s_0 \\
& - 2\phi \big(\cos k_x - \cos k_y\big) s_z - \mu,
\end{split}
\end{equation}
defined in the basis $c_{\boldsymbol{k}} = (c_{\boldsymbol{k}\uparrow A},  c_{\boldsymbol{k}\downarrow B})^{T}$. When the Fermi surface lies near the $\Gamma$ point, the continuum limit can be obtained by expanding the dispersion around $\boldsymbol{k} = 0$ using $\cos k_x \approx 1 - \tfrac{1}{2}k_x^2$ and $\cos k_y \approx 1 - \tfrac{1}{2}k_y^2$, with $k_x = k_F \cos\theta$ and $k_y = k_F \sin\theta$. Substituting these into $\mathcal{H}_0(\boldsymbol{k})$ yields
\begin{equation}
\begin{split}
\mathcal{H}_0(\boldsymbol{k}) 
&\approx -4t_2 - \mu + t_2 k_F^2 
      + \phi k_F^2 \cos(2\theta)s_z \\
&\simeq \xi - \delta \cos(l_{\alpha}\theta)s_z ,\label{eq:NSN}
\end{split}
\end{equation}
where $\xi$ is the isotropic part of the band dispersion and $\delta$ measures the amplitude of the spin splitting. The resulting structure is formally equivalent to the nematic–spin–nematic $\alpha$ phase with angular momentum $l_{\alpha} = 2$.

\section{Superconducting susceptibilities of spin-split Fermi surfaces}
\label{app:A}
This appendix derives the superconducting susceptibilities of a generic spin-split Fermi surface using the NSN framework reviewed in Appendix \ref{app:A0}. Superconducting instabilities in the $s$- and $d$-wave channels emerging from the NSN phase were investigated in Ref.~\cite{Soto-Garrido2014}. The analysis reveals a rich phase diagram featuring both uniform superconducting states and various pair-density-wave (PDW) phases (see also \cite{parthenios2025}). Here, we extend this analysis to the $p$-wave channel and to generic spin-split Fermi surfaces described by Eq. \ref{eq:NSN}. Generally, the splitting between the spin-up and spin-down Fermi surfaces naturally favors equal-spin $p$-wave pairing when $l_{\alpha}$ is even. To analyze the superconducting instabilities in a system with arbitrary $l_{\alpha}$ (even $l_\alpha$, corresponding to altermagnets or odd $l_\alpha$, corresponding to odd-parity magnets), we evaluate the superconducting susceptibility. A general Hamiltonian that captures both the $\alpha$ and $\beta$ phases is written as
\begin{equation}
H = H_0 - g H_p,
\end{equation}
where $H_0$ describes the normal-state band structure and $H_p$ encodes the effective pairing interaction with coupling strength $g$.

The non-interacting Hamiltonian is written as
\begin{equation}
H_0 = \sum_{\boldsymbol{k},\alpha} c_{\boldsymbol{k},\alpha}^\dagger \big( \epsilon_{\boldsymbol{k}} \sigma_0 + \boldsymbol{B}_\theta \cdot \boldsymbol{\sigma} \big) c_{\boldsymbol{k},\alpha},
\end{equation}
where the components of the effective field $\boldsymbol{B}_\theta$ are
\begin{equation*}
B_{\theta,z} = -\delta \cos(l_{\alpha}\theta),~
B_{\theta,x} = -\bar{n} \cos(l_{\beta}\theta),~
B_{\theta,y}= -\bar{n} \sin(l_{\beta}\theta),
\end{equation*}
and its magnitude is
\begin{equation}
|\boldsymbol{B}_\theta| = \sqrt{\bar{n}^{2} + \big[\delta \cos(l_{\alpha}\theta)\big]^{2}}.
\end{equation}
In the following, we keep the discussion general and allow for both $\alpha$ and $\beta$ phases. We note, however, that in a two-dimensional tetragonal altermagnetic system, the constraints imposed by $C_4 \mathcal{T}$ symmetry enforce $\bar{n}=0$, corresponding to a pure $\alpha$ phase. The altermagnetic $\beta$ phase can emerge when spin-orbit coupling is present and if either additional symmetries are broken or for lattices in which the $d_{x^2-y^2}$ and $d_{xy}$ altermagnetic order parameters transform as the same irreducible representation, such as hexagonal lattices. 

For mixed-spin pairing, the interaction term reads
\begin{align}
H_p^{\uparrow\downarrow}
&= \sum_{\boldsymbol{k},\boldsymbol{k}',\boldsymbol{q}}
\gamma_{\boldsymbol{k}}\gamma_{\boldsymbol{k}'}
c_{\boldsymbol{k}+\frac{\boldsymbol{q}}{2},\uparrow}^\dagger
c_{-\boldsymbol{k}+\frac{\boldsymbol{q}}{2},\downarrow}^\dagger
c_{-\boldsymbol{k}'+\frac{\boldsymbol{q}}{2},\downarrow}
c_{\boldsymbol{k}'+\frac{\boldsymbol{q}}{2},\uparrow}
+ (\uparrow \leftrightarrow \downarrow).
\end{align}
For equal-spin pairing, it becomes
\begin{align}
H_p^{\uparrow\uparrow}
&= \sum_{\boldsymbol{k},\boldsymbol{k}',\boldsymbol{q}}
\gamma_{\boldsymbol{k}}\gamma_{\boldsymbol{k}'}
c_{\boldsymbol{k}+\frac{\boldsymbol{q}}{2},\uparrow}^\dagger
c_{-\boldsymbol{k}+\frac{\boldsymbol{q}}{2},\uparrow}^\dagger
c_{-\boldsymbol{k}'+\frac{\boldsymbol{q}}{2},\uparrow}
c_{\boldsymbol{k}'+\frac{\boldsymbol{q}}{2},\uparrow}
+ (\uparrow \leftrightarrow \downarrow).
\end{align}
The form factor $\gamma_{\boldsymbol{k}}$ satisfies
\begin{equation}
\gamma_{-\boldsymbol{k}} = (-1)^{l_\gamma} \gamma_{\boldsymbol{k}},
\end{equation}
where $l_\gamma$ is the orbital angular momentum of the pairing channel. Typical basis functions are
\begin{equation}
\gamma_{\boldsymbol{k}} = 
\begin{cases}
1, & s\text{-wave},\\
\sqrt{2}\cos 2\theta, & d\text{-wave},\\
\sqrt{2}\cos \theta, & p_x\text{-wave}
\end{cases}
\end{equation}
The superconducting susceptibility is given by
\begin{align}
\chi_{\boldsymbol{Q}}= \int \frac{d^2 k}{(2\pi)^2}
|\gamma_{\boldsymbol{k}}|^2
\frac{1 - n_F[\epsilon(\boldsymbol{k} + \boldsymbol{Q}/2)]
        - n_F[\epsilon(-\boldsymbol{k} + \boldsymbol{Q}/2)]}
     {\epsilon(\boldsymbol{k} + \boldsymbol{Q}/2)
      + \epsilon(-\boldsymbol{k} + \boldsymbol{Q}/2)} .
\end{align}
To simplify the integration near the Fermi surface, we transform
\begin{equation}
\int \frac{d^2 k}{(2\pi)^2} \rightarrow N(E_F) \int_{-\Lambda}^{\Lambda} d\xi \int_0^{2\pi} \frac{d\theta}{2\pi},
\end{equation}
where $N(E_F)$ is the density of states at the Fermi level and $\Lambda$ is an effective pairing cutoff scale. Here $\xi = \epsilon_k - \epsilon_F$ denotes the energy measured from the Fermi level, $\theta$ is the angle of $\boldsymbol{k}$, and $\phi_Q$ is the angle of $\boldsymbol{Q}$.

We first perform the integration over $\xi$. Using
\begin{equation}
\begin{aligned}
E_1 \left(\boldsymbol{k}+\tfrac{\boldsymbol{Q}}{2}\right) &= \xi - \sqrt{\bar{n}^2 + [\delta \cos(l_\alpha \theta)]^2} + \tfrac{Q}{2}\cos(\theta - \phi_Q),\\
E_1 \left(-\boldsymbol{k}+\tfrac{\boldsymbol{Q}}{2}\right) &= \xi - \sqrt{\bar{n}^2 + [\delta \cos(l_\alpha \theta)]^2} - \tfrac{Q}{2}\cos(\theta - \phi_Q),\\
E_2 \left(\boldsymbol{k}+\tfrac{\boldsymbol{Q}}{2}\right) &= \xi + \sqrt{\bar{n}^2 + [\delta \cos(l_\alpha \theta)]^2} + \tfrac{Q}{2}\cos(\theta - \phi_Q),\\
E_2 \left(-\boldsymbol{k}+\tfrac{\boldsymbol{Q}}{2}\right) &= \xi + \sqrt{\bar{n}^2 + [\delta \cos(l_\alpha \theta)]^2} - \tfrac{Q}{2}\cos(\theta - \phi_Q),
\end{aligned}
\end{equation}
we define
\begin{equation}
\chi_{ab} = \int_{-\Lambda}^{\Lambda} d\xi
\frac{1 - n_F[E_a(-\boldsymbol{k} + \boldsymbol{Q}/2)] - n_F[E_b(\boldsymbol{k} + \boldsymbol{Q}/2)]}
{E_a(-\boldsymbol{k} + \boldsymbol{Q}/2) + E_b(\boldsymbol{k} + \boldsymbol{Q}/2)}.
\label{chiab}
\end{equation}

At zero temperature, we obtain
\begin{align}
\chi_{11} = \chi_{22}
&= \tfrac{1}{2} \Big[
\ln \left| \frac{\Lambda + \sqrt{\bar{n}^2 + (\delta \cos l_\alpha \theta)^2}}
                 {\tfrac{Q}{2}\cos(\theta - \phi_Q)} \right| \nonumber\\
&\quad
+ \ln \left| \frac{\Lambda - \sqrt{\bar{n}^2 + (\delta \cos l_\alpha \theta)^2}}
                 {\tfrac{Q}{2}\cos(\theta - \phi_Q)} \right|
\Big],
\end{align}
and
\begin{align}
\chi_{12}(\boldsymbol{Q}) &= \chi_{21}(-\boldsymbol{Q})\nonumber\\
&= \ln \left|
\frac{\Lambda}
{\sqrt{\bar{n}^2 + (\delta \cos l_\alpha \theta)^2}
 - \tfrac{Q}{2}\cos(\theta - \phi_Q)}
\right|.
\end{align}

The difference between $\chi_{12}$ and $\chi_{21}$ vanishes after integrating over $\theta$. The term $\chi_{11}$ contributes to the logarithmic divergence at $Q = 0$, whereas $\chi_{12}$ and $\chi_{21}$ produce finite peaks at nonzero momentum.

\subsection{Mixed-spin pairing}
\begin{figure}[h]
    \centering
    \includegraphics[width=0.49\linewidth]{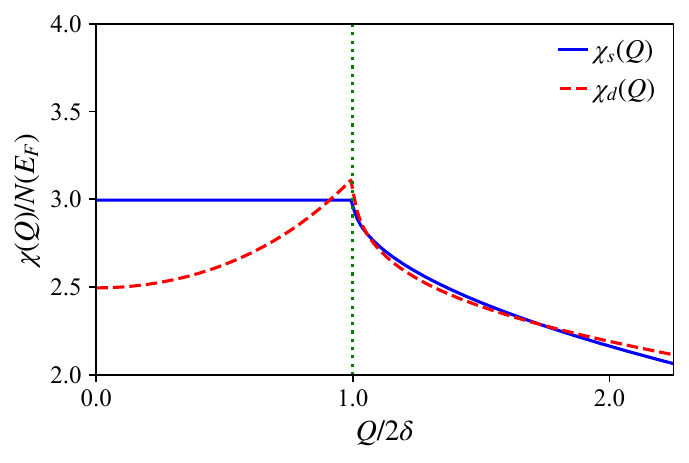}
    \includegraphics[width=0.49\linewidth]{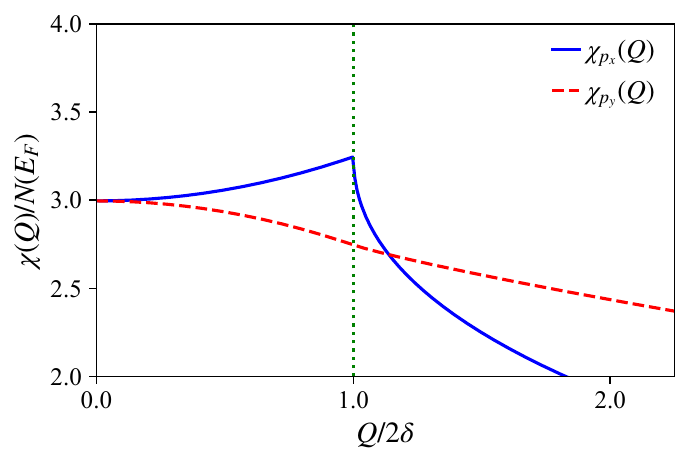}
    \caption{Mixed-spin superconducting susceptibility $\chi(Q)/N(E_F)$ versus pair momentum $Q$ in the $\alpha$ phase ($\bar n=0$, $l_\alpha=2$). Left: $s$- and $d$-wave form factors. Right: $p$-wave channels ($p_x$, $p_y$, and chiral $p_x \pm i p_y$). Both panels show a finite-momentum peak at $Q=2\delta$ (vertical dotted line) arising from the lack of perfect nesting on spin-split Fermi surfaces; here $\phi_Q=0$ and $Q$ is measured in units of $2\delta$}
    \label{fig:alpha-mixed-pair}
\end{figure}

Let us define $L=l_\alpha+l_\beta+l_\gamma$. For mixed-spin pairing with even $L$, the superconducting susceptibility at momentum $\boldsymbol{Q}$ is
\begin{equation}
\frac{\chi_{\boldsymbol{Q}}}{N(E_F)}=\int \frac{d\theta}{2\pi} |\gamma_{\boldsymbol{k}}|^{2} \chi_{12}.
\end{equation}
For odd $L$ it takes the form
\begin{align}
\frac{\chi_{\boldsymbol{Q}}}{N(E_F)}
&= \int \frac{d\theta}{2\pi}
|\gamma_{\boldsymbol{k}}|^{2} \notag \\
&\times \left[
\frac{\bar{n}^{2}}{\bar{n}^{2}+(\delta\cos l_\alpha\theta)^{2}}\chi_{11}
+\frac{(\delta\cos l_\alpha\theta)^{2}}{\bar{n}^{2}+(\delta\cos l_\alpha\theta)^{2}}\chi_{12}
\right].
\end{align}
In the $\alpha$ phase ($\bar{n}=0$), even and odd values of $L$ yield identical results. The mixed-spin susceptibility reduces to
\begin{equation}
\frac{\chi_{\boldsymbol{Q}}}{N(E_F)}=\int \frac{d\theta}{2\pi} |\gamma_{\boldsymbol{k}}|^{2} \ln\left|\frac{\Lambda}{\delta\cos(l_\alpha\theta)-\tfrac{Q}{2}\cos(\theta-\phi_Q)}\right|,
\end{equation}
which coincides with Ref.~\cite{Soto-Garrido2014}. Setting $l_\alpha=2$ yields the curves shown in Fig.~\ref{fig:alpha-mixed-pair}.

\begin{figure}[h]
    \centering
    \includegraphics[width=0.49\linewidth]{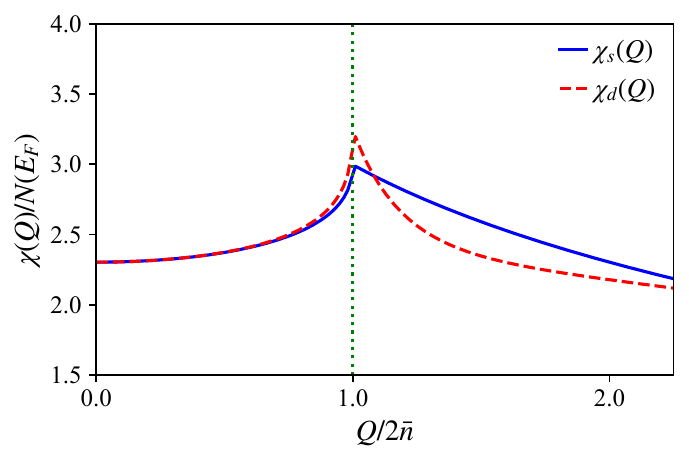}
    \includegraphics[width=0.49\linewidth]{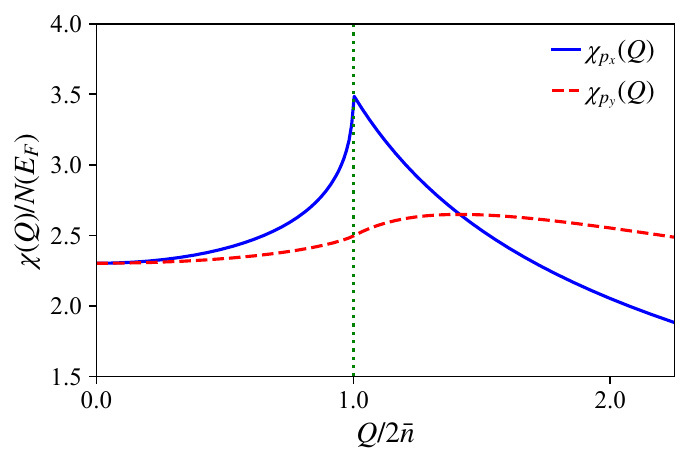}
    \caption{Mixed-spin superconducting susceptibility $\chi(Q)/N(E_F)$ versus pair momentum $Q$ in the $\beta$ phase ($\delta=0$) with even $L=l_\beta+l_\gamma$. Left: $s$- and $d$-wave form factors ($l_\beta$ even, $l_\gamma$ even). Right: $p$-wave channels ($l_\beta$ odd, $l_\gamma$ odd). Both panels show a finite-momentum peak at $Q=2\bar{n}$ (vertical dotted line); here $\phi_Q=0$ and $Q$ is measured in units of $2\bar{n}$.}
    \label{fig:beta-even-pair}
\end{figure}

In the $\beta$ phase ($\delta=0$) and for even $L=l_\beta+l_\gamma$, the mixed-spin susceptibility reduces to
\begin{equation}
\frac{\chi_{\boldsymbol{Q}}}{N(E_F)}=\int \frac{d\theta}{2\pi} |\gamma_{\boldsymbol{k}}|^{2} \ln\left|\frac{\Lambda}{\bar{n}-\tfrac{Q}{2}\cos(\theta-\phi_Q)}\right|,
\end{equation}
which matches Ref.~\cite{Soto-Garrido2014}. The two allowed parity assignments are shown in Fig.~\ref{fig:beta-even-pair}.

\begin{figure}[h]
    \centering
    \includegraphics[width=0.49\linewidth]{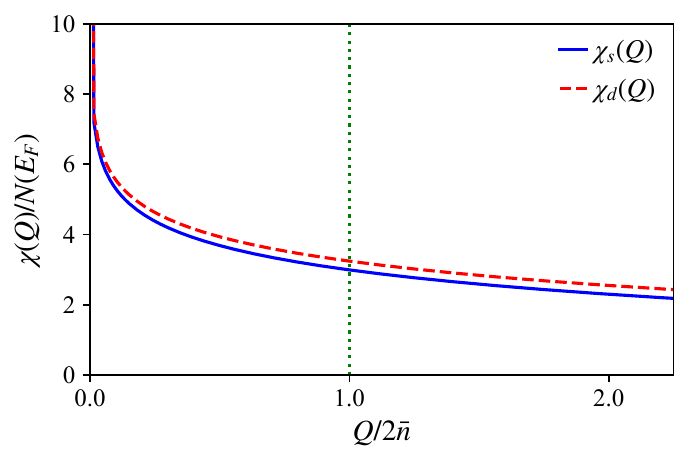}
    \includegraphics[width=0.49\linewidth]{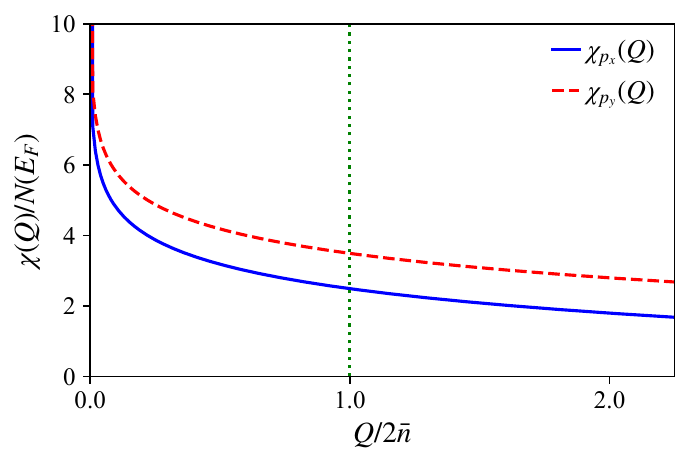}
    \caption{Mixed-spin superconducting susceptibility $\chi(Q)/N(E_F)$ versus pair momentum $Q$ in the $\beta$ phase ($\delta=0$) with odd $L=l_\beta+l_\gamma$. Left: $s$- and $d$-wave form factors ($l_\beta$ odd, $l_\gamma$ even). Right: $p$-wave channels ($l_\beta$ even, $l_\gamma$ odd). Both panels show a peak at $Q=0$ arising from the logarithmic enhancement in the odd-$L$ sector; here $\phi_Q=0$ and $Q$ is measured in units of $2\bar{n}$.}
    \label{fig:beta-odd-pair}
\end{figure}

In the $\beta$ phase ($\delta=0$) and for odd $L$ the mixed-spin susceptibility is
\begin{align}
\frac{\chi_{\boldsymbol{Q}}}{N(E_F)}
&= \int \frac{d\theta}{2\pi}
|\gamma_{\boldsymbol{k}}|^{2} \notag \\
&\times \frac{1}{2}\left[
\ln\left|\frac{\Lambda+\bar{n}}{\tfrac{Q}{2}\cos(\theta-\phi_Q)}\right|
+\ln\left|\frac{\Lambda-\bar{n}}{\tfrac{Q}{2}\cos(\theta-\phi_Q)}\right|
\right].
\end{align}
The two allowed parity assignments are shown together in Fig.~\ref{fig:beta-odd-pair}.

\subsection{Equal-spin pairing}
For equal-spin pairing, only the spin-triplet channel is allowed, implying that $l_\gamma$ must be odd. The superconducting susceptibility at momentum $\boldsymbol{Q}$ is given by

\begin{align}
\frac{\chi(\mathbf{Q})}{N(E_F)}
= \int \frac{d\theta}{2\pi} |\gamma_{\mathbf{k}}|^2 \Bigg[\chi_{11} \notag &+ 
\frac{(\delta \cos l_\alpha \theta)^2}{\bar{n}^2 + (\delta \cos l_\alpha \theta)^2}\chi_{11} \\
&+ \frac{\bar{n}^2}{\bar{n}^2 + (\delta \cos l_\alpha \theta)^2}\chi_{12}
\Bigg].
\end{align}
In the $\alpha$ phase ($\bar n=0$), the susceptibility is
\begin{align}
\frac{\chi_{\boldsymbol{Q}}}{N(E_F)}
&= \int \frac{d\theta}{2\pi} |\gamma_{\boldsymbol{k}}|^{2} \notag \\
&\times \left[
\ln\left|\frac{\Lambda+\delta\cos(l_\alpha\theta)}{\tfrac{Q}{2}\cos(\theta-\phi_Q)}\right|
+\ln\left|\frac{\Lambda-\delta\cos(l_\alpha\theta)}{\tfrac{Q}{2}\cos(\theta-\phi_Q)}\right|
\right],
\label{eq:alpha-equal-spin}
\end{align}
which yields a logarithmic divergence as $Q\to 0$. In the $\beta$ phase ($\delta=0$), the susceptibility reads
\begin{align}
&\frac{\chi_{\boldsymbol{Q}}}{N(E_F)}\nonumber\\
&= \int \frac{d\theta}{2\pi} |\gamma_{\boldsymbol{k}}|^{2}
\frac{1}{2}
\left[
\ln\left|\frac{\Lambda+\bar n}{\tfrac{Q}{2}\cos(\theta-\phi_Q)}\right|
+\ln\left|\frac{\Lambda-\bar n}{\tfrac{Q}{2}\cos(\theta-\phi_Q)}\right|
\right] \notag \\
&\quad + \int \frac{d\theta}{2\pi} |\gamma_{\boldsymbol{k}}|^{2}
\ln\left|
\frac{\Lambda}{
\bar n-\tfrac{Q}{2}\cos(\theta-\phi_Q)}
\right|,
\label{eq:beta-equal-spin}
\end{align}
so that the first term also produces a divergence when $Q\to 0$. Meanwhile, the second term generates a finite-momentum feature near $Q=2\bar n$.

\begin{figure}[h]
\centering
\includegraphics[width=0.49\linewidth]{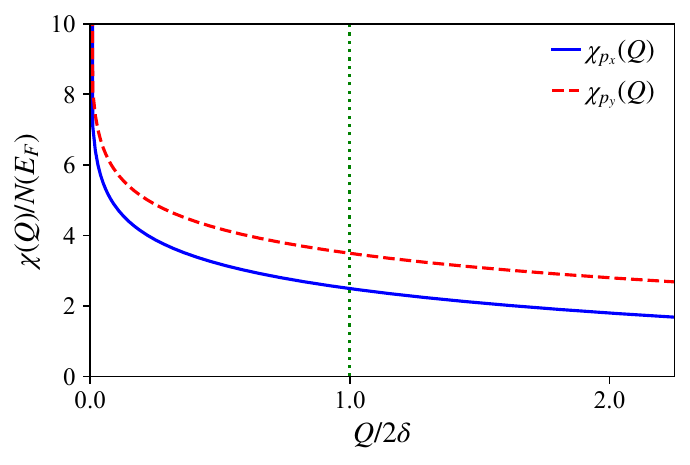}
\includegraphics[width=0.49\linewidth]{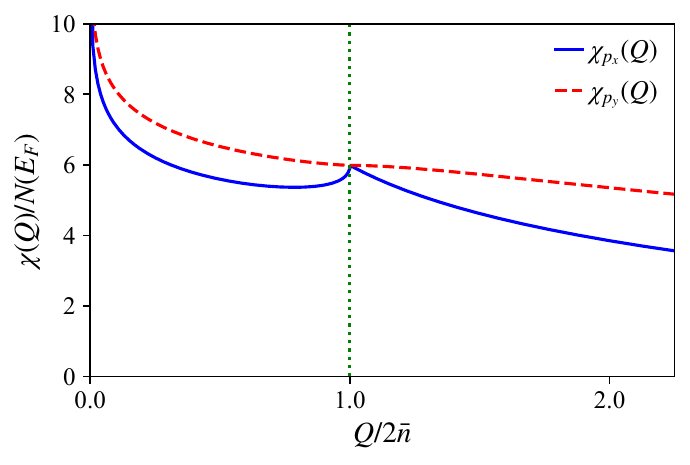}
\caption{Equal-spin (triplet) $p$-wave superconducting susceptibility $\chi(Q)/N(E_F)$ versus pair momentum $Q$ for the $p_x$, $p_y$, and chiral $p_x\pm i p_y$ form factors. Left: $\alpha$ phase ($\bar n=0$, representative $l_\alpha=2$) showing the logarithmic divergence as $Q\to 0$ implied by Eq.~\eqref{eq:alpha-equal-spin}. Right: $\beta$ phase ($\delta=0$) illustrating the combination of a $Q\to 0$ divergence and a finite-momentum feature near $Q=2\bar n$ from Eq.~\eqref{eq:beta-equal-spin}. Here $\phi_Q=0$ and $Q$ is measured in units of the indicated splitting.}
\label{fig:equal-spin-p-pair}
\end{figure}

\section{Mean field theory of the SC states}
\label{app:B}

\subsection{Spinless model}
\label{app:spinless}

In the absence of spin–orbit coupling and fluctuation effects, equal-spin pairing in the spin-up and spin-down sectors are decoupled. As a warm-up, we first analyze a spinless model that captures an individual  equal-spin sector; 
in the next subsection we incorporate couplings between the two spin Fermi surfaces.

For the spinless case, the non-interacting Hamiltonian is
\begin{equation}
H_0=\sum_{\boldsymbol{k}}\big[-2t_{x}\cos k_x-2t_{y}\cos k_y-\mu\big] c_{\boldsymbol{k}}^\dagger c_{\boldsymbol{k}}
=\sum_{\boldsymbol{k}}\xi_{\boldsymbol{k}} c_{\boldsymbol{k}}^\dagger c_{\boldsymbol{k}},
\end{equation}
where $\xi_{\boldsymbol{k}}=-2t_{x}\cos k_x-2t_{y}\cos k_y-\mu$.
The interaction is taken separable in the $p$-wave basis,
\begin{equation}
H_{\text{int}}=\frac{1}{2}\sum_{\boldsymbol{k},\boldsymbol{k}'}V(\boldsymbol{k}-\boldsymbol{k}') c_{\boldsymbol{k}}^\dagger c_{-\boldsymbol{k}}^\dagger c_{-\boldsymbol{k}'}c_{\boldsymbol{k}'},
\end{equation}
with $V(\boldsymbol{k}-\boldsymbol{k}')=-2V_x g_x(\boldsymbol{k})g_x^*(\boldsymbol{k}')-2V_y g_y(\boldsymbol{k})g_y^*(\boldsymbol{k}')$, $g_x(\boldsymbol{k})=\sin k_x$ and $g_y(\boldsymbol{k})=\sin k_y$. The gap function is
\begin{equation}
\Delta(\boldsymbol{k})=-\frac{1}{N_L}\sum_{\boldsymbol{k}'}V(\boldsymbol{k}-\boldsymbol{k}') \langle c_{\boldsymbol{k}'}c_{-\boldsymbol{k}'}\rangle,
\end{equation}
where $N_L$ is the number of lattice sites. Introducing order parameters
\begin{equation}
\Delta^{\eta}=-\frac{2}{N_L}\sum_{\boldsymbol{k}'}V_{\eta} g_{\eta}^*(\boldsymbol{k}') \langle c_{\boldsymbol{k}'}c_{-\boldsymbol{k}'}\rangle,\quad \eta=x,y,
\end{equation}
the form of the gap is
\begin{equation}
\Delta(\boldsymbol{k})=-\Delta^{x}g_x(\boldsymbol{k})-\Delta^{y}g_y(\boldsymbol{k}).
\end{equation}
Within mean-field theory,
\begin{equation}
H_{\text{int}}= \frac{1}{2}\sum_{\boldsymbol{k}}\Big[\Delta^*(\boldsymbol{k}) c_{-\boldsymbol{k}}c_{\boldsymbol{k}}+\Delta(\boldsymbol{k}) c_{\boldsymbol{k}}^\dagger c_{-\boldsymbol{k}}^\dagger\Big]+K,
\end{equation}
with $K=-\frac{1}{2}\sum_{\boldsymbol{k},\boldsymbol{k}'}V(\boldsymbol{k}-\boldsymbol{k}') \langle c_{\boldsymbol{k}}^\dagger c_{-\boldsymbol{k}}^\dagger\rangle\langle c_{-\boldsymbol{k}'}c_{\boldsymbol{k}'}\rangle$.
In the Nambu basis $\Psi_{\boldsymbol{k}}=(c_{\boldsymbol{k}}, c_{-\boldsymbol{k}}^\dagger)^{T}$ the Bogoliubov–de Gennes Hamiltonian is
\begin{equation}
H=\frac{1}{2}\sum_{\boldsymbol{k}}\Psi_{\boldsymbol{k}}^\dagger
\begin{pmatrix}
\xi_{\boldsymbol{k}} & \Delta(\boldsymbol{k})\\
\Delta^*(\boldsymbol{k}) & -\xi_{\boldsymbol{k}}
\end{pmatrix}
\Psi_{\boldsymbol{k}}
+\frac{1}{2}\sum_{\boldsymbol{k}}\xi_{\boldsymbol{k}}+K.
\end{equation}
The Bogoliubov transformation is given by
\begin{eqnarray}
\left(
\begin{array}{c}
c_{\boldsymbol{k}}\\
c_{\boldsymbol{-k}}^\dagger 
\end{array}
\right)
=
\left(
\begin{array}{cc}
u_{\boldsymbol{k}} & -v_{\boldsymbol{k}} \\
v_{\boldsymbol{k}}^* & u_{\boldsymbol{k}}^* 
\end{array}
\right)
\left(
\begin{array}{c}
\gamma_{\boldsymbol{k}}\\
\gamma_{\boldsymbol{-k}}^\dagger 
\end{array}
\right).
\end{eqnarray}
After diagonalization, the Hamiltonian becomes
\begin{eqnarray}
H=\frac{1}{2}\sum_{\boldsymbol{k}}E_{\boldsymbol{k}}\big(\gamma_{\boldsymbol{k}}^\dagger \gamma_{\boldsymbol{k}}-\gamma_{\boldsymbol{-k}} \gamma_{\boldsymbol{-k}}^\dagger\big)+\frac{1}{2}\sum_{\boldsymbol{k}}\xi_{\boldsymbol{k}}+K,
\end{eqnarray}
where $E_{\boldsymbol{k}}=\sqrt{\xi_{\boldsymbol{k}}^2+\Delta^*(\boldsymbol{k})\Delta(\boldsymbol{k})}$, and the Bogoliubov coefficients are
$u_{\boldsymbol{k}}=\frac{E_{\boldsymbol{k}}+\xi_{\boldsymbol{k}}}{\sqrt{(E_{\boldsymbol{k}}+\xi_{\boldsymbol{k}})^2+|\Delta(\boldsymbol{k})|^2}}, \quad
v_{\boldsymbol{k}}=\frac{\Delta(\boldsymbol{k})}{\sqrt{(E_{\boldsymbol{k}}+\xi_{\boldsymbol{k}})^2+|\Delta(\boldsymbol{k})|^2}}.$

The self-consistent equation is
\begin{eqnarray}
\Delta^\eta=\frac{1}{N_L}\sum_{\boldsymbol{k}}g_\eta^* V_\eta \frac{\tanh(\frac{E_{\boldsymbol{k}}}{2T})}{E_{\boldsymbol{k}}}\big(\Delta^x g_x+\Delta^y g_y\big),
\end{eqnarray}
and the free energy is
\begin{align}
F_s &= \langle H\rangle - TS = \frac{1}{2}\sum_{\boldsymbol{k}}
\frac{\Delta^*(\boldsymbol{k})\Delta(\boldsymbol{k})}{2E_{\boldsymbol{k}}}
\tanh\left(\frac{E_{\boldsymbol{k}}}{2T}\right) \notag \\
&\quad + \frac{1}{2}\sum_{\boldsymbol{k}}
\big(\xi_{\boldsymbol{k}} - E_{\boldsymbol{k}}\big)
+ T\sum_{\boldsymbol{k}}\ln\big(1 - f_{\boldsymbol{k}}\big),
\end{align}
where $f_{\boldsymbol{k}}=\frac{1}{e^{E_{\boldsymbol{k}}/T}+1}$. At $T=0$, the self-consistent equation becomes
\begin{eqnarray}
\Delta^\eta=\frac{1}{N_L}\sum_{\boldsymbol{k}} V_\eta \frac{\sin k_{\eta}}{E_{\boldsymbol{k}}}\big(\Delta^x \sin k_x+\Delta^y \sin k_y\big),
\end{eqnarray}
and the condensation energy is
\begin{align}
E_{\text{cond}} &= E_s - E_n \notag \\
&= \frac{1}{2}\sum_{\boldsymbol{k}}
\big(|\xi_{\boldsymbol{k}}| - E_{\boldsymbol{k}}\big) + \frac{1}{2}\sum_{\boldsymbol{k}}
\frac{\Delta^*(\boldsymbol{k})\Delta(\boldsymbol{k})}{2E_{\boldsymbol{k}}}.
\end{align}
We parametrize $(\Delta^{x},\Delta^{y})=\Delta e^{i\theta} (\cos\alpha, e^{i\beta}\sin\alpha)$ with $\theta\in[0,2\pi)$, $\alpha\in[0,\pi/2]$ and $\beta\in[-\pi,\pi)$. Minimizing the condensation energy over $(\Delta,\alpha,\beta)$ yields $\beta=\pm\pi/2$ generically, except when one component vanishes ($\alpha=0$ or $\pi/2$). Because the band has only $C_2$ (not $C_4$) symmetry, equal weights for the two $p$-wave components are not required.

\begin{figure}
\centering
\includegraphics[width=1.0\linewidth]{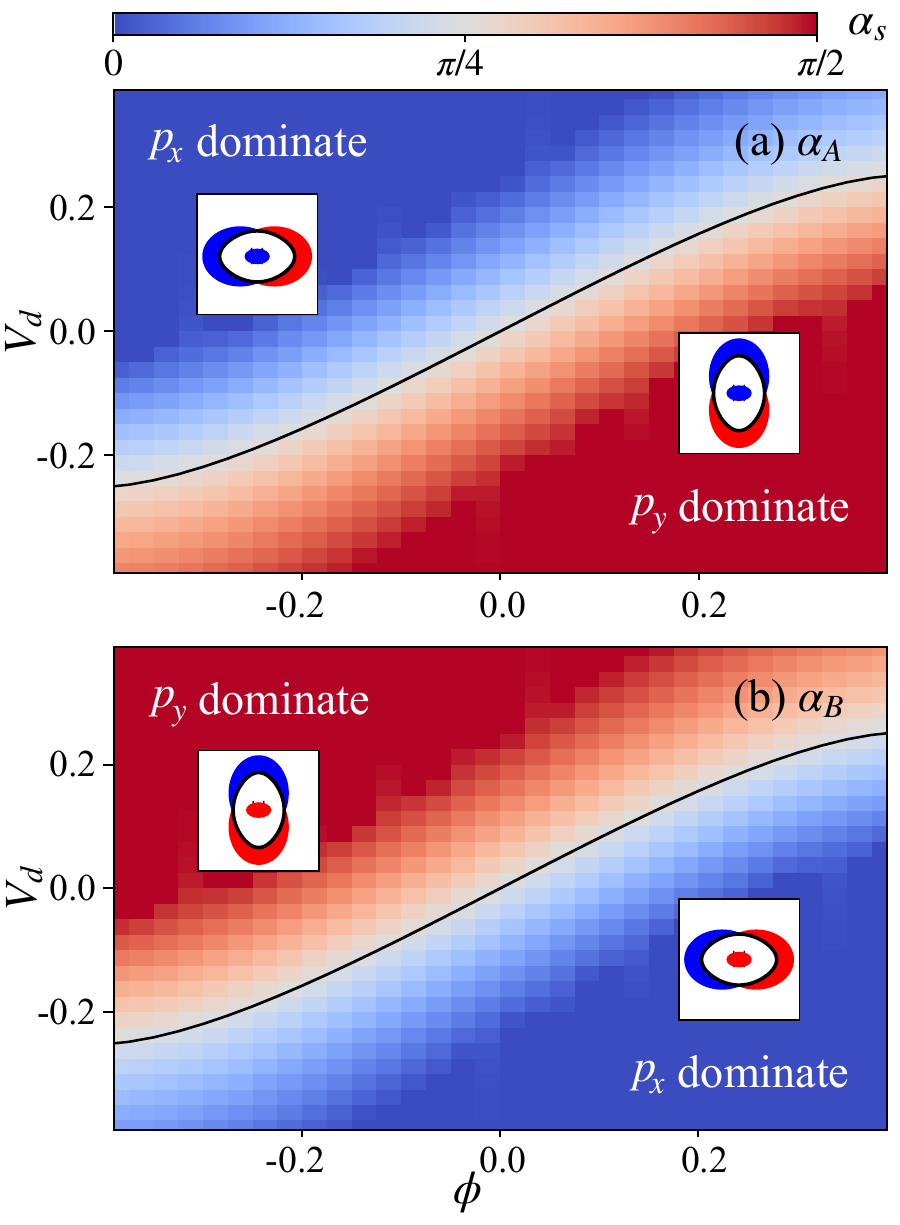}  
\caption{
Anisotropic superconducting ground states in altermagnetic systems without additional fluctuations obtained for $t_2=0.75$, $V_2=2.25$ and $\mu=-2.1$. 
The angle $\alpha_s=\tan^{-1} \big(|\Delta_s^y|/|\Delta_s^x|\big)$ is shown as a function of hopping difference $\phi$ and pairing anisotropy $V_d$ for (a) the $A$ and (b) the $B$ bands. 
Insets: schematic $p_x$ and $p_y$ configurations for $\phi<0$ and $\phi>0$.
The relation $\alpha_A+\alpha_B=\pi/2$ holds throughout, consistent with $C_4\mathcal{T}$ symmetry. 
The solid line marks $\alpha_s=\pi/4$.
}
\label{fig3}
\end{figure}

We now include both spin sectors while keeping them decoupled. The Hamiltonian for the second spin sector has the same form as that of the first, with $t_x \leftrightarrow t_y$ and $V_x \leftrightarrow V_y$. Within this mean-field theory, we determine the superconducting ground states of the altermagnetic system. For $t_2=(t_x+t_y)/2=0.75$, $V_2=(V_x+V_y)/2=2.25$, and $\mu=-2.1$, we obtain the superconducting ground states shown in Fig.~\ref{fig3}.
Panels (a) and (b) show the angle $\alpha_s=\tan^{-1} \big(|\Delta_s^y|/|\Delta_s^x|\big)$ on the $A$ and $B$ bands, respectively, as functions of the effective hopping anisotropy $\phi=(t_y-t_x)/2$ and pairing anisotropy $V_d=(V_y-V_x)/2$. In both panels, blue and red regions correspond to $p_x$-dominated and $p_y$-dominated superconducting states.

The two bands exhibit complementary behavior: $\alpha_A$ and $\alpha_B$ are related by $\alpha_A+\alpha_B=\pi/2$ throughout the entire parameter space, reflecting that the superconducting state preserves the combined $C_4\mathcal{T}$ symmetry of the AM state. The solid curve marks the locus $\alpha_s=\pi/4$, where the $p_x$ and $p_y$ components contribute equally and the order parameter reduces to the conventional $p_x\pm i p_y$ form. The $\alpha_s$ generally deviates from $\pi/4$, varying smoothly between $0$ and $\pi/2$ across the parameter space. The insets illustrate the corresponding $p_x$ and $p_y$ configurations for representative values of $\phi<0$ and $\phi>0$.

\subsection{Spinful model}
For an altermagnetic system with two spin sectors, the Bogoliubov–de Gennes Hamiltonian in the Nambu basis $\Psi_{\boldsymbol{k}}=(c_{\boldsymbol{k}A}, c_{\boldsymbol{k}B}, c^\dagger_{-\boldsymbol{k}A}, c^\dagger_{-\boldsymbol{k}B})^{T}$ is
\begin{equation}
H_{\text{BdG}}(\boldsymbol{k})=
\begin{pmatrix}
\xi^A_{\boldsymbol{k}} & h_{ab}(\boldsymbol{k}) & \Delta_A(\boldsymbol{k}) & 0 \\
h_{ab}^*(\boldsymbol{k}) & \xi^B_{\boldsymbol{k}} & 0 & \Delta_B(\boldsymbol{k}) \\
\Delta_A^*(\boldsymbol{k}) & 0 & -\xi^A_{\boldsymbol{k}} & -h_{ab}^*(-\boldsymbol{k}) \\
0 & \Delta_B^*(\boldsymbol{k}) & -h_{ab}(-\boldsymbol{k}) & -\xi^B_{\boldsymbol{k}}
\end{pmatrix}.
\end{equation}
The superconducting gaps are taken in the $p$-wave basis as
\begin{align}
\Delta_A(\boldsymbol{k}) &= -\Delta_A^x \sin k_x - \Delta_A^y \sin k_y, \notag \\
\Delta_B(\boldsymbol{k}) &= -\Delta_B^x \sin k_x - \Delta_B^y \sin k_y,
\end{align}
and the normal-state dispersions are
\begin{align}
\xi^A &= -2(t_{2}+\phi)\cos k_x - 2(t_{2}-\phi)\cos k_y - \mu, \notag \\
\xi^B &= -2(t_{2}-\phi)\cos k_x - 2(t_{2}+\phi)\cos k_y - \mu.
\end{align}
Defining the Green's function $G^{-1}(\boldsymbol{k},i\omega_n)=i\omega_n-H_{\text{BdG}}(\boldsymbol{k})$, where $\omega_n = (2n+1)\pi T$ is the Matsubara frequency, the free energy is given by
\begin{align}
F = &-\frac{T}{2N_L}
\sum_{\boldsymbol{k},n}
\ln\det G^{-1}(\boldsymbol{k},i\omega_n) \notag \\
&\quad + \frac{|\Delta_A^{x}|^2}{2V_{2b}}
+ \frac{|\Delta_A^{y}|^2}{2V_{2a}} + \frac{|\Delta_B^{x}|^2}{2V_{2a}}
+ \frac{|\Delta_B^{y}|^2}{2V_{2b}}.\label{aux:Landau}
\end{align}
Equivalently, the determinant can be written explicitly as
\begin{equation}
\begin{aligned}
F= &-\frac{T}{2N_L}\sum_{\boldsymbol{k},n}\ln\Big[\big(\omega_n^2+(\xi_{\boldsymbol{k}}^{A})^2+|\Delta_A(\boldsymbol{k})|^2\big)\\
&\times\big(\omega_n^2+(\xi_{\boldsymbol{k}}^{B})^2+|\Delta_B(\boldsymbol{k})|^2\big)\\
&+|h_{ab}(\boldsymbol{k})|^4+|h_{ab}(\boldsymbol{k})|^2\big(2\omega_n^2-2\xi_{\boldsymbol{k}}^{A}\xi_{\boldsymbol{k}}^{B}\big)\\
&+h_{ab}^*(\boldsymbol{k})h_{ab}^*(-\boldsymbol{k})\Delta_A(\boldsymbol{k})\Delta_B^*(\boldsymbol{k})\\
&+h_{ab}(\boldsymbol{k})h_{ab}(-\boldsymbol{k})\Delta_A^*(\boldsymbol{k})\Delta_B(\boldsymbol{k})\Big]\\
&+\frac{|\Delta_A^{x}|^2}{2V_{2b}}+\frac{|\Delta_A^{y}|^2}{2V_{2a}}+\frac{|\Delta_B^{x}|^2}{2V_{2a}}+\frac{|\Delta_B^{y}|^2}{2V_{2b}}.
\end{aligned}
\end{equation}
When $h_{ab}=0$, which is the case of the model introduced in the main text, the two spin sectors decouple and the free energy separates into independent $A$ and $B$ contributions.

\section{Ginzburg–Landau free energy}\label{app:C}
\subsection{Microscopic derivation of the Landau coefficients}\label{ref:app-SC}
Near the superconducting transition, we expand the free energy in Eq. (\ref{aux:Landau}) in the order parameters $\Delta_A^x,\Delta_A^y,\Delta_B^x,\Delta_B^y$. The quadratic contribution is obtained from the standard loop expansion,
\begin{align}
F^{(2)}
&= \frac{1}{2} k_B T \sum_{\boldsymbol{k},n}
\mathrm{Tr}\big[
G_p(i\omega_n,\boldsymbol{k})
\hat{\Delta}(\boldsymbol{k})
G_h(i\omega_n,\boldsymbol{k})
\hat{\Delta}^\dagger(\boldsymbol{k})
\big] \notag \\
&\quad + N_L\Big(
\frac{|\Delta_A^x|^2}{2V_{2b}}
+ \frac{|\Delta_A^y|^2}{2V_{2a}}
+ \frac{|\Delta_B^x|^2}{2V_{2a}}
+ \frac{|\Delta_B^y|^2}{2V_{2b}}
\Big).
\end{align}
where $\hat{\Delta}(\boldsymbol{k})=\mathrm{diag}\big(\Delta_A(\boldsymbol{k}), \Delta_B(\boldsymbol{k})\big)$ and $\Delta_{A,B}(\boldsymbol{k})=-\Delta_{A,B}^x\sin k_x-\Delta_{A,B}^y\sin k_y$. The quartic terms follow analogously from
\begin{equation}
F^{(4)}=\frac{1}{4}k_B T \sum_{\boldsymbol{k},n} \mathrm{Tr} \big[(G_p\hat{\Delta}G_h\hat{\Delta}^\dagger)^2\big].
\end{equation}
The inverse particle and hole Green’s functions are
\begin{align}
G_{p}^{-1}(i\omega_n,\boldsymbol{k}) &=
\begin{pmatrix}
i\omega_n-\xi^A_{\boldsymbol{k}} & -h_{ab}(\boldsymbol{k}) \\
-h_{ab}^*(\boldsymbol{k}) & i\omega_n-\xi^B_{\boldsymbol{k}}
\end{pmatrix}, \notag \\
G_{h}^{-1}(i\omega_n,\boldsymbol{k}) &=
\begin{pmatrix}
i\omega_n+\xi^A_{\boldsymbol{k}} & h_{ab}^*(-\boldsymbol{k}) \\
h_{ab}(-\boldsymbol{k}) & i\omega_n+\xi^B_{\boldsymbol{k}}
\end{pmatrix}.
\end{align}
Correspondingly,
\begin{align}
G_{p}(i\omega_n,\boldsymbol{k}) &=
\frac{1}{
(i\omega_n-\xi^A_{\boldsymbol{k}})
(i\omega_n-\xi^B_{\boldsymbol{k}})
-|h_{ab}(\boldsymbol{k})|^2} \notag \\
&\times
\begin{pmatrix}
i\omega_n-\xi^B_{\boldsymbol{k}} & h_{ab}(\boldsymbol{k}) \\
h_{ab}^*(\boldsymbol{k}) & i\omega_n-\xi^A_{\boldsymbol{k}}
\end{pmatrix}.
\end{align}
\begin{align}
G_{h}(i\omega_n,\boldsymbol{k}) &=
\frac{1}{
(i\omega_n+\xi^A_{\boldsymbol{k}})
(i\omega_n+\xi^B_{\boldsymbol{k}})
-|h_{ab}(-\boldsymbol{k})|^2} \notag \\
&\times
\begin{pmatrix}
i\omega_n+\xi^B_{\boldsymbol{k}} & -h_{ab}^*(-\boldsymbol{k}) \\
-h_{ab}(-\boldsymbol{k}) & i\omega_n+\xi^A_{\boldsymbol{k}}
\end{pmatrix}.
\end{align}

For later Matsubara summations it is convenient to define
\begin{align}
E_{1,2}^2
&= \frac{1}{2}\big[(\xi^A)^2 + (\xi^B)^2 + 2|h_{ab}|^2\big]\\
&\mp \frac{1}{2}\sqrt{
\big[(\xi^A)^2 - (\xi^B)^2\big]^2
+ 4|h_{ab}|^2(\xi^A + \xi^B)^2
}.
\end{align}
Also for later convenience, we introduce $g_2(k)$, which corresponds to one of the three form factors:
\begin{equation}
\sin^2 k_x,~
\sin^2 k_y,~
\sin k_x\sin k_y,
\end{equation}
and the function $g_4(k)$, which can assume the forms:
\begin{equation}
\sin^4 k_x,~\sin^4 k_y,~ \sin^2 k_x \sin k^2_y.
\end{equation}
The quadratic contribution reads
\begin{equation}
\begin{aligned}
F_2 &= r_1\big(|\Delta_A^x|^2+|\Delta_B^y|^2\big)
+ r_2\big(|\Delta_A^y|^2+|\Delta_B^x|^2\big)\\
&+ \big[r_{11}\Delta_A^{x*}\Delta_B^x+\text{h.c.}\big]
+ \big[r_{22}\Delta_A^{y*}\Delta_B^y+\text{h.c.}\big]\\
&+ \big[r_{12}(\Delta_A^{x*}\Delta_B^y+\Delta_A^{y*}\Delta_B^x)+\text{h.c.}\big].
\end{aligned}
\end{equation}
After the Matsubara summation, a generic quadratic coefficient takes the form
\begin{align}
r &= \frac{1}{2N_L}\sum_{k} g_2(k)
\left[
-\frac{A(k)}{2E_1}\tanh\Big(\frac{E_1}{2T}\Big)
-\frac{B(k)}{2E_2}\tanh\Big(\frac{E_2}{2T}\Big)
\right] \notag \\
&\quad + \frac{1}{2V},
\label{coeff_r}
\end{align}
where $A(k)$ and $B(k)$ depend on the specific channel. Explicitly:
\begin{equation}
\begin{aligned}
r_1:&\quad g_2(k)=\sin^2 k_x,\ V=V_{2b},\ \\ &\quad A(k)=\frac{(\xi^B)^2-E_1^2}{E_2^2-E_1^2},\ B(k)=\frac{E_2^2-(\xi^B)^2}{E_2^2-E_1^2},\\
r_2:&\quad g_2(k)=\sin^2 k_y,\ V=V_{2a},\ \\ &\quad A(k)=\frac{(\xi^B)^2-E_1^2}{E_2^2-E_1^2},\ B(k)=\frac{E_2^2-(\xi^B)^2}{E_2^2-E_1^2},\\
r_{11}:&\quad g_2(k)=\sin^2 k_x,\ \\ &\quad A(k)=\frac{h_{ab}(k) h_{ab}(-k)}{E_2^2-E_1^2},\ B(k)=\frac{- h_{ab}(k) h_{ab}(-k)}{E_2^2-E_1^2}.
\end{aligned}
\end{equation}
The remaining coefficients follow by the replacements $g_2(k)$ with the appropriate symmetry factors.
Specifically, $r_{22}$ involves $\sin^2 k_y$, whereas $r_{12}$ involves $\sin k_x \sin k_y$.

For the quartic coefficients we illustrate the contribution from the same-sector terms $|\Delta_A|^4$ and $|\Delta_B|^4$, which are the only ones that survive when $h_{ab}=0$. Other quartic coefficients are obtained analogously.

\begin{equation}
\begin{aligned}
F_4\big(|\Delta_A|^4,|\Delta_B|^4\big)
&= u_1\big(|\Delta_A^x|^4+|\Delta_B^y|^4\big)
+ u_2\big(|\Delta_A^y|^4+|\Delta_B^x|^4\big)\\
&+ v_{xy}\big(|\Delta_A^x|^2|\Delta_A^y|^2+|\Delta_B^x|^2|\Delta_B^y|^2\big)\\
&+ \frac{v_{xy}}{4}\Big[(\Delta_A^x \Delta_A^{y*})^2+(\Delta_B^x \Delta_B^{y*})^2+\text{h.c.}\Big],
\end{aligned}
\end{equation}
A typical quartic coefficient takes the form:
\begin{equation}
\begin{aligned}
u = \frac{1}{4N_L}\sum_{k} g_{4}(k)\Bigl[
  A(k)\bigl(\tfrac{1}{4E_1^3}\tanh \tfrac{E_1}{2T} 
  - \tfrac{1}{8T E_1^2}\operatorname{sech}^2 \tfrac{E_1}{2T}\bigr)&
\\
  + B(k)\bigl(\tfrac{1}{4E_2^3}\tanh \tfrac{E_2}{2T}
  - \tfrac{1}{8T E_2^2}\operatorname{sech}^2 \tfrac{E_2}{2T}\bigr)&
\\
  + C(k)\bigl(-\tfrac{1}{2E_1}\tanh \tfrac{E_1}{2T}\bigr)
  + D(k)\bigl(-\tfrac{1}{2E_2}\tanh \tfrac{E_2}{2T}\bigr)
\Bigr],
\end{aligned}
\end{equation}
where $A(k)$, $B(k)$, $C(k)$, and $D(k)$ are specified by the particular quartic channel. For $u_1$:
\begin{equation}
\begin{aligned}
g_4(k)&=\sin^4 k_x,\\
A(k)&=\left(\frac{(\xi^B)^2-E_1^2}{E_2^2-E_1^2}\right)^2,\qquad
B(k)=\left(\frac{E_2^2-(\xi^B)^2}{E_2^2-E_1^2}\right)^2,\\
C(k)&=\frac{2\big[(\xi^B)^2-E_1^2\big]\big[(\xi^B)^2-E_2^2\big]}{(E_2^2-E_1^2)^3},\quad
D(k)=-C(k).
\end{aligned}
\end{equation}
The coefficients $u_2$ and $v_{xy}$ can be obtained by the replacements $g_{4}$ with the appropriate symmetry factors.

\subsection{Evaluation of the Landau coefficients in the continuum limit}

To gain analytic insight, we take the continuum limit of the model by expanding around small momenta. It is convenient to use the generic parametrization of spin-split Fermi surfaces in terms of spin-triplet $\alpha$-phase and $\beta$-phase order parameters introduced in Eq. (\ref{eq:nem-spin-nem}) . We then rewrite the dispersions, inter-sector coupling, and form factors as
\begin{equation}
\begin{aligned}
\xi^A_{\boldsymbol{k}}&=\xi-\delta\cos(l_\alpha\theta_k),\qquad
\xi^B_{\boldsymbol{k}}=\xi+\delta\cos(l_\alpha\theta_k),\\
h_{ab}(\boldsymbol{k})&=-\bar{n}\big[\cos(l_\beta\theta_k)-i\sin(l_\beta\theta_k)\big]
=-\bar{n} e^{-i l_\beta \theta_k},\\
g_x(\boldsymbol{k})&=\sqrt{2}\cos\theta_k,\qquad
g_y(\boldsymbol{k})=\sqrt{2}\sin\theta_k,
\end{aligned}
\end{equation}
where $\xi=k^2/2m - \mu$ is an isotropic dispersion, $l_\alpha=2$ for the $d$-wave altermagnetic phase, and $\theta_k$ is the polar angle of $\boldsymbol{k}$ (henceforth we drop the subscript and write $\theta$). We keep the parity of the $\beta$ phase order parameter, $l_\beta$, arbitrary for now to allow us to study the effects of symmetry-breaking fields. But we emphasize that in the lattice model introduced in the main text, $\bar{n}=0$.

We introduce the normal-state branches
\begin{equation}
E_{1,2}(\boldsymbol{k})=\xi\mp\sqrt{\bar{n}^2+\big[\delta\cos(l_\alpha\theta)\big]^2}.
\end{equation}
The general quadratic coefficient $r$ has the structure given in Eq.~\ref{coeff_r}. 
Near the Fermi surface we use
\begin{equation}
\frac{1}{N_L}\sum_{\boldsymbol{k}} \to \int\frac{d^2k}{(2\pi)^2}
 \to N(E_F)\int_{-\Lambda}^{\Lambda}d\xi\int\frac{d\theta}{2\pi},
\end{equation}
with $N(E_F)$ the density of states at the Fermi level and $\Lambda$ the cutoff.

For later convenience we decompose
\begin{equation}
A(k)=A^{11}+A^{12}\frac{2E_1}{E_1+E_2},\qquad
B(k)=B^{11}+B^{12}\frac{2E_2}{E_1+E_2},
\end{equation}
and define
\begin{equation}
C^{11}=A^{11}+B^{11},\qquad C^{12}=A^{12}+B^{12}.
\end{equation}
The particle-particle bubble is given by
\begin{equation}
\chi_{ab}=\int_{-\Lambda}^{\Lambda} d\xi \frac{1-n_F(E_a)-n_F(E_b)}{E_a+E_b},
\end{equation}
which coincides with the definition in Eq.~\ref{chiab} at $Q=0$. As a result, we obtain
\begin{equation}
r=-\frac{1}{2}N(E_F)\int\frac{d\theta}{2\pi} g_2(\theta) \big[C^{11}\chi_{11}+C^{12}\chi_{12}\big]+\frac{1}{2V}.\label{aux:r}
\end{equation}
For $r_1$ we use
\begin{equation}
\begin{aligned}
g_2(\theta)&=2\cos^2\theta,\qquad V=V_{2b},\\
A(k)&=\frac{(\xi^B)^2-E_1^2}{E_2^2-E_1^2},\quad
B(k)=\frac{E_2^2-(\xi^B)^2}{E_2^2-E_1^2},
\end{aligned}
\end{equation}
and
\begin{equation}
\begin{aligned}
C^{11}=\frac{1}{2}+\frac{1}{2}\frac{(\delta\cos l_\alpha\theta)^2}{\bar{n}^2+(\delta\cos l_\alpha\theta)^2},~
C^{12}=\frac{1}{2}\frac{\bar{n}^2}{\bar{n}^2+(\delta\cos l_\alpha\theta)^2}.
\end{aligned}
\end{equation}
Substituting these into the general expression for $r$ gives
\begin{equation}
r_1=\frac{1}{2V_{2b}}-\frac{1}{2}N(E_F)\int_0^{2\pi}\frac{d\theta}{2\pi} 2\cos^2\theta \big[C^{11}\chi_{11}+C^{12}\chi_{12}\big],
\end{equation}
while the $y$ component follows by replacing $\cos^2\theta\to\sin^2\theta$ and $V_{2b}\to V_{2a}$:
\begin{equation}
r_2=\frac{1}{2V_{2a}}-\frac{1}{2}N(E_F)\int_0^{2\pi}\frac{d\theta}{2\pi} 2\sin^2\theta \big[C^{11}\chi_{11}+C^{12}\chi_{12}\big].
\end{equation}
The second terms in $r_{1}$ and $r_{2}$ are minus the superconducting susceptibilities and, as $T\to0$, exhibit the familiar logarithmic diverging.

At first sight it may seem that $r_1=r_2$ when $V_{2a}=V_{2b}$, since the integrands appear invariant under $\theta\to\theta+\pi/2$ even for $\delta\ne0$. However, this is an artifact of the continuum model. Explicit lattice calculations show that $r_1\ne r_2$ whenever $\phi\neq 0$. The mismatch arises because a finite $\delta$ distorts the Fermi surface and hence the angular density of states, producing inequivalent $\cos^2\theta$ and $\sin^2\theta$ averages and, consequently, $r_1\ne r_2$.

For $r_{11}$ we use Eq. (\ref{aux:r}) with
\begin{equation}
\begin{aligned}
g_2(\theta)&=2\cos^2\theta,\\
A(k)&=-B(k)= -\frac{1}{E_2^2-E_1^2},
\end{aligned}
\end{equation}
together with
\begin{equation}
\begin{aligned}
C^{11}=-\frac{1}{2}\frac{1}{\bar{n}^2+(\delta\cos l_\alpha\theta)^2},~
C^{12}=\frac{1}{2}\frac{1}{\bar{n}^2+(\delta\cos l_\alpha\theta)^2}.
\end{aligned}
\end{equation}
Recall that $h_{ab}(\boldsymbol{k})=-\bar{n} e^{-i l_\beta\theta}$, hence
\begin{equation}
h_{ab}(\boldsymbol{k}) h_{ab}(-\boldsymbol{k})=(-1)^{l_\beta}\bar{n}^2 e^{-i2l_\beta\theta}.
\end{equation}
Combining these ingredients, the mixed-sector quadratic kernels read
\begin{align}
r_{11},r_{12}
=\frac{1}{2}\int_0^{2\pi}\frac{d\theta}{2\pi} 
g_2(\theta) (-1)^{l_\beta}\bar n^{ 2} e^{-i2l_\beta\theta}C(\theta)
\end{align}
where
\begin{equation}
g_2(\theta)=
\begin{cases}
2\cos^2\theta,& \text{for } r_{11},\\ 
2\cos\theta \sin\theta,& \text{for } r_{12},
\end{cases}
\end{equation}
and $C(\theta)=C^{11}\big(\chi_{11}-\chi_{12}\big)$.

Under $\theta\to\pi-\theta$, corresponding to a reflection with respect to the mirror perpendicular to the $x$-axis, the functions $\chi_{11}$, $\chi_{12}$, $C^{11}$, $C^{12}$ are invariant, while
\begin{align}
e^{-i2l_\beta\theta} &= \cos(2l_\beta\theta) - i\sin(2l_\beta\theta), \notag \\
e^{-i2l_\beta(\pi-\theta)} &= \cos(2l_\beta\theta) + i\sin(2l_\beta\theta).
\end{align}
By changing the integral variable $\theta \to \pi - \theta$, terms that change sign cancel each other, leaving us with:
\begin{equation}
\begin{aligned}
r_{11}&=\frac{1}{2}\int_0^{2\pi}\frac{d\theta}{2\pi} 2\cos^2\theta [(-1)^{l_\beta}\bar{n}^2\cos(2l_\beta\theta)] C(\theta),\\
r_{22}&=\frac{1}{2}\int_0^{2\pi}\frac{d\theta}{2\pi} 2\sin^2\theta [(-1)^{l_\beta}\bar{n}^2\cos(2l_\beta\theta)] C(\theta),\\
r_{12}&=\frac{-i}{2}\int_0^{2\pi}\frac{d\theta}{2\pi} \sin(2\theta) [(-1)^{l_\beta}\bar{n}^2\sin(2l_\beta\theta)] C(\theta).
\end{aligned}
\end{equation}
For $l_\alpha=2$, a further shift $\theta\to\theta+\pi/2$ gives
\begin{equation}
r_{22}=(-1)^{l_\beta} r_{11}.
\end{equation}
and also that $r_{12} =0$ if $l_{\beta}$ is even. Thus, the parity of $l_\beta$ determines the relative chirality once the phases $\beta_{A,B}$ in Eq.~\ref{parametrize} are pinned at $\beta_{A,B}=\pm\pi/2$. The phase-coupling term then fixes $\beta_B-\beta_A$: for even $l_\beta$, $r_{11}=r_{22}$ and the minimum favors $\beta_B=\beta_A$, corresponding to the same chirality in both sectors (chiral $p_x+ip_y$ in $A$ and $p_x+ip_y$ in $B$, or both with opposite chirality). For odd $l_\beta$, $r_{22}=-r_{11}$ and the minimum favors $\beta_B=-\beta_A$, corresponding to opposite chiralities (helical $p_x+ip_y$ in one sector and $p_x-ip_y$ in the other).
As an example, Rashba spin–orbit coupling corresponds to $l_\beta=1$ and implies $r_{22}=-r_{11}$, consistent with Refs.~\cite{Zhu2023,Heung2025}.

\subsection{Evaluation of the Landau coefficients in the lattice model}
The free-energy of the tetragonal model used in the main text can be obtained in a straightforward way from Eq. (\ref{aux:Landau}) by setting $h_{ab}(\boldsymbol{k})=0$. The free-energy becomes:
\begin{equation}
\begin{aligned}
F_{\text{SC}} = &
  r_1 (|\Delta_A^x|^2 + |\Delta_B^y|^2)
  + r_2 (|\Delta_A^y|^2 + |\Delta_B^x|^2) \\
&+ u_1 (|\Delta_A^x|^4 + |\Delta_B^y|^4)
  + u_2 (|\Delta_A^y|^4 + |\Delta_B^x|^4) \\
&+ v_{xy} (|\Delta_A^x|^2|\Delta_A^y|^2 + |\Delta_B^x|^2|\Delta_B^y|^2)\\
 & + w_{xy}\left[(\Delta_A^x\Delta_A^{y*})^2 + (\Delta_B^x\Delta_B^{y*})^2 + \text{c.c.}\right],
\end{aligned}
\label{eq:F_SC}
\end{equation}
with the quadratic coefficients:
\begin{equation}
\begin{aligned}
r_1&=\frac{1}{2V_{2b}}-\frac{T}{2N_L}\sum_{k,n}(\sin k_x)^2 \frac{1}{(i\omega_n-\xi_k^A)(-i\omega_n-\xi_k^A)}\\
&=\frac{1}{2V_{2b}}- \frac{1}{2N_L}\sum_{k} (\sin k_x)^2 
\frac{1}{2\xi_k^A}\tanh\Big(\frac{\xi_k^A}{2T}\Big),\\
&=\frac{1}{2V_{2b}}- \frac{1}{2N_L}\sum_{k} (\sin k_y)^2 
\frac{1}{2\xi_k^B}\tanh\Big(\frac{\xi_k^B}{2T}\Big),\\
r_2&=\frac{1}{2V_{2a}}-\frac{T}{2N_L}\sum_{k,n}(\sin k_y)^2 \frac{1}{(i\omega_n-\xi_k^A)(-i\omega_n-\xi_k^A)}\\
&=\frac{1}{2V_{2a}}- \frac{1}{2N_L}\sum_{k} (\sin k_y)^2 
\frac{1}{2\xi_k^A}\tanh\Big(\frac{\xi_k^A}{2T}\Big), \label{aux:r1r2}
\end{aligned}
\end{equation}

and the quartic coefficients:
\begin{equation}
\begin{aligned}
u_1&=\frac{T}{4N_L}\sum_{k,n}(\sin k_x)^4\left[ \frac{1}{(i\omega_n-\xi_k^A)(i\omega_n+\xi_k^A)}\right]^2,\\
u_2&=\frac{T}{4N_L}\sum_{k,n}(\sin k_y)^4\left[ \frac{1}{(i\omega_n-\xi_k^A)(i\omega_n+\xi_k^A)}\right]^2 ,\\
v_{xy}&=\frac{T}{N_L}\sum_{k,n}(\sin k_x \sin k_y)^2\left[ \frac{1}{(i\omega_n-\xi_k^A)(i\omega_n+\xi_k^A)}\right]^2\\
w_{xy}&=\frac{T}{4N_L}\sum_{k,n}(\sin k_x \sin k_y)^2\left[ \frac{1}{(i\omega_n-\xi_k^A)(i\omega_n+\xi_k^A)}\right]^2
\end{aligned}
\end{equation}
Since $w_{xy}>0$, the last term of the free-energy expansion is minimized by a relative phase $\beta_{A,B}=\pm\pi/2$ between the $p_x$ and $p_y$ components.
Since $h_{ab}(\boldsymbol{k})=0$, the free energy is decoupled in the $A$ and $B$ band sectors.  Inter-sector coupling emerges either by explicitly breaking the lattice symmetries, such as by including inversion-symmetry breaking terms (Rashba coupling)  $\bar{n}\neq0$ and $l_\beta=1$ in the Hamiltonian, or by considering the effects of fluctuations of nematic fluctuations and spin current-loop fluctuations.

\section{Splitting between the two superconducting transition temperatures}
\label{Tc_splitting}

As discussed in the main text, even when the nearest-neighbor interactions are isotropic, the transition temperatures $T_1^c$ and $T_2^c$ associated with the condensation of two sets of two-component order parameters $(\Delta_A^x,\Delta_B^y)$ and $(\Delta_A^y,\Delta_B^x)$ are different. We now discuss how the critical-temperature splitting $|T_1^c-T_2^c|$ depends on the AM order parameter.

Using the results of Eq. (\ref{aux:r1r2}), the quadratic coefficients are of the form
\begin{equation}
r_1(T)=\frac{1}{2}[1/V_{2b}-\chi_1(T)],\qquad
r_2(T)=\frac{1}{2}[1/V_{2a}-\chi_2(T)],
\end{equation}
where the pairing susceptibilities are
\begin{equation}
\chi_{1,2}(T)=\frac{1}{N_L}\sum_{\bm k}\sin^2 k_{x,y} F(\xi_{\bm k},T),
\end{equation}
and
\begin{equation}
F(\xi,T)\equiv \frac{1}{2\xi}\tanh\left(\frac{\xi}{2T}\right).
\end{equation}
We decompose the dispersion into a $C_4$-symmetric part $\xi_0(\bm k)=-2t_2(\cos k_x+\cos k_y)-\mu$ and a $C_4$-breaking part $\delta\xi(\bm k)=-2 \phi(\cos k_x-\cos k_y)$:
\begin{equation}
\xi_{\bm k}=\xi_0(\bm k)+\delta\xi(\bm k),
\label{app:xi_decompose}
\end{equation}
where $ \phi$ denotes the effective AM order parameter in the projected low-energy two-band description.
Note that $\xi_0(k_x,k_y)=\xi_0(-k_y,k_x)$ is symmetric under $C_4$ rotation, while $\delta\xi(k_x,k_y)=-\delta\xi(-k_y,k_x)$ is antisymmetric.

To expose the leading dependence on the $T_c$ splitting on $\phi$, we expand $F(\xi_{\bm k},T)=F(\xi_0+\delta\xi,T)$ in $\delta\xi$:
\begin{equation}
F(\xi_0+\delta\xi,T)=F(\xi_0,T)+F'(\xi_0,T) \delta\xi+O(\delta\xi^2).
\label{app:F_expand}
\end{equation}
The susceptibility difference becomes
\begin{equation}
\begin{aligned}
\chi_1(T)&-\chi_2(T)
=\frac{1}{N_L}\sum_{\bm k}\big(\sin^2k_x-\sin^2k_y\big) F(\xi_0,T)\\
&\quad+\frac{1}{N_L}\sum_{\bm k}\big(\sin^2k_x-\sin^2k_y\big) F'(\xi_0,T) \delta\xi(\bm k)\\ 
&\quad+O(\delta\xi^2).
\end{aligned}
\end{equation}
The first term vanishes identically by $C_4$ symmetry: the factor $(\sin^2k_x-\sin^2k_y)$ is antisymmetric, while $F(\xi_0,T)$ is symmetric, hence the Brillouin-zone sum cancels. Therefore the leading nonzero contribution is
\begin{equation}
\begin{aligned}
\chi_1(T)-\chi_2(T)
&=\frac{1}{N_L}\sum_{\bm k}\big(\sin^2k_x-\sin^2k_y\big) F'(\xi_0,T)\\ 
&\times \big[-2 \phi(\cos k_x-\cos k_y)\big]
+O(\phi^{ 2}),
\label{app:dchi_linear}
\end{aligned}
\end{equation}
which is linear in the effective AM order parameter $\phi$.

Let $T_{0}^c$ denote the reference transition temperature in the limit $\tilde \phi=0$ and $V_{2a}=V_{2b}\equiv V_2$, defined by
\begin{equation}
r_0(T_0^c)\equiv \frac{1}{2}[1/V_2-\chi_0(T_0^c)]=0,
\end{equation}
with $\chi_0\equiv \chi_1=\chi_2 $ evaluated at $ \phi=0$.
For weak anisotropies (small $ \phi$ and small interaction difference $V_{2d}\equiv (V_{2b}-V_{2a})/2$), the shifts of the two transition temperatures follow from linearizing $r_{1,2}(T^c_{1,2})=0$ around $T^c_{0}$:
\begin{equation}
\delta T^c_{1,2}\equiv T^c_{1,2}-T^c_{0}=-\frac{\delta r_{1,2}(T^c_{0})}{r_0'(T^c_{0})},
\end{equation}
where $\delta r_{1,2}(T)=r_{1,2}(T)-r_0(T)$ and $r_0'(T)=-\chi_0'(T)$.
Keeping only terms linear in the AM order parameter, one obtains the splitting
\begin{equation}
T^c_{1}-T^c_{2}
=
-\frac{1}{2r_0'(T^c_{0})}
\left[
\left(\frac{1}{V_{2b}}-\frac{1}{V_{2a}}\right)
-\big(\chi_1-\chi_2\big)\Big|_{T^c_{0}}
\right].
\label{app:dTc_general}
\end{equation}
Using \eqref{app:dchi_linear} it follows that, to leading order,
\begin{equation}
T^c_{1}-T^c_{2}
=
A V_d + B  \phi + O(\phi^{ 2},V_d^{ 2}, \phi V_d).
\label{app:dTc_linear}
\end{equation}
This result shows that the dominant control parameters for the two-$T_c$ splitting are the AM order parameter $ \phi$ and the interaction difference $V_d$.

In the projected low-energy two-band description, the effective AM order parameter $ \phi$ depends on $N_{\rm am}$.
When $|N_{\rm am}|\ll 4|t_1|$, corresponding to weak AM order, a low-energy expansion near the $\Gamma$ point yields an effective $d$-wave spin splitting $\phi \simeq |N_{\rm am} |t_d/(4t_1)$~\cite{jungwirth2024b}. In contrast, deep in the AM phase where the order is strong and the low-energy states are nearly spin-sublattice polarized, the magnitude of the splitting is $|\phi|=|t_d|$, while its sign reverses under $N_{\rm am}\to -N_{\rm am}$ (together with the corresponding change of spin–sublattice locking), i.e. $\phi \simeq t_d \mathrm{sgn}(N_{\rm am})$. As a consequence, for isotropic interactions ($V_d=0$), the $T_c$ splitting is linear in $N_{\rm am}$ at small $|N_{\rm am}|$, while it saturates (up to a sign) at large $|N_{\rm am}|$.

\section{Evaluation of the Landau coefficients generated by nematic fluctuations}
\label{ref:app-nematic}

As explained in the main text, the free energy in the presence of nematic fluctuations is $F=F_{\text{SC}}+F_{\text{nem}}+F_{n\text{-}SC}$, with a fluctuating charge–nematic field $\varphi$ described by
\begin{equation}
F_{\text{nem}}=\frac{\varphi^2}{2\chi_{\text{nem}}},\label{aux:Fnem}
\end{equation}
and the leading symmetry-allowed coupling to superconductivity
\begin{equation}
F_{n\text{-}SC}=\varphi(\lambda_1 S_1+\lambda_2 S_2),
\end{equation}
with $S_1=|\Delta_A^x|^2-|\Delta_B^y|^2, 
S_2=|\Delta_A^y|^2-|\Delta_B^x|^2$.
Here $\chi_{\text{nem}}>0$ in the altermagnetic phase, and $\lambda_{1,2}$ is related to the inter-sublattice interaction $V_1 n_{i,1}n_{i,2}$, as we show below.

Eliminating $\varphi$ by minimization yields $\varphi=-\chi_{\text{nem}}(\lambda_1 S_1+\lambda_2 S_2)$ and
\begin{equation}
F_{\text{nem}}+F_{n\text{-}SC}=-\frac{\chi_{\text{nem}}}{2}(\lambda_1 S_1+\lambda_2 S_2)^2.
\end{equation}
Expanding the square and grouping terms one finds that nematic fluctuations (i) renormalize the intra-band quartic terms and (ii) generate inter-band biquadratic couplings:
\begin{equation}
\begin{aligned}
F_{\text{nem}}+F_{\text{n-SC}}
&=-\frac{\chi_{\text{nem}}\lambda_1^2}{2}\big(|\Delta_A^x|^4+|\Delta_B^y|^4\big)\\
  &-\frac{\chi_{\text{nem}}\lambda_2^2}{2}\big(|\Delta_A^y|^4+|\Delta_B^x|^4\big)\\
&-\chi_{\text{nem}}\lambda_1\lambda_2\big(|\Delta_A^x|^2|\Delta_A^y|^2+|\Delta_B^x|^2|\Delta_B^y|^2\big)\\
&+\chi_{\text{nem}}\lambda_1\lambda_2\big(|\Delta_A^x|^2|\Delta_B^x|^2+|\Delta_A^y|^2|\Delta_B^y|^2\big)\\
&+\chi_{\text{nem}}\lambda_1^2|\Delta_A^x|^2|\Delta_B^y|^2
+\chi_{\text{nem}}\lambda_2^2|\Delta_A^y|^2|\Delta_B^x|^2.
\end{aligned}
\end{equation}
The first line reduces the effective $u_1$ and $u_2$ and the mixed intra-band term reduces $v_{xy}$ when combined with the free-energy terms of the bare model. The remaining contributions generate the positive inter-band couplings (as in Eq.~(4.1) of the main text),
\begin{equation}
v_{AB}=\chi_{\text{nem}}\lambda_1\lambda_2,\qquad
v_{AB}^{xy}=\chi_{\text{nem}}\lambda_1^2,\qquad
v_{AB}^{yx}=\chi_{\text{nem}}\lambda_2^2,
\end{equation}
which promote competition between the gap components and thus enhance tendencies toward nematic superconductivity. No phase-locking term of the form $(\Delta_A^{x*}\Delta_B^x\Delta_A^y\Delta_B^{y*}+\text{c.c.})$ is produced by nematic fluctuations alone.

We now turn to a microscopic evaluation of the nematic–SC couplings $\lambda_{1,2}$. We begin with a generic density–density interaction between nearest neighbors on opposite sublattices,
\begin{equation}
H_{\text{inter}} = \sum_{\langle i,j \rangle} V_{\sigma \sigma'}(i,j)   n_{i\sigma} n_{j\sigma'} = \frac{1}{N_L} \sum_{\boldsymbol{q}} V_{\sigma \sigma'}(\boldsymbol{q})   n_{-\boldsymbol{q}\sigma}   n_{\boldsymbol{q}\sigma'},
\end{equation}
where \( n_{\boldsymbol{q}\sigma} = \sum_{\boldsymbol{k}} c^{\dagger}_{\boldsymbol{k+q}\sigma} c_{\boldsymbol{k}\sigma} \), and \( N_L \) is the total number of lattice sites.

We focus on the nearest-neighbor repulsive interaction, which favors uniform charge nematic order. In this case, the interaction reduces to
\begin{equation}
H_{\text{inter}} \sim \frac{V_1}{N_L} n_{1} n_{2} =  \frac{V_1}{4N_L} \left[ (n_{1} + n_{2})^2 - (n_{1} - n_{2})^2 \right],
\end{equation}
where \( V_1 > 0 \) is the strength of the repulsive interaction, \(n_{1}\) and \(n_{2}\) are the local densities on sublattices 1 and 2, respectively.
The nematic contribution to the interaction energy takes the form
\begin{equation}
H_{\text{nem}} \sim -\frac{V_1}{4N_L} (n_{1} - n_{2})^2.
\end{equation}
To decouple this quartic term, we perform a Hubbard-Stratonovich transformation:
\begin{equation}
e^{\frac{V_1}{4N_L} (n_1 - n_2)^2} \propto \int \mathcal{D}[\varphi]   e^{-N_LV_1 \varphi^2 + V_1 \varphi (n_1 - n_2)}.
\end{equation}
Deep in the altermagnetic phase, we use the effective spin-band locked basis $(c_{\boldsymbol{k}\uparrow A}, c_{\boldsymbol{k}\downarrow B})$, whose components predominantly reside on sublattices 1 and 2, respectively.
This yields the following contribution to the fermionic action:
\begin{equation}
S_{\text{nem}} = \int d\tau \sum_{\boldsymbol{k}} \left[ -V_1\varphi \left( c^\dagger_{\boldsymbol{k}A} c_{\boldsymbol{k}A} - c^\dagger_{\boldsymbol{k}B} c_{\boldsymbol{k}B} \right) + V_1\varphi^2 \right],
\end{equation}
where \( \varphi \) is the Hubbard-Stratonovich field corresponding to the nematic order parameter. 

The Nambu Green’s function in the presence of nematic ($\varphi$) and superconducting ($\Delta$) fields reads
\begin{equation}
G^{-1}(\boldsymbol{k},i\omega_n)=G_0^{-1}(\boldsymbol{k},i\omega_n)-\Sigma_{\varphi}-\Sigma_{\Delta}(\boldsymbol{k}).
\end{equation}
Adopting the basis $(c_{\boldsymbol{k}A}, c_{\boldsymbol{k}B}, c^{\dagger}_{-\boldsymbol{k}A}, c^{\dagger}_{-\boldsymbol{k}B})^{T}$ the matrices take the form (spin indices omitted for brevity, $\uparrow$ on $A$, $\downarrow$ on $B$):
\begingroup\setlength{\arraycolsep}{4pt}\renewcommand{\arraystretch}{1.0}
\begin{equation}
\begin{aligned}
G_0^{-1}&=
\begin{pmatrix}
i\omega_n-\xi^A_{\boldsymbol{k}} & 0 & 0 & 0\\
0 & i\omega_n-\xi^B_{\boldsymbol{k}} & 0 & 0\\
0 & 0 & i\omega_n+\xi^A_{-\boldsymbol{k}} & 0\\
0 & 0 & 0 & i\omega_n+\xi^B_{-\boldsymbol{k}}
\end{pmatrix},
\\
\Sigma_{\varphi}&=
\begin{pmatrix}
- V_1\varphi & 0 & 0 & 0\\
0 & + V_1\varphi & 0 & 0\\
0 & 0 & + V_1\varphi & 0\\
0 & 0 & 0 & - V_1\varphi
\end{pmatrix},
\\
\Sigma_{\Delta}&=
\begin{pmatrix}
0 & 0 & \Delta_A(\boldsymbol{k}) & 0\\
0 & 0 & 0 & \Delta_B(\boldsymbol{k})\\
\Delta_A^*(\boldsymbol{k}) & 0 & 0 & 0\\
0 & \Delta_B^*(\boldsymbol{k}) & 0 & 0
\end{pmatrix},
\end{aligned}
\end{equation}
\endgroup
where $\Delta_{A/B}(\boldsymbol{k})=-\Delta_{A/B}^x\sin k_x-\Delta_{A/B}^y\sin k_y$.

The trilinear coupling term, which couples the nematic field linearly to the square of the superconducting order parameter, is given by
\begin{equation}
F_{\text{n-SC}} = \frac{1}{2} T \sum_{\boldsymbol{k},n} \mathrm{Tr} \left[ G_0 \Sigma_\varphi G_0 \Sigma_\Delta G_0 \Sigma_\Delta \right].
\end{equation}
After integrating out the fermionic degrees of freedom, the coefficients $\lambda_{1,2}$ can be expressed as
\begin{align}
\lambda_{1,2}
=\frac{TV_1}{2N_L} \sum_{k,n} g_{x,y}^2(k)
\Bigg[
&-\frac{1}{(i\omega_n-\xi_k^A)^2(i\omega_n+\xi_k^A)}
\nonumber \\
&+\frac{1}{(i\omega_n+\xi_k^A)^2(i\omega_n-\xi_k^A)}
\Bigg].
\end{align}

Performing the Matsubara summation yields
\begin{align}
\lambda_{1,2} =& \frac{V_1}{4N_L} \sum_{\mathbf{k}} g_{x,y}^{2}(\mathbf{k}) 
\nonumber \\
&\times\Bigg[
\frac{2n_{F}(\xi_{\mathbf{k}}^{A}) - 1}{(\xi_{\mathbf{k}}^{A})^{2}} 
+\frac{2\beta}{\xi_{\mathbf{k}}^{A}} n_{F}(\xi_{\mathbf{k}}^{A}) \left(1 - n_{F}(\xi_{\mathbf{k}}^{A})\right) 
\Bigg],
\end{align}
where $n_F(\xi)$ is the Fermi-Dirac distribution function and $\beta = 1/T$.

\section{Evaluation of the Landau coefficients generated by spin current-loop fluctuations}
\label{ref:app-loop}

\subsection{Derivation of the microscopic coupling}

We now consider spin current-loop fluctuations arising from a nearest-neighbor density–density repulsion between opposite sublattices. The interaction is
\begin{equation}
H_{\text{int}}
= V_1 \sum_{\langle i,j \rangle} n_{1,i,\sigma} n_{2,j,\sigma'}
= V_1\sum_{\langle i,\delta \rangle}
c^\dagger_{1,i,\sigma} c_{1,i,\sigma} 
c^\dagger_{2,i+\delta,\sigma'} c_{2,i+\delta,\sigma'} .
\end{equation}

Following Ref. \cite{Tsai2015}, we define the inter-sublattice operators:
\[
\mathcal{O}^a_{i,\delta} = \frac{1}{2} D_\delta   c^\dagger_{1,i,\sigma}   s^a_{\sigma\sigma'}   c_{2,i+\delta,\sigma'}
\]

where $s^a$ are Pauli matrices for $a = x, y, z$ and the identity matrix $s^0$ for $a=0$. The form factor $D_\delta$ is defined as:
\[
D_\delta = 
\begin{cases}
+1 & \text{if } \delta = \pm(\tfrac{1}{2}\hat{x} + \tfrac{1}{2}\hat{y}) \\
-1 & \text{if } \delta = \pm(\tfrac{1}{2}\hat{x} - \tfrac{1}{2}\hat{y})
\end{cases}
\]

We define the real and imaginary parts of the order parameters:

\[
X^a_{i,\delta} = \mathcal{O}^a_{i,\delta} + (\mathcal{O}^a_{i,\delta})^\dagger, \quad 
Y^a_{i,\delta} = i \left[ (\mathcal{O}^a_{i,\delta})^\dagger - \mathcal{O}^a_{i,\delta} \right]
\]

The interaction term can then be rewritten approximately as:

\[
H_{\text{int}} \approx - \frac{V_1}{2} \sum_{i,\delta,a} \left[ (X^a_{i,\delta})^2 + (Y^a_{i,\delta})^2 \right]
\]
Applying a HS transformation to decouple the interaction, we obtain the mean-field Hamiltonian:

\begin{align}
H_{\text{MF}} =\sum_{i,\delta,a} 
2V_1\left[
D_\delta \left( Q^a - i\Phi^a \right)   c^\dagger_{1,i,\sigma} s^a_{\sigma\sigma'} c_{2,i+\delta,\sigma'} + \text{h.c.}
\right] \nonumber \\
+ 8V \sum_{i,a} \left( |Q^a|^2 + |\Phi^a|^2 \right)
\end{align}
The order parameters $Q^a$ and $\Phi^a$ are
\begin{equation}
\begin{aligned}
Q^a & =\frac{1}{8} \operatorname{Re} \sum_\delta D_\delta\left\langle c_{1, i, \sigma}^{\dagger} s_{\sigma \sigma^{\prime}}^a c_{2, i+\delta, \sigma^{\prime}}\right\rangle ,\\
\Phi^a & =\frac{1}{8} \operatorname{Im} \sum_\delta D_\delta\left\langle c_{1, i, \sigma}^{\dagger} s_{\sigma \sigma^{\prime}}^a c_{2, i+\delta, \sigma^{\prime}}\right\rangle.
\end{aligned}
\end{equation}
In the AM phase, the low-energy degrees of freedom corresponds to spin-up electrons on band $A$ and spin-down electrons on band $B$. Since bands $A$ and $B$ are nearly sublattice polarized in our model, we replace $(1,2)$ by $(A,B)$ and consider only spin up in $A$ and spin down in $B$. As a result, we leave the spin labels implicit. 
We introduce a complex spin current-loop order parameter $\phi_l$ and its coupling to the superconducting order parameters. The mean-field Hamiltonian is
\begin{equation}
H_{\text{MF}}
= 2V_1 \sum_{i,\delta}\Big[ D_{\delta} \phi_l  c^{\dagger}_{A,i} c_{B,i+\delta}
+ \text{h.c.}\Big]
+ 8V_1 \sum_i |\phi_l|^{2}.
\end{equation}
Transforming to momentum space, the fermionic Green’s function in the presence of both $\phi_l$ and the superconducting pairing field $\Delta$ is given by
\begin{equation}
G^{-1}_{\boldsymbol{k},n} = G_0^{-1} - \Sigma_{\phi _l}- \Sigma_\Delta.
\end{equation}
In the reduced Nambu basis
\[
\Psi^\dagger=\big(c^\dagger_{A},\ c^\dagger_{B},\ c_{A},\ c_{B}\big),
\]
the ingredients read as follows.

\begingroup\setlength{\arraycolsep}{4pt}\renewcommand{\arraystretch}{1.0}
{Bare Green’s function:}
\begin{align}
G_0^{-1}&(\boldsymbol{k},i\omega_n)= \nonumber \\
&\begin{pmatrix}
i\omega_n-\xi^{A}_{\boldsymbol{k}} & 0 & 0 & 0\\
0 & i\omega_n-\xi^{B}_{\boldsymbol{k}} & 0 & 0\\
0 & 0 & i\omega_n+\xi^{A}_{-\boldsymbol{k}} & 0\\
0 & 0 & 0 & i\omega_n+\xi^{B}_{-\boldsymbol{k}}
\end{pmatrix}.
\end{align}

{Spin current-loop part:}
\begin{equation}
\Sigma_{\phi_l}(\boldsymbol{k})=
\begin{pmatrix}
0 & \phi_{\boldsymbol{k}} & 0 & 0\\
\phi_{\boldsymbol{k}}^{*} & 0 & 0 & 0\\
0 & 0 & 0 & -\phi_{\boldsymbol{k}}^{*}\\
0 & 0 & -\phi_{\boldsymbol{k}} & 0
\end{pmatrix},
\end{equation}
with $\phi_{\boldsymbol{k}}=-8V_1 \sin \frac{k_x}{2} \sin \frac{k_y}{2} \phi_l.$

{Superconducting part:}
\begin{equation}
\Sigma_{\Delta}(\boldsymbol{k})=
\begin{pmatrix}
0 & 0 & \Delta_A(\boldsymbol{k}) & 0\\
0 & 0 & 0 & \Delta_B(\boldsymbol{k})\\
\Delta_A^{*}(\boldsymbol{k}) & 0 & 0 & 0\\
0 & \Delta_B^{*}(\boldsymbol{k}) & 0 & 0
\end{pmatrix}.
\end{equation}
\endgroup

In this basis, the spin current-loop field couples electrons between bands as
\[
\phi _l\sim c^\dagger_{A} c_{B},
\]
while the superconducting order parameters describe intra-band, spin-polarized triplet pairing,
\[
\Delta_{A}\sim c^\dagger_{A}c^\dagger_{A},\qquad
\Delta_{B}\sim c^\dagger_{B}c^\dagger_{B}.
\]

To examine the possibility of a cubic coupling, we compute the leading third-order correction to the free energy:
\[
F^{(3)} \sim T \sum_{\boldsymbol{k},n} \mathrm{Tr} \left[ G_0 \Sigma_{\phi_l} G_0 \Sigma_\Delta G_0 \Sigma_\Delta \right].
\]
However, due to the fermionic structure,  $\Sigma_{\phi_l}$ and $\Sigma_\Delta$ act on orthogonal sectors of Nambu space: $\Sigma_{\phi_l}$ connects particles on different bands, while $\Sigma_\Delta$ couples particle-particle or hole-hole sectors within the same band. As a result, their product has no nonzero diagonal components, and the trace vanishes.
The cubic coupling term between $\phi_l$ and $\Delta$ is strictly forbidden due to the incompatibility of spin and band indices. This implies that the leading-order coupling between spin current-loop fluctuations and superconductivity emerges only at quartic order, as we show below.

We find two distinct quartic coupling terms given by:
\begin{align}
F^{(4)}_1 &\sim T \sum_{\boldsymbol{k},n} \mathrm{Tr} \left[ G_0 \Sigma_{\phi_l}  G_0 \Sigma_{\phi_l} G_0 \Sigma_\Delta G_0 \Sigma_\Delta \right], \nonumber \\ 
F^{(4)}_2 &\sim T \sum_{\boldsymbol{k},n} \mathrm{Tr} \left[ G_0 \Sigma_{\phi_l} G_0 \Sigma_\Delta G_0 \Sigma_{\phi_l} G_0 \Sigma_\Delta \right] .
\end{align}
Therefore, the effective coupling between $\phi_l$ and the superconducting order parameters takes the form
\begin{equation}
\begin{aligned}
F_{\phi_l\text{–SC}}
&= \gamma_1 |\phi_l|^2\big(|\Delta_A^x|^2+|\Delta_B^y|^2\big)
 + \gamma_2 |\phi_l|^2\big(|\Delta_A^y|^2+|\Delta_B^x|^2\big)\\
&\quad + \Big[\gamma_3 \phi_l^2\big(\Delta_A^{x*}\Delta_B^{x}+\Delta_A^{y*}\Delta_B^{y}\big)+\text{c.c.}\Big]\\
&\quad + \Big[\gamma_4 \phi_l^2\big(\Delta_A^{x*}\Delta_B^{y}+\Delta_A^{y*}\Delta_B^{x}\big)+\text{c.c.}\Big].
\end{aligned}
\label{F_phi_sc}
\end{equation}
The coefficients $\gamma_{1,2}$ are given by
\begin{equation}
\gamma_{1,2} =\frac{T}{N_L}\sum_{k,n}g_{x,y}^2(k)D_k^2\frac{1}{(i\omega_n-\xi^A_k)^2(i\omega_n-\xi^B_k)(i\omega_n+\xi^A_k)}
\end{equation}
where $D_k=-8V_1\sin\frac{k_x}{2}\sin\frac{k_y}{2}$.

After performing the Matsubara frequency summation, this becomes
\begin{equation}
\begin{aligned}
\gamma_{1,2} = \frac{1}{N_L} &\sum_{\boldsymbol{k}} \frac{g_{x,y}^2(\boldsymbol{k}) D_{\boldsymbol{k}}^2}{2 \xi^A_{\boldsymbol{k}} (\xi^A_{\boldsymbol{k}} - \xi^B_{\boldsymbol{k}})} \\
\times \bigg[ &\frac{1}{T} n_F(\xi^A_{\boldsymbol{k}}) \left( n_F(\xi^A_{\boldsymbol{k}}) - 1 \right) + \frac{1}{2\xi^A_{\boldsymbol{k}}} \left( 1 - 2n_F(\xi^A_{\boldsymbol{k}}) \right) \\
& - \frac{n_F(\xi^B_{\boldsymbol{k}}) - n_F(\xi^A_{\boldsymbol{k}})}{\xi^B_{\boldsymbol{k}} - \xi^A_{\boldsymbol{k}}}
+ \frac{n_F(\xi^B_{\boldsymbol{k}}) - n_F(-\xi^A_{\boldsymbol{k}})}{\xi^B_{\boldsymbol{k}} + \xi^A_{\boldsymbol{k}}} \bigg],
\end{aligned}
\end{equation}
where $n_F(\xi)$ is the Fermi-Dirac distribution function.

Similarly, the coefficients $\gamma_{3,4}$ with $g_3=\sin k_x\sin k_x, g_4=\sin k_x\sin k_y$ are given by
\begin{equation}
\begin{aligned}
\gamma_{3,4} &= -\frac{T}{2N_L} \sum_{\boldsymbol{k},n} g_{3,4} D_{\boldsymbol{k}}^2 
\frac{1}{\left[ (i\omega_n)^2 - (\xi^A_{\boldsymbol{k}})^2 \right] \left[ (i\omega_n)^2 - (\xi^B_{\boldsymbol{k}})^2 \right]} \\
&= \frac{1}{2N_L} \sum_{\boldsymbol{k}} g_{3,4} D_{\boldsymbol{k}}^2 \frac{1}{(\xi^A_{\boldsymbol{k}})^2 - (\xi^B_{\boldsymbol{k}})^2}
\\
& \qquad \times \Big[  
\frac{\tanh \left( \xi^A_{\boldsymbol{k}} / 2T \right)}{2\xi^A_{\boldsymbol{k}}} 
- \frac{\tanh \left( \xi^B_{\boldsymbol{k}} / 2T \right)}{2\xi^B_{\boldsymbol{k}}}  \Big].
\end{aligned}
\end{equation}
Note that $\gamma_4 = 0$ due to mirror symmetry $M_y: k_x \to -k_x$. We retain the $\gamma_4$ term here for completeness.

\subsection{Renormalized free-energy}
To derive the Landau coefficients generated by the coupling between spin current-loop fluctuations and superconductivity, we begin by considering the regime near the superconducting transition temperature, where only the components $\Delta_A^x$ and $\Delta_B^y$ remain finite.  
The relevant part of the free energy, including the coupling to the spin current-loop order parameter $\phi_l$, reads:
\begin{equation}
\begin{aligned}
F(\phi_l)
&= F_{\phi_l} + F_{\phi_l\text{–SC}} \\ 
&= (1/2\chi_{\text{lp}}) |\phi_l|^2 
  + \gamma_1 |\phi_l|^2 \left( |\Delta_A^x|^2 + |\Delta_B^y|^2 \right) \\
&\quad + \left[ \gamma_4  \phi_l^2  \Delta_A^{x*}\Delta_B^y + \text{c.c.} \right],
\end{aligned}
\end{equation}
where $\chi_{\text{lp}}$ is the spin current-loop susceptibility.

We rewrite this as
\begin{equation}
F(\phi_l)  = a|\phi_l|^2 + b\phi_l^2 + b^*\phi_l^{*2},
\end{equation}
where
\begin{equation}
\begin{aligned}
a &= 1/2\chi_{\text{lp}}+ \gamma_1 \left( |\Delta_A^x|^2 + |\Delta_B^y|^2 \right), \\
b &= \gamma_{4} \Delta_A^{x*} \Delta_B^y.
\end{aligned}
\end{equation}
We assume there is no preexisting spin current-loop order and superconductivity does not induce an instability in $\phi_l$, i.e. $a>2|b|$. 

We now decompose $\phi_l$ into its real and imaginary parts: $\phi_l = \phi_1 + i \phi_2$. The free energy becomes
\begin{equation}
F(\phi_l) = (a + 2\text{Re}(b))\phi_1^2 + (a - 2\text{Re}(b))\phi_2^2 - 4\text{Im}(b)\phi_1 \phi_2,
\end{equation}
which we can write in matrix form:
\begin{align}
F(\phi_l) &=
\begin{pmatrix}
\phi_1 & \phi_2
\end{pmatrix}
\begin{pmatrix}
a + 2\text{Re}(b) & -2\text{Im}(b) \\
-2\text{Im}(b) & a - 2\text{Re}(b)
\end{pmatrix}
\begin{pmatrix}
\phi_1 \\ \phi_2
\end{pmatrix}
\nonumber \\
&\equiv 
\boldsymbol{\phi}^\top \mathbf{M} \boldsymbol{\phi}.
\end{align}
We integrate out the $\phi$ field to obtain the partition function:
\begin{equation}
Z = \int d\phi_1  d\phi_2 \exp\left( -\beta \boldsymbol{\phi}^\top \mathbf{M} \boldsymbol{\phi} \right)
= \frac{\pi}{\beta \sqrt{a^2 - 4|b|^2}}.
\end{equation}
Therefore, the effective free energy is
\begin{equation}
F_{\text{eff}} = -\frac{1}{\beta} \ln Z = \frac{1}{2\beta} \ln\left(a^2 - 4|b|^2\right) + \text{const}.
\end{equation}
To extract the superconducting contribution, we expand using $\ln(1 + x) \approx x - \frac{x^2}{2}$:
\begin{equation}
x = \frac{\gamma_1 U/\chi_{\text{lp}} + \gamma_1^2 U^2 - 4|\gamma_{4}|^2 V}{(1/2\chi_{\text{lp}})^2},
\end{equation}
where
\begin{equation}
U = |\Delta^x_A|^2 + |\Delta^y_B|^2, \qquad
V = |\Delta^x_A|^2  |\Delta^y_B|^2.
\end{equation}

Then the effective free energy becomes
\begin{equation}
\begin{aligned}
F_{\text{eff}} &= \frac{2\chi_{\text{lp}}\gamma_1}{\beta } \left( |\Delta^x_A|^2 + |\Delta^y_B|^2 \right) 
- \frac{2\chi_{\text{lp}}^2\gamma_1^2}{\beta} \left( |\Delta^x_A|^2 + |\Delta^y_B|^2 \right)^2 \\
&\quad - \frac{8\chi_{\text{lp}}^2\gamma_{4}^2}{\beta } |\Delta^x_A|^2  |\Delta^y_B|^2 + \mathcal{O}(\Delta^6).
\end{aligned}
\end{equation}
Thus, the coefficient of the inter-band coupling term $|\Delta^x_A|^2 |\Delta^y_B|^2$ is
\begin{equation}
-\frac{4\chi_{\text{lp}}^2}{\beta } \left( \gamma_1^2 + 2\gamma_{4}^2 \right),
\end{equation}
which is negative, and hence favors coexistence between the two superconducting components.

We now consider the low-temperature regime, where all four superconducting components,
$\Delta_A^x$, $\Delta_A^y$, $\Delta_B^x$, and $\Delta_B^y$, acquire finite magnitudes.

The coupling between the spin current-loop order parameter $\phi_l$ and the superconducting order parameters takes the general form given in Eq.~\ref{F_phi_sc}.

We rewrite this as
\begin{equation}
F(\phi_l)  = a|\phi_l|^2 + b\phi_l^2 + b^*\phi_l^{*2},
\end{equation}
where
\begin{equation}
\begin{aligned}
a &= 1/2\chi_{\text{lp}}
+ \gamma_1 \left( |\Delta_A^x|^2 + |\Delta_B^y|^2 \right)
+ \gamma_2 \left( |\Delta_A^y|^2 + |\Delta_B^x|^2 \right), \\
b &= \gamma_3 \left( \Delta_A^{x*}\Delta_B^x + \Delta_A^{y*}\Delta_B^y \right)
+ \gamma_4 \left( \Delta_A^{x*}\Delta_B^y + \Delta_A^{y*}\Delta_B^x \right).
\end{aligned}
\end{equation}
We assume there is no preexisting spin current-loop order.  The effective free energy is
\begin{equation}
F_{\text{eff}} = -\frac{1}{\beta} \ln Z = \frac{1}{2\beta} \ln\left(a^2 - 4|b|^2\right) + \text{const}.
\end{equation}
To extract the superconducting contribution, we expand using $\ln(1 + x) \approx x - \frac{x^2}{2}$.

The resulting quadratic terms in the superconducting free energy, together with their coefficients, are summarized as follows.

\paragraph*{Diagonal amplitude quadratic terms:}
\begin{equation}
\begin{aligned}
|\Delta_A^x|^2 + |\Delta_B^y|^2 &  \longrightarrow   \frac{2\chi_{\text{lp}}\gamma_1}{\beta}, \\ 
|\Delta_A^y|^2 + |\Delta_B^x|^2 &  \longrightarrow   \frac{2\chi_{\text{lp}}\gamma_2}{\beta}.
\end{aligned}
\end{equation}

The induced quartic terms in the superconducting free energy, with their corresponding coefficients, are summarized below.
\paragraph*{Diagonal amplitude quartic terms:}
\begin{equation}
\begin{aligned}
|\Delta_A^x|^4 + |\Delta_B^y|^4 
&  \longrightarrow   -\frac{2\chi_{\text{lp}}^2\gamma_1^2}{\beta}, \\ 
|\Delta_A^y|^4 + |\Delta_B^x|^4 
&  \longrightarrow   -\frac{2\chi_{\text{lp}}^2\gamma_2^2}{\beta}.
\end{aligned}
\end{equation}

\paragraph*{Mixed amplitude terms:}
\begin{equation}
\begin{aligned}
|\Delta_A^x|^2|\Delta_A^y|^2 + |\Delta_B^x|^2|\Delta_B^y|^2
&  \longrightarrow   -\frac{4\chi_{\text{lp}}^2\gamma_1\gamma_2}{\beta}, \\ 
|\Delta_A^x|^2|\Delta_B^y|^2
&  \longrightarrow   -\frac{2\chi_{\text{lp}}^2}{\beta}\bigl(2\gamma_1^2 + 4\gamma_4^2\bigr), \\ 
|\Delta_A^y|^2|\Delta_B^x|^2
&  \longrightarrow   -\frac{2\chi_{\text{lp}}^2}{\beta}\bigl(2\gamma_2^2 + 4\gamma_4^2\bigr), \\ 
|\Delta_A^x|^2|\Delta_B^x|^2 + |\Delta_A^y|^2|\Delta_B^y|^2
&  \longrightarrow   -\frac{2\chi_{\text{lp}}^2}{\beta}\bigl(2\gamma_1\gamma_2 + 4\gamma_3^2\bigr).
\end{aligned}
\end{equation}
\paragraph*{Cross-Terms:}
\begin{equation}
\begin{aligned}
\Re(\Delta_A^{x*} \Delta_B^x \Delta_A^y \Delta_B^{y*})
&\quad \longrightarrow \quad -\frac{2\chi_{\text{lp}}^2}{ \beta } \cdot 8 \gamma_{3}^2, \\ 
\Re(|\Delta_A^x|^2 \Delta_B^x \Delta_B^{y*})
&\quad \longrightarrow \quad -\frac{2\chi_{\text{lp}}^2}{ \beta } \cdot 8 \gamma_{3} \gamma_{4}, \\ 
\Re(\Delta_A^{x*} \Delta_A^y |\Delta_B^x|^2)
&\quad \longrightarrow \quad -\frac{2\chi_{\text{lp}}^2}{ \beta } \cdot 8 \gamma_{3} \gamma_{4}, \\ 
\Re(\Delta_A^{y*} \Delta_A^x |\Delta_B^y|^2)
&\quad \longrightarrow \quad -\frac{2\chi_{\text{lp}}^2}{ \beta }\cdot 8 \gamma_{3} \gamma_{4}, \\ 
\Re(|\Delta_A^y|^2 \Delta_B^y \Delta_B^{x*})
&\quad \longrightarrow \quad -\frac{2\chi_{\text{lp}}^2}{ \beta } \cdot 8 \gamma_{3} \gamma_{4}, \\ 
\Re(\Delta_A^{x*} \Delta_B^y \Delta_A^y \Delta_B^{x*})
&\quad \longrightarrow \quad -\frac{2\chi_{\text{lp}}^2}{ \beta } \cdot 8 \gamma_{4}^2.
\end{aligned}
\end{equation}
As mentioned earlier, $\gamma_4 = 0$ due to symmetry. However, it becomes relevant when certain symmetries are explicitly broken—for instance, by Rashba spin-orbit coupling, which breaks mirror symmetry.

In this work, we focus on the case without any additional symmetry breaking. 
Therefore, we retain only the term $\Re(\Delta_A^{x*} \Delta_B^x \Delta_A^y \Delta_B^{y*})$. 
As discussed in the mean-field analysis, the superconducting ground state on each band satisfies a relative phase 
$\beta^{A,B} = \arg(\Delta_{A,B}^y) - \arg(\Delta_{A,B}^x) = \pm \pi/2$, 
corresponding to four degenerate configurations of the form $(p \pm i\alpha  p) \otimes (\alpha p \pm i p)$.

The coupling term $\Re(\Delta_A^{x*} \Delta_B^x \Delta_A^y \Delta_B^{y*})$ 
introduces a phase-dependent contribution proportional to $-\cos(\beta^A - \beta^B)$, 
which energetically favors alignment of the phase differences between the two bands, i.e., $\beta^A = \beta^B$. 
This interaction lifts the degeneracy and stabilizes the states 
$(p + i \alpha p) \otimes (\alpha p + i p)$ and $(p - i \alpha p) \otimes (\alpha p - i p)$ as the ground states.

\section{Phase Diagram for finite $V_d$}\label{app:D}
In the main text, we focused on the case where the next-nearest-neighbor repulsion is the same along $x$ and $y$ , i.e. $V_d=0$. In this Appendix, we consider the more general situation with finite $V_d$. We first summarize the effects of nematic and spin current-loop fluctuations on the free energy. We then present the phase diagrams for finite $V_d$ at different chemical potentials, considering each type of fluctuation separately, and finally show the phase diagram in the presence of both types of fluctuations.

\subsection{Total Free energy}
Using the results of the previous Appendices, the bare, nematic, and spin current-loop contributions combine into the total superconducting free energy
\begin{equation}
\begin{aligned}
F &= 
  r_1 (|\Delta_A^x|^2 + |\Delta_B^y|^2)
  + r_2 (|\Delta_A^y|^2 + |\Delta_B^x|^2) \\
&+ u_1 (|\Delta_A^x|^4 + |\Delta_B^y|^4)
  + u_2 (|\Delta_A^y|^4 + |\Delta_B^x|^4) \\
&+ v_{xy} (|\Delta_A^x|^2|\Delta_A^y|^2 + |\Delta_B^x|^2|\Delta_B^y|^2)\\
 & + w_{xy}\left[(\Delta_A^x\Delta_A^{y*})^2 + (\Delta_B^x\Delta_B^{y*})^2 + \text{c.c.}\right] \\
&+ v_{AB} (|\Delta_A^x|^2|\Delta_B^x|^2 + |\Delta_A^y|^2|\Delta_B^y|^2)\\
  &+ v_{AB}^{xy} |\Delta_A^x|^2|\Delta_B^y|^2
  +  v_{AB}^{yx}|\Delta_A^y|^2|\Delta_B^x|^2 \\
&+w_{AB} (\Delta_A^{x*}\Delta_B^x\Delta_A^y\Delta_B^{y*} + \text{c.c.}).
\end{aligned}
\label{app:FreeEnergy}
\end{equation}

\paragraph*{Quadratic terms.}
\begin{equation}
\begin{aligned}
r_1 &= r_1^{(0)} + \frac{2\chi_{\text{lp}}\gamma_1}{\beta},\\
r_2 &= r_2^{(0)} + \frac{2\chi_{\text{lp}}\gamma_2}{\beta},
\end{aligned}
\end{equation}
where \(r_1^{(0)}\) and \(r_2^{(0)}\) are the bare quadratic coefficients obtained from the superconducting susceptibility.
The additional terms proportional to \(\gamma_{1,2}\) originate from the coupling to spin current-loop fluctuations.

\paragraph*{Intra-band quartic terms.}
\begin{equation}
\begin{aligned}
u_1 &= u_1^{(0)} - \frac{\chi_{\text{nem}}\lambda_1^2}{2} - \frac{2\chi_{\text{lp}}^2\gamma_1^2}{\beta},\\
u_2 &= u_2^{(0)} - \frac{\chi_{\text{nem}}\lambda_2^2}{2} - \frac{2\chi_{\text{lp}}^2\gamma_2^2}{\beta},\\
v_{xy} &= v_{xy}^{(0)} - {\chi_{\text{nem}}\lambda_1\lambda_2} - \frac{4\chi_{\text{lp}}^2\gamma_1\gamma_2}{\beta},\\
w_{xy} &= w_{xy}^{(0)} .
\end{aligned}
\end{equation}
Here the first corrections (\(\propto \lambda_i\)) come from nematic fluctuations,
while the second corrections (\(\propto \gamma_i\)) arise from spin current-loop fluctuations.

\paragraph*{Inter-band quartic terms.}
\begin{equation}
\begin{aligned}
v_{AB} &= {\chi_{\text{nem}}\lambda_1\lambda_2} - \frac{4\chi_{\text{lp}}^2(\gamma_1\gamma_2 + 2\gamma_3^2)}{\beta},\\
v_{AB}^{xy} &= {\chi_{\text{nem}}\lambda_1^2} - \frac{4\chi_{\text{lp}}^2\gamma_1^2}{\beta},\\
v_{AB}^{yx} &= {\chi_{\text{nem}}\lambda_2^2} - \frac{4\chi_{\text{lp}}^2\gamma_2^2}{\beta},\\
w_{AB} &= -\frac{8\chi_{\text{lp}}^2\gamma_3^2}{\beta}.
\end{aligned}
\end{equation}
The inter-band terms $v_{AB}$, $v_{AB}^{xy}$, and $v_{AB}^{yx}$ arise from nematic fluctuations and are positive, favoring competition between superconducting components and promoting nematic superconducting phases. 
The $w_{AB}$ term vanishes if only nematic fluctuations are taken into account, but becomes finite when spin current-loop fluctuations are present, selecting a chiral configuration and reducing the ground-state degeneracy.

We parametrize the order parameters as
\begin{align}
(\Delta^A_x,\Delta^A_y) 
= \Delta_A e^{i\theta_A}\big(\cos\alpha_A,  e^{i\beta_A}\sin\alpha_A\big), 
\nonumber \\
(\Delta^B_x,\Delta^B_y) 
= \Delta_B e^{i\theta_B}\big(\cos\alpha_B,  e^{i\beta_B}\sin\alpha_B\big).
\end{align}
where $\theta_s\in[0,2\pi)$, $\beta_s\in[-\pi,\pi)$, and $\alpha_s\in[0,\pi/2]$, with $s\in\{A,B\}$.

The resulting free energy takes the form
\begin{equation}
\begin{aligned}
F &=
r_1 \big(\Delta_A^2 \cos^2\alpha_A + \Delta_B^2 \sin^2\alpha_B\big) \\
&+ r_2 \big(\Delta_A^2 \sin^2\alpha_A + \Delta_B^2 \cos^2\alpha_B\big) \\
&+ u_1\big(\Delta_A^4 \cos^4\alpha_A + \Delta_B^4 \sin^4\alpha_B\big) \\
&+ u_2\big(\Delta_A^4 \sin^4\alpha_A + \Delta_B^4 \cos^4\alpha_B\big) \\
&+ v_{xy}
   \big(\Delta_A^4 \cos^2\alpha_A \sin^2\alpha_A  + \Delta_B^4 \cos^2\alpha_B \sin^2\alpha_B\big) \\
&+2w_{xy}\Delta_A^4 \cos^2\alpha_A \sin^2\alpha_A \cos(2\beta_A)\\
&+2w_{xy}\Delta_B^4 \cos^2\alpha_B \sin^2\alpha_B \cos(2\beta_B) \\
&+ \Delta_A^2 \Delta_B^2\big[v_{AB}
   \big( \cos^2\alpha_A \cos^2\alpha_B +  \sin^2\alpha_A \sin^2\alpha_B\big) \\
&\quad+ v_{AB}^{xy} \cos^2\alpha_A \sin^2\alpha_B
+v_{AB}^{yx} \sin^2\alpha_A \cos^2\alpha_B \\
&\quad+ w_{AB}   \cos\alpha_A \sin\alpha_A \cos\alpha_B \sin\alpha_B \cos(\beta_B - \beta_A)\big].
\end{aligned}
\end{equation}

In this representation, there are four $U(1)$ fields $\theta_A$,$\theta_B$, $\beta_A$ and  $\beta_B$.
The interaction terms constrain the relative phases $\beta_A$ and $\beta_B$.
The $w_{xy}$ term energetically selects $ \beta_A = \pm \tfrac{\pi}{2}, \beta_B = \pm \tfrac{\pi}{2}$, enforcing a relative phase of $\pi/2$ between the $x$ and $y$ components within each band sector.  As a result, only two continuous symmetries remain, $U(1)^s:\theta_s\to\theta_s+\theta^s_{0}$.

When the inter-band coupling $w_{AB}$ vanishes, corresponding to the absence of spin current-loop fluctuations, the free energy exhibits two independent spinless time-reversal symmetries, $\tilde{\mathcal T}^s:\ (\Delta_x^s,\Delta_y^s)\to(\Delta_x^{s*},\Delta_y^{s*})$.
A nonvanishing $w_{AB}$ term further locks the relative chiral phases between the two sectors by requiring $\beta_A=\beta_B$, reducing the symmetry to a single spinless time-reversal operation $\tilde{\mathcal T}$ acting simultaneously on both $A$ and $B$ sectors.

The altermagnetic normal state preserves the combined antiunitary symmetry $C_4\mathcal T$.
At the level of the Landau free energy, the theory is invariant under the discrete transformation $(\Delta_x^A,\Delta_y^A)\leftrightarrow (\Delta_y^B,\Delta_x^B)$, which exchanges the two band sectors together with the $x$ and $y$ components. This transformation squares to the identity and therefore constitutes a discrete $Z_2$ exchange symmetry. In the phases where the amplitudes obey $|\Delta_x^A|=|\Delta_y^B|$ and $|\Delta_y^A|=|\Delta_x^B|$, this $Z_2$ symmetry is unbroken. Strong nematic fluctuations enhance the competition between the four components and drive a spontaneous imbalance, $|\Delta_x^A|\neq|\Delta_y^B|$ and $|\Delta_y^A|\neq|\Delta_x^B|$, corresponding to a $C_4\mathcal T$-breaking ($Z_2$-breaking) superconducting state.

Consequently, the full symmetry structure relevant for the superconducting phase transitions in our system is $U(1)^A\times U(1)^B\times \tilde{\mathcal T}\times Z_2$.
The distinct phases appearing in the phase diagram can therefore be classified according to the symmetries that are spontaneously broken in each phase, which are summarized as follows.

\begin{figure}[t]
    \centering
    \includegraphics[width=0.9\linewidth]{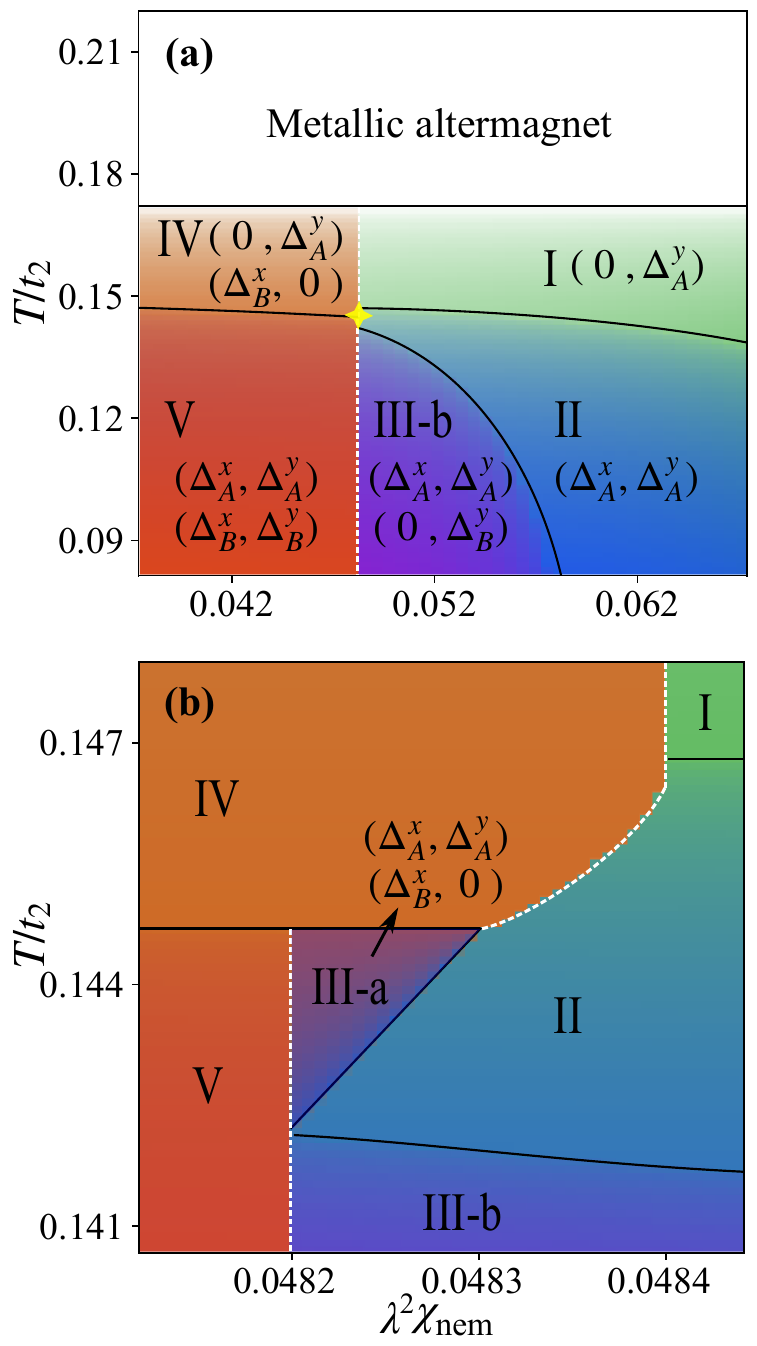}
\caption{
Superconducting phase diagrams as a function of the dimensionless temperature $T/t_2$ and the nematic susceptibility $\lambda^2 \chi_{\text{nem}}$ obtained for $\phi=-0.25$ and $V_{d}=-0.25$ at $\mu=-2.1$, where $\lambda$ denotes the average of $\lambda_1$ and $\lambda_2$.
Black solid lines indicate continuous transitions, while the white dashed line marks a first-order transition. 
Panel (b) is a zoom in the region around the yellow star in panel (a), revealing a small range where phase~III-a is realized.
Phase III-b is absent at $V_d=0$.
}
    \label{nem_Vd_2d1}
\end{figure}

\begin{figure}[t]
    \centering
    \includegraphics[width=0.9\linewidth]{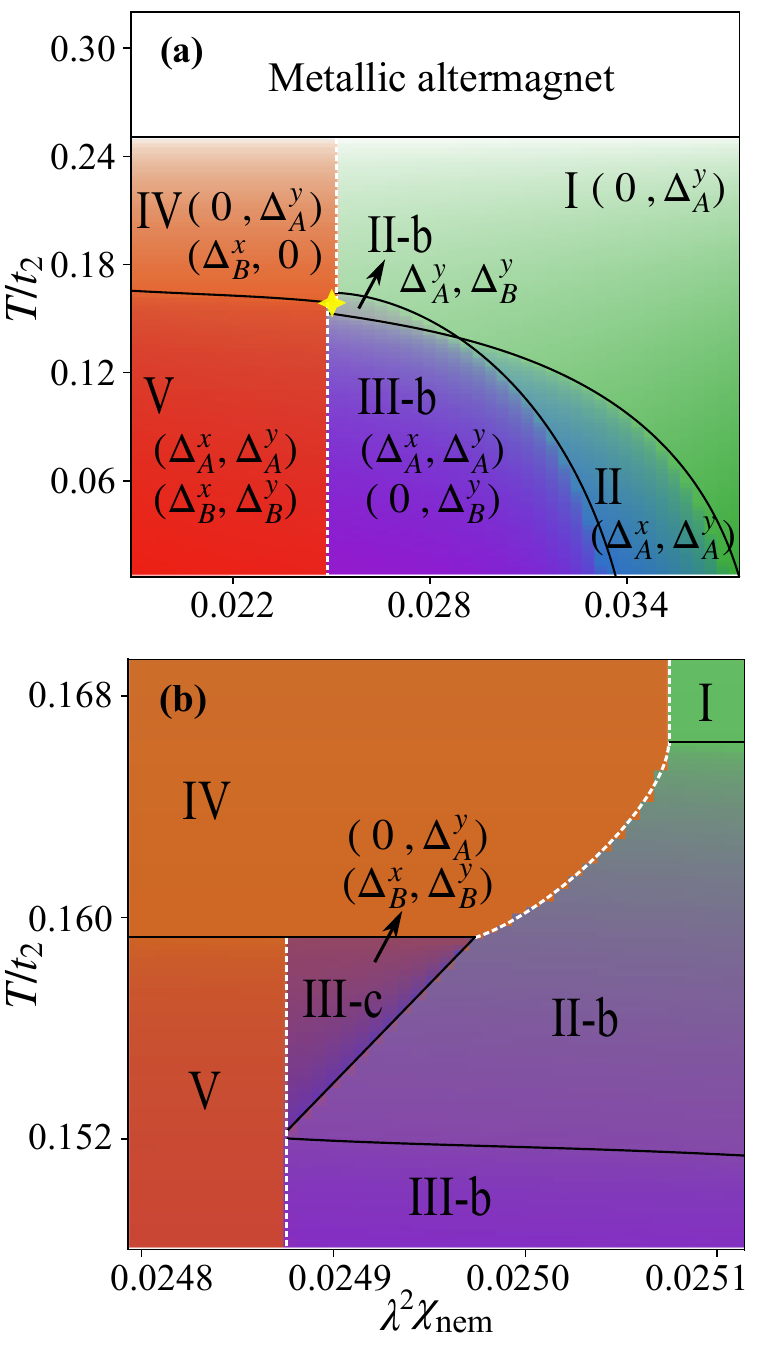}
\caption{
Same as in Fig.~\ref{nem_Vd_2d1}, but for chemical potential $\mu=-1.9$. 
In addition to the emergence of phases III-b and III-c, another superconducting phase II-b emerges.
}
    \label{nem_1d9}
\end{figure}

\begin{itemize}
    \item \textbf{With nematic fluctuations:}
    \begin{itemize}
        \item Phase~I: broken $U(1)^A \times Z_2$.
        \item Phase~II: broken $U(1)^A \times \tilde{\mathcal{T}} \times Z_2$.
        \item Phase~II-b: broken $U(1)^A \times U(1)^B \times Z_2$.
        \item Phases~III (a--d): broken $U(1)^A \times U(1)^B \times \tilde{\mathcal{T}} \times Z_2$.
        \item Phase~IV: broken $U(1)^A \times U(1)^B$.
        \item Phase~V: broken $U(1)^A \times U(1)^B \times \tilde{\mathcal{T}}$.
    \end{itemize}
    \item \textbf{With spin current-loop fluctuations:}
    \begin{itemize}
        \item Phase V-chiral: broken $U(1)^A \times U(1)^B \times \tilde{\mathcal{T}}$.
        \item Phase IV: broken $U(1)^A \times U(1)^B$.
    \end{itemize}
\end{itemize}

Here we describe the phases in terms of broken $U(1)$ symmetries; strictly speaking, in two dimensions this corresponds to quasi–long-range order of Berezinskii–Kosterlitz–Thouless type rather than true long-range order.

\begin{figure}[t]
    \centering
    \includegraphics[width=0.9\linewidth]{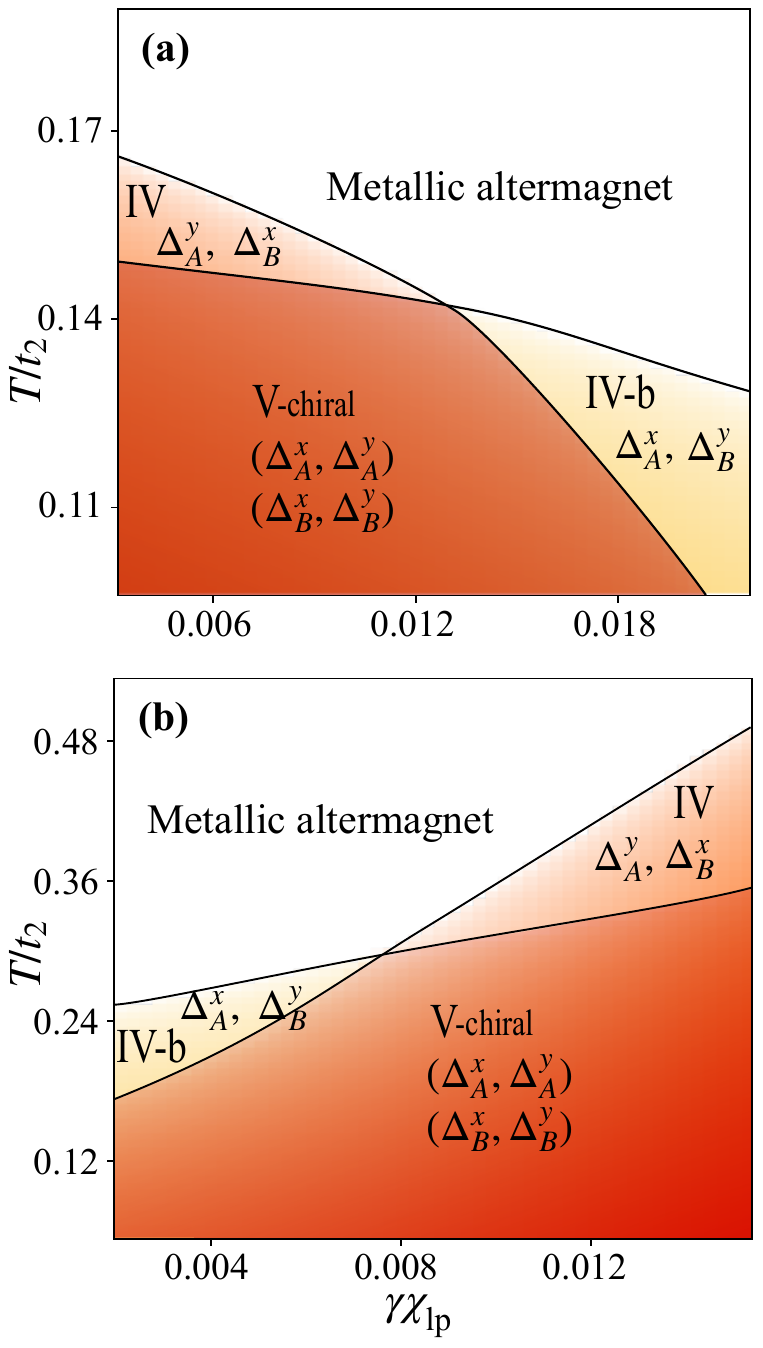}
    \caption{
    Superconducting phase diagrams including spin current-loop fluctuations for $\phi=-0.25$, and $V_{d}=-0.25$, where $\gamma$ denotes the average of $\gamma_1$ and $\gamma_2$. (a) Original parameter set: $\mu=-2.1$, $V_2=2.25$.
    (b) Alternative parameter set: $\mu=-2.8$, $V_2=3.75$.
    In (a) [(b)], spin current-loop fluctuations suppress (enhance) the superconducting transition temperature $T_c$. At low temperatures, all four superconducting components coexist. In (a) [(b)], the $(\Delta_A^x,\Delta_B^y)$ components vanish first upon increasing the temperature in the regime of small [large] $\chi_{\mathrm{lp}}$, whereas $(\Delta_A^y,\Delta_B^x)$ components vanish first in the regime of large [small] $\chi_{\mathrm{lp}}$. Phase V-chiral represents a chiral state, corresponding to either $(\epsilon p + i p)_A \otimes (p + i \epsilon p)_B$ or $(\epsilon p - i p)_A \otimes (p - i \epsilon p)_B$.
    }
    \label{lp_Vd_2d1}
\end{figure}

\begin{figure}[t]
    \centering
    \includegraphics[width=0.9\linewidth]{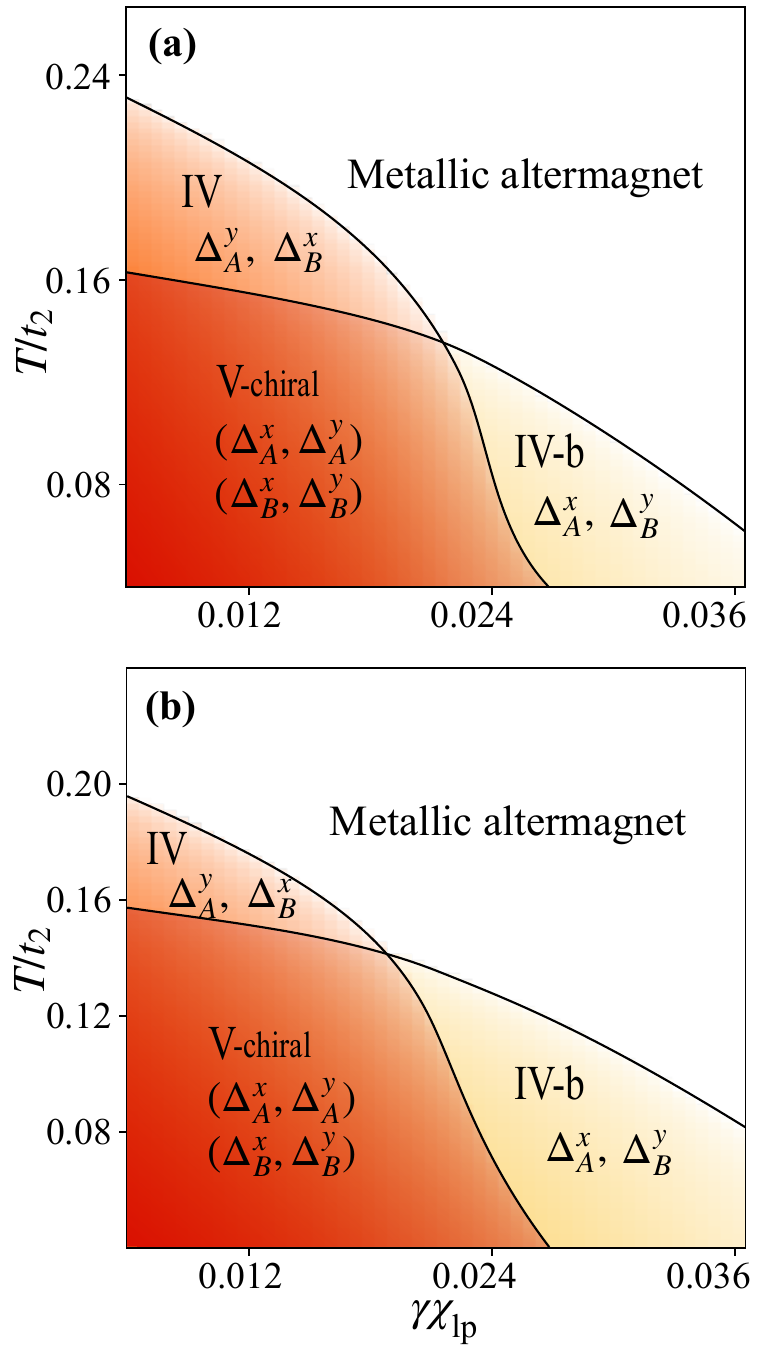}
    \caption{
    Superconducting phase diagrams in the presence of spin current-loop fluctuations for $\phi=-0.25$ and $V_d=-0.25$, where $\gamma$ denotes the average of $\gamma_1$ and $\gamma_2$. Panels (a) and (b) correspond to chemical potentials $\mu=-1.9$ and $\mu=-2.0$, respectively. The phase diagrams exhibit qualitatively similar behavior to the $\mu=-2.1$ case shown in Fig. \ref{lp_Vd_2d1} (a). For small $\chi_{\mathrm{lp}}$, the $(\Delta_A^y,\Delta_B^x)$ components persist to higher temperatures, whereas for large $\chi_{\mathrm{lp}}$, the $(\Delta_A^x,\Delta_B^y)$ components dominate at higher temperatures.
    }
    \label{lp_mu}
\end{figure}

\subsection{Phase diagrams at different chemical potentials}
\label{app:chempot}

We now illustrate how these symmetry-distinct phases are realized in concrete phase diagrams for finite pairing interaction anisotropy $V_d$ at different chemical potentials, first in the presence of nematic fluctuations and subsequently including spin current-loop fluctuations.

We begin by considering the effects of nematic fluctuations alone, which give rise to the phase diagram shown in Fig. \ref{nem_Vd_2d1}. The phase diagram is plotted as a function of temperature $T$ and nematic susceptibility $\chi_{\mathrm{nem}}$ for finite $V_d\neq 0$. Compared to the $V_d=0$ case discussed in the main text, an additional superconducting phase, labeled phase III-b and closely related to phase III-a, appears together with the previously identified phases. Black solid lines denote continuous phase transitions, while the white dashed line indicates a first-order transition separating distinct superconducting states. The yellow star marks a region where several superconducting phases converge. Figure \ref{nem_Vd_2d1} (b) shows a magnified view of this region, revealing a narrow phase III-a that is not visible on the scale of panel (a).
Importantly, the appearance of more than one type of superconducting phase~III within the same phase diagram is robust against small variations of model parameters. By tuning the chemical potential, we obtain a qualitatively different phase diagram, shown in Fig.~\ref{nem_1d9}. In addition to the emergence of phases~III-b and III-c, another superconducting state also appears, characterized by nonvanishing $(\Delta_A^y,\Delta_B^y)$ components and denoted as phase II-b.

\begin{figure}[t]
    \centering
    \includegraphics[width=0.9\linewidth]{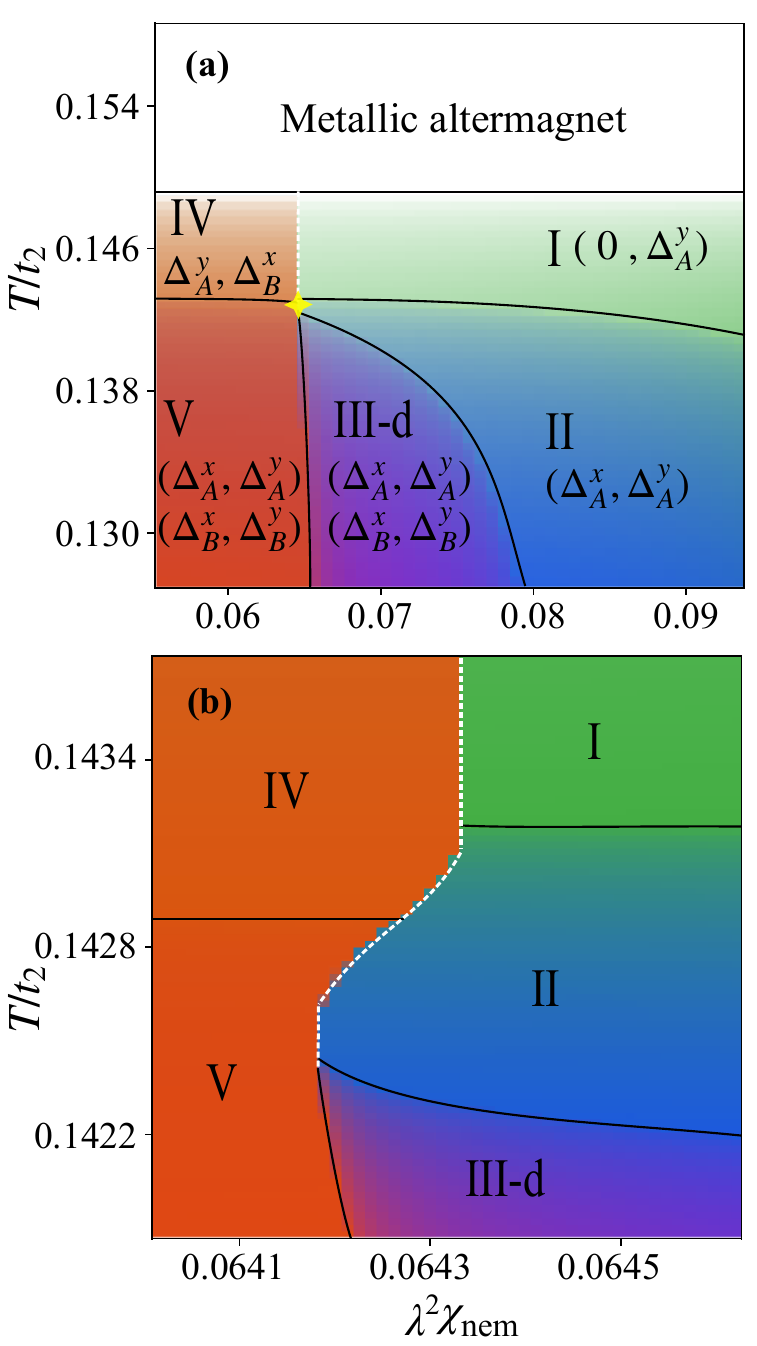}
    \caption{
    Superconducting phase diagram as a function of the dimensionless temperature $T/t_2$ and the nematic susceptibility $\lambda^2 \chi_{\text{nem}}$ for $\phi=-0.25$ and $V_{d}=-0.25$ at $\mu = -2.1$ with finite spin current-loop fluctuations $\chi_{\text{lp}}= 1/6$, where $\lambda$ denotes the average of $\lambda_1$ and $\lambda_2$.
    Black solid lines denote continuous transitions, while the white dashed line marks a first-order transition. The yellow star indicates the region where several phases meet; a zoomed-in view of this region is shown in panel (b). In the presence of finite spin current-loop fluctuations, the three-component phase disappears and is replaced by a four-component phase (labeled as Phase III-d). This phase differs from Phase V, as the exchange $Z_2$ symmetry is broken in Phase III-d, i.e., $|\Delta_A^x| \ne |\Delta_B^y|$ and $|\Delta_A^y| \ne |\Delta_B^x|$.
    }
    \label{nem1}
\end{figure}

\begin{figure}[t]
    \centering
    \includegraphics[width=0.9\linewidth]{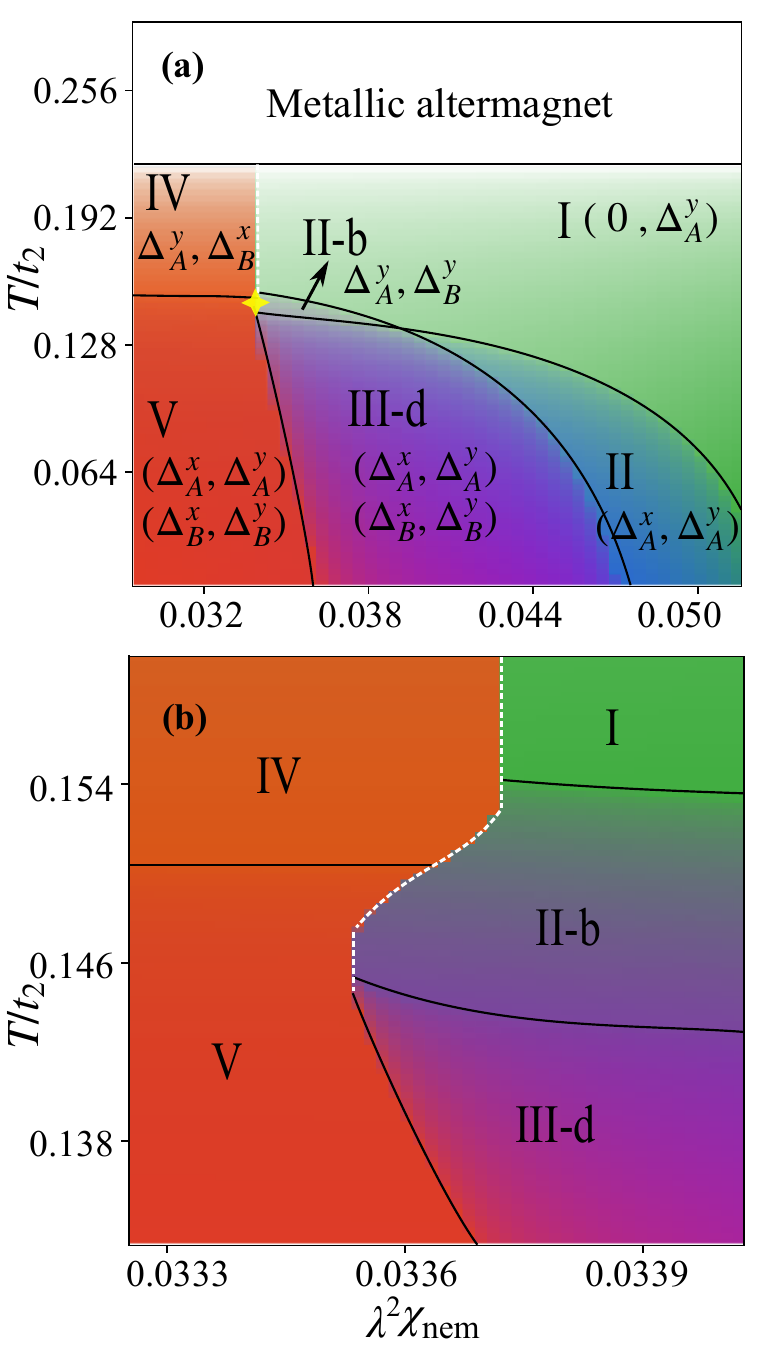}
    \caption{ Superconducting phase diagram as a function of the dimensionless temperature $T/t_2$ and the nematic susceptibility $\lambda^2 \chi_{\text{nem}}$ at $\mu = -1.9$ and $\chi_{\text{lp}}=1/6$. The yellow star indicates the region where several phases meet; a zoomed-in view of this region is shown in panel (b). 
    }
    \label{nem2}
\end{figure}

We next consider the effects of spin current-loop fluctuations at finite $V_d$. Figure \ref{lp_Vd_2d1} presents superconducting phase diagrams in the presence of spin current-loop fluctuations for finite $V_d$, with panels (a) and (b) corresponding to two different parameter sets. In panel (a), spin current-loop fluctuations suppress the superconducting transition temperature $T_c$, whereas in panel (b) they enhance it. In contrast to the $V_d=0$ case, the phase-transition lines intersect at finite $V_d$. Nevertheless, the overall structure of the phase diagram remains qualitatively similar upon varying parameters. In contrast to the nematic case, where $\mu = -2.1$ and $\mu = -1.9$ yield qualitatively different phase diagrams, the spin current-loop case exhibits similar qualitative behavior across nearby chemical potentials. 
For completeness, we provide the phase diagrams for $\mu = -1.9$ and $-2.0$ in Fig.~\ref{lp_mu}.

\subsection{Coexistence of nematic and spin current-loop fluctuations}
In the presence of only nematic fluctuations, the system can exhibit a sequence of three-step transitions upon increasing temperature: from a state with three nonzero superconducting components, to a two-component state, then to a one-component state, and finally to the normal phase.  
We now turn to the case in which nematic and spin current-loop fluctuations coexist. As we show below, the superconducting state with three nonzero components is no longer stable in the presence of spin current-loop fluctuations.

A phase with three nonzero superconducting components is only possible if the coupling term $w_{AB}$ vanishes.  
To see this, consider a representative configuration with $\rho^x_B = 0$ but $\rho_x^A \rho^y_A \rho^y_B \neq 0$,  
where $\rho_s^{x(y)}=|\Delta_s^{x(y)}|$ denotes the amplitude of the superconducting components.  
The stationarity condition is then
\[
\left.\frac{\partial F}{\partial \rho^x_B}\right|_{\rho^x_B=0} 
\sim w_{AB} \rho^x_A\rho^y_A\rho^y_B .
\]
For the three-component phase to be stable, this derivative must vanish, which requires $w_{AB} = 0$.  
This condition is satisfied in the absence of spin current-loop fluctuations, i.e., in the limit $\chi_{\text{lp}} \to 0$.

As illustrated in Fig.~\ref{nem1} and Fig.~\ref{nem2}, 
the system can no longer sustain a state with exactly three nonzero superconducting components. 
In particular, Phase~III, which previously contained three components (with $\Delta_B^x = 0$), is replaced by a phase in which all four components become finite, with $\Delta_B^x$ small but nonzero. 
Furthermore, the first-order transition between the four- and three-component states is replaced by a continuous transition within the four-component manifold, evolving from a state with $|\Delta_A^x| = |\Delta_B^y|$ and $|\Delta_A^y| = |\Delta_B^x|$ to one with $|\Delta_A^x| \ne |\Delta_B^y|$ and $|\Delta_A^y| \ne |\Delta_B^x|$.

\bibliography{ref}

\end{document}